\documentclass[twocolumn]{aastex631}
\usepackage{graphicx} 
\usepackage{amsmath}
\usepackage{afterpage}

\definecolor{ps}{HTML}{ad50ff}

\usepackage{placeins}

\newcommand{\HI}{H\tiny{ }\footnotesize{I}\normalsize{ }}

\newcommand{\kms}{\ifmmode\,{\rm km}\,{\rm s}^{-1}\else km$\,$s$^{-1}$\fi}
\newcommand{\Lya}{\ifmmode\,{\rm Ly}{\rm \alpha}\else Ly$\alpha$\fi}
\newcommand{\Msun}{\mathrm{M}_{\sun}}

\begin{document}

\title{JWST absorption line spectroscopy with SPURS: ISM covering fractions and kinematics in individual galaxies at $z=5-9$} 
\shortauthors{Vasan G.C. et al.}

\author[0000-0002-2645-679X]{Keerthi Vasan G.C.}
\affiliation{The Observatories of the Carnegie Institution for Science, 813 Santa Barbara Street, Pasadena, CA 91101, USA}
\correspondingauthor{Keerthi Vasan G.C.}
\email{kvch153@gmail.com}

\author[0000-0002-9132-6561]{Peter Senchyna}
\affiliation{The Observatories of the Carnegie Institution for Science, 813 Santa Barbara Street, Pasadena, CA 91101, USA}

\author[0000-0002-3407-1785]{Charlotte A. Mason}
\affiliation{Cosmic Dawn Center (DAWN)}
\affiliation{Niels Bohr Institute, University of Copenhagen, Jagtvej 128, 2200 Copenhagen N, Denmark}

\author[0000-0002-2178-5471]{Zuyi Chen}
\affiliation{Cosmic Dawn Center (DAWN)}
\affiliation{Niels Bohr Institute, University of Copenhagen, Jagtvej 128, 2200 Copenhagen N, Denmark}

\author[0000-0001-6106-5172]{Daniel P. Stark}
\affiliation{Department of Astronomy, University of California, Berkeley, Berkeley, CA 94720, USA}

\author[0000-0001-5860-3419]{Tucker Jones}
\affiliation{Department of Physics and Astronomy, University of California, Davis, 1 Shields Avenue, Davis, CA 95616, USA}

\author[0000-0003-1432-7744]{Lily Whitler}
\affiliation{Kavli Institute for Cosmology, University of Cambridge, Madingley Road, Cambridge, CB3 0HA, UK}
\affiliation{Cavendish Laboratory, University of Cambridge, JJ Thomson Avenue, Cambridge, CB3 0US, UK}

\author[0000-0002-4453-5870]{Kelsey S. Glazer}
\affiliation{Department of Physics and Astronomy, University of California, Davis, 1 Shields Avenue, Davis, CA 95616, USA}

\author[0000-0002-6290-3198]{Manuel Aravena}
\affiliation{Instituto de Estudios Astrof\'{\i}cos, Facultad de Ingenier\'{\i}a y Ciencias, Universidad Diego Portales, Av. Ej\'ercito 441, Santiago, Chile}
\affiliation{Millenium Nucleus for Galaxies (MINGAL)}

\author[0000-0003-3926-1411]{Jorge González-López}
\affiliation{Instituto de Astrofísica, Facultad de Física, Pontificia Universidad Católica de Chile, Santiago 7820436, Chile}

\author[0000-0003-4564-2771]{Ryan Endsley}
\affiliation{Department of Astronomy, The University of Texas at Austin, Austin, TX 78712, USA}
\affiliation{Cosmic Frontier Center, The University of Texas at Austin, Austin, TX 78712, USA}  

\author[0000-0001-5487-0392]{Viola Gelli}
\affiliation{Cosmic Dawn Center (DAWN)}
\affiliation{Niels Bohr Institute, University of Copenhagen, Jagtvej 128, 2200 Copenhagen N, Denmark}

\author[0000-0001-5940-338X]{Mengtao Tang}
\affiliation{Tsung-Dao Lee Institute, Shanghai Jiao Tong University, 1 Lisuo Road, Shanghai 201210, People’s Republic of China}
\affiliation{School of Physics and Astronomy, Shanghai Jiao Tong University, 800 Dongchuan Road, Shanghai 200240, People’s Republic of China}
\affiliation{State Key Laboratory of Dark Matter Physics, Shanghai Jiao Tong University, 1 Lisuo Road, Shanghai 201210, People’s Republic of China}

\author[0000-0001-8426-1141]{Michael W. Topping}
\affiliation{Steward Observatory, University of Arizona, 933 N Cherry Avenue, Tucson, AZ 85721, USA}

\begin{abstract}
We present deep rest-ultraviolet (UV) spectra of six luminous $z=5$ -- 9  galaxies in the Abell-2744 field taken as part of the JWST Cycle 4 Large Program SPURS. 
The individual galaxy spectra show unambiguous detections of interstellar medium (ISM) metal absorption lines from low- and high-ionization states of enriched gas, which we use to probe the ISM gas porosity and kinematics. 
We find a striking diversity in the absorption profiles.
We find low-ionization gas covering fractions ranging from 0.2 to 0.9, indicating a heterogeneous and patchy neutral ISM.
The low-ionization kinematics also show a large diversity, with velocity centroid values ranging from $+$70 to a significantly blueshifted $-140$ km$\,$s$^{-1}$, while the high-ion gas shows mostly blueshifted absorption, indicating the presence of multiphase outflows.
While all sources show outflow signatures in blueshifted wings, we also find that half of our sample, in particular those with the lowest stellar masses and highest sSFRs, have low-ionization velocity centroids close to systemic velocities.
This is in contrast to near-ubiquitous bulk low-ionization gas outflows at lower redshifts.
We suggest that this diversity of kinematics may be due to the bulk of the cold gas having low outflow velocities in the lowest mass and highest sSFR systems, potentially due to inefficient entrainment and/or an unresolved infalling component.
These spectra reveal a metal-enriched ISM with complex gas geometry and kinematics, and highlight the potential of deep JWST grating spectroscopy to reveal the properties of the ISM during the reionization era.  
\end{abstract}


\section{Introduction}

Understanding the formation and evolution of the first galaxies, and how they drove the reionization of intergalactic hydrogen in the first billion years, is a long-standing goal of extra-galactic astronomy \citep{Robertson2022_ARA,starkObservationsFirstGalaxies2025}.
Feedback from massive stars and the cycling of baryons between galaxies and their surroundings are thought to be critical for driving these processes \citep[see e.g.,][for recent reviews]{erbFeedbackLowmassGalaxies2015,tumlinson_CGM_review,peroux2020,ClaudeAndre2023}. 
Feedback-driven outflows and gas inflows from the cosmic web shape the structure of the interstellar medium (ISM), and are therefore thought to play a central role in the regulation of star formation and the escape of ionizing photons in early galaxies \citep[e.g.,][]{Kimm2019,Ma2020_FIRE2,Kakiichi_lyC_escape}.
At the peak of cosmic star formation ($z\sim2$), outflows are almost ubiquitous in star-forming galaxies  \citep[e.g.,][]{shapley2003, steidel2010}, meanwhile the average ionizing photon escape fraction -- a key unknown for understanding reionization -- is generally low \citep[$<10\%$; e.g.][]{steidelKeckLymanContinuum2018,pahlUncontaminatedMeasurementEscaping2021,begleyVANDELSSurveyMeasurement2022,jungConstraintsLymanContinuum2024}.
If similar ISM conditions persist at higher redshifts, this implies a dominant contribution from faint galaxies below our current detection limits to complete reionization by $z\sim5.5$ \citep[e.g.][]{finkelsteinConditionsReionizingUniverse2019,atekMostPhotonsThat2024,gelliImpactMassdependentStochasticity2024,starkObservationsFirstGalaxies2025}.
However, direct constraints on the baryon cycle and ISM structure in galaxies {\it during} the Epoch of Reionization ($z>5$) have long been out of reach observationally.

The James Webb Space Telescope (JWST) has opened a new window on galaxy formation in this era, providing our first direct insights into the earliest galaxies.
As spectroscopic samples of $z>5$ galaxies have grown, early results have revealed several surprises relative to expectations from lower redshift, including the confirmation of a significant number of UV-luminous galaxies at $z>10$ \citep[e.g.,][]{Castellano2023_glass,Carniani2024,momz14_rohan}, well in excess of pre-JWST theoretical models \citep[e.g.,][]{Mason2015_UVluminosity,tacchellaRedshiftindependentEfficiencyModel2018,Yung2020}; detections of Lyman-$\alpha$ (\Lya) emission at $z>10$, when the Intergalactic Medium (IGM) was expected to be almost fully neutral \citep{bunker_gnz11,Witstok2024}; and a large diversity of emission line strengths, indicating extremely `bursty' star formation \citep{Endsley2024,Endsley_2025_diversity_sfh,Mengtao_z_9_14}.
The origin of these results remains debated, but they suggest physical conditions and feedback processes that may differ substantially from what is typically seen at lower redshifts \citep[e.g.,][]{fire2_stochastic_SFR,density_modeulated_SFE_rachel_somerville}.
If star formation conditions and feedback processes evolve in the first billion years, we may also expect to see evolution in the ISM structure, which can be directly probed with JWST spectroscopy.

Rest-frame ultra-violet (UV) spectroscopy provides a powerful probe of ISM conditions and the baryon cycle via absorption lines, which are shaped by feedback-driven outflows and inflowing gas \citep[e.g.,][]{shapley2003}.
Outflowing gas produces blueshifted absorption, which is seen in nearly all $z\sim2$ star-forming galaxies across both low- and high-ionization metal species \citep[e.g.,][]{steidel2010}.
Work at low redshifts has demonstrated outflows are linked to high star formation rate surface densities \citep[$\Sigma_{\rm SFR} \gtrsim 0.1 M_\odot{\rm yr}^{-1}{\rm kpc}^{-2}$; e.g.,][]{heckman1990, Bron_duvet_outflows}, as would be expected if stellar (supernova) feedback drives outflows.
Luminous galaxies ($M_{UV}\lesssim-20$) at $z>5$ exceed this star formation rate surface density threshold \citep[e.g.,][]{Morishita2024_sizeanalysis, Yoshiaki2024_sizemorphology, FRESCO_slitless_SFR} and thus we might expect them to drive outflows, but the character of feedback in such early systems remains a significant source of debate \citep[e.g.][]{dekelEfficientFormationMassive2023,boylan-kolchinAcceleratedDarkMatter2025,ferraraNoBlueRed2026}.

In addition to tracing gas kinematics, low-ionization absorption lines also constrain the porosity of neutral gas in the ISM, therefore providing one of our best indirect probes of the escape of hydrogen-ionizing (Lyman continuum, LyC) photons \citep[e.g.,][]{jones2012,nicha2016-escape-fraction,Pahl_2020}, which remains a key uncertainty in understanding the reionization process \citep[e.g.][]{naiduRapidReionizationOligarchs2020,robertsonGalaxyFormationReionization2022,jaskotIonizingRadiationEscape2025}.
Direct measurements of Lyman continuum photons are essentially impossible during the reionization era, due to the high opacity of the IGM at $z\gtrsim4$ \citep[e.g.,][]{Inoue2014}. However, significant effort to establish indirect indicators of LyC leakage using $z\sim0-3$ samples has demonstrated that the covering fraction of low-ionization absorption lines is among the best tracers of LyC leakage \citep[e.g.,][]{Reddy2016,Reddy2022,Gazagnes2018,Saldana-Lopez2022}.

However, until recently, these absorption lines have been inaccessible for individual galaxies in the reionization era, limiting our understanding of ISM kinematics and porosity in the first billion years.
Pre-JWST, continuum spectroscopy of $z>5$ galaxies has been impossible from the ground, and most existing JWST spectroscopy has been taken with the low resolution NIRSpec prism and/or only shallow grating spectroscopy. 
Recently, \citet{pancakez_kelsey} demonstrated the potential of JWST to reveal absorption lines in $z\geq6$ galaxies using stacked grating spectra. 
Their analysis demonstrated that the average EW of low-ionization absorption lines is lower than seen at $z\sim2-5$ \citep[e.g.,][]{jones2012, du2018, Pahl_2020}, suggesting a reduced average covering fraction of ISM gas and consequently a potentially higher ionizing photon escape fraction. 
Meanwhile, the peak absorption line velocity in their stack was found to be lower than typically observed at $z<6$, potentially implying weaker or less prevalent outflows. 
However, these stacking analyses are limited in their ability to capture the diversity in ISM properties between galaxies, and their correlations with other physical properties, which are important for understanding what shapes the structure of the ISM and escape of ionizing photons.

Progress requires deeper rest-frame UV grating spectroscopy to detect ISM absorption lines in \textit{individual galaxies}. Motivated by this, we are conducting the SPectroscopic Ultra-deep Reionization-era Survey (SPURS, GO 9214, PIs Mason and Stark), to obtain the first ultra-deep (29\,hr) $R\sim1000$ grating observations of the rest-frame UV of reionization-era galaxies, using the JWST/NIRSpec micro-shutter assembly (MSA)  \citep{NIRSpec_Jakobsen, NIRSpec_boker}.
In this work, we use observations in the first SPURS field, Abell 2744, to characterize ISM absorption lines in individual $z>5$ galaxies for the first time. In particular, we present SPURS observations of six luminous galaxies selected for their high achieved continuum signal-to-noise (SNR) and investigate the ISM gas covering fractions and kinematics revealed by their absorption lines.

This paper is organized as follows.
Section~\ref{sec:data} describes the new SPURS observations and ancillary data used for our investigation. Section~\ref{sec:kinematics-and-coveringfraction} presents our methods and characterization of the absorption lines. 
We present our results in Section~\ref{sec:trends} and discuss the implications in Section~\ref{sec:discussion}. We summarize our conclusions in Section~\ref{sec:conclusion}.
Throughout this paper, we adopt $\Lambda$CDM cosmology with $\Omega_m=0.3$, $\Omega_\Lambda=0.7$ and $H_0=70$ \kms Mpc$^{-1}$.
All the magnitudes are reported in the AB system \citep{oke_gunn_ab_magnitude}. 

\begin{figure*}[!ht]
    \centering
    \includegraphics[width=\linewidth]{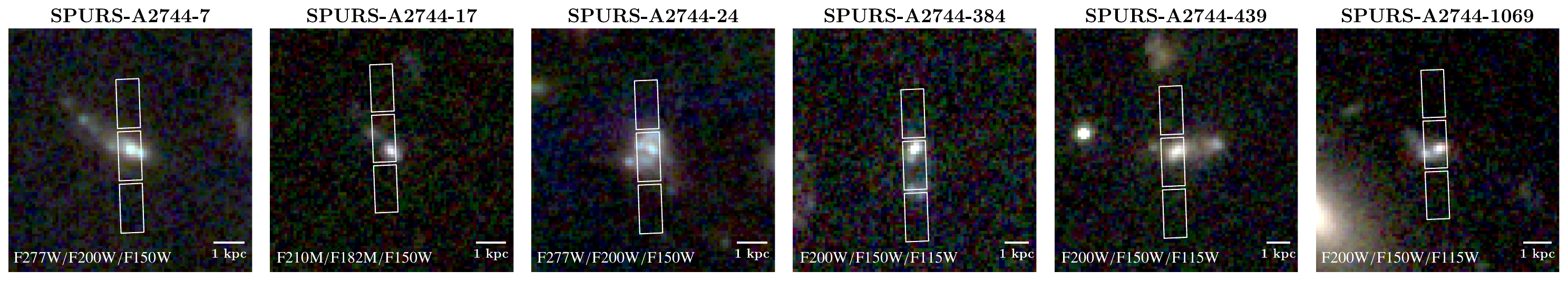}
    \begin{tabular}{cc}
     & \\ 
       \includegraphics[width=0.485\linewidth]{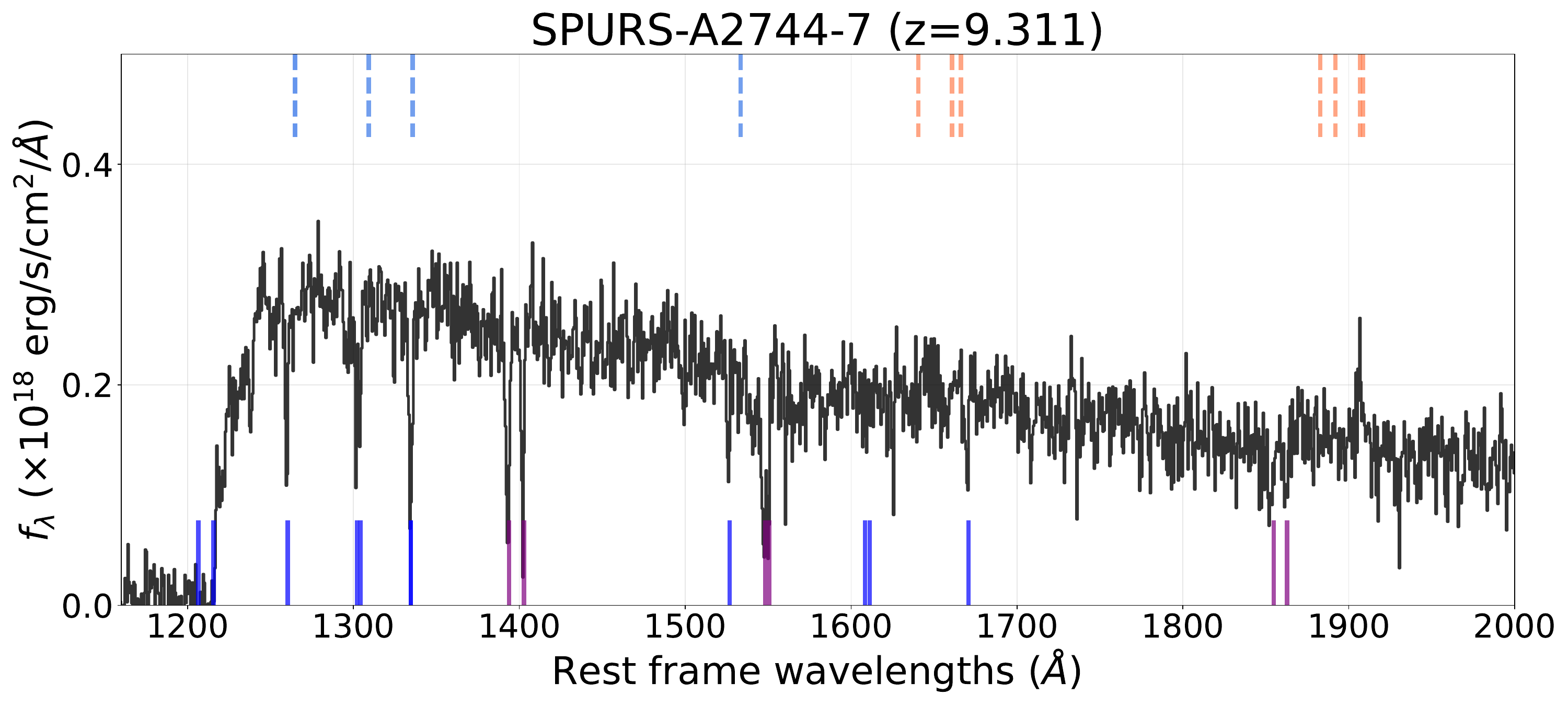}  & \includegraphics[width=0.485\linewidth]{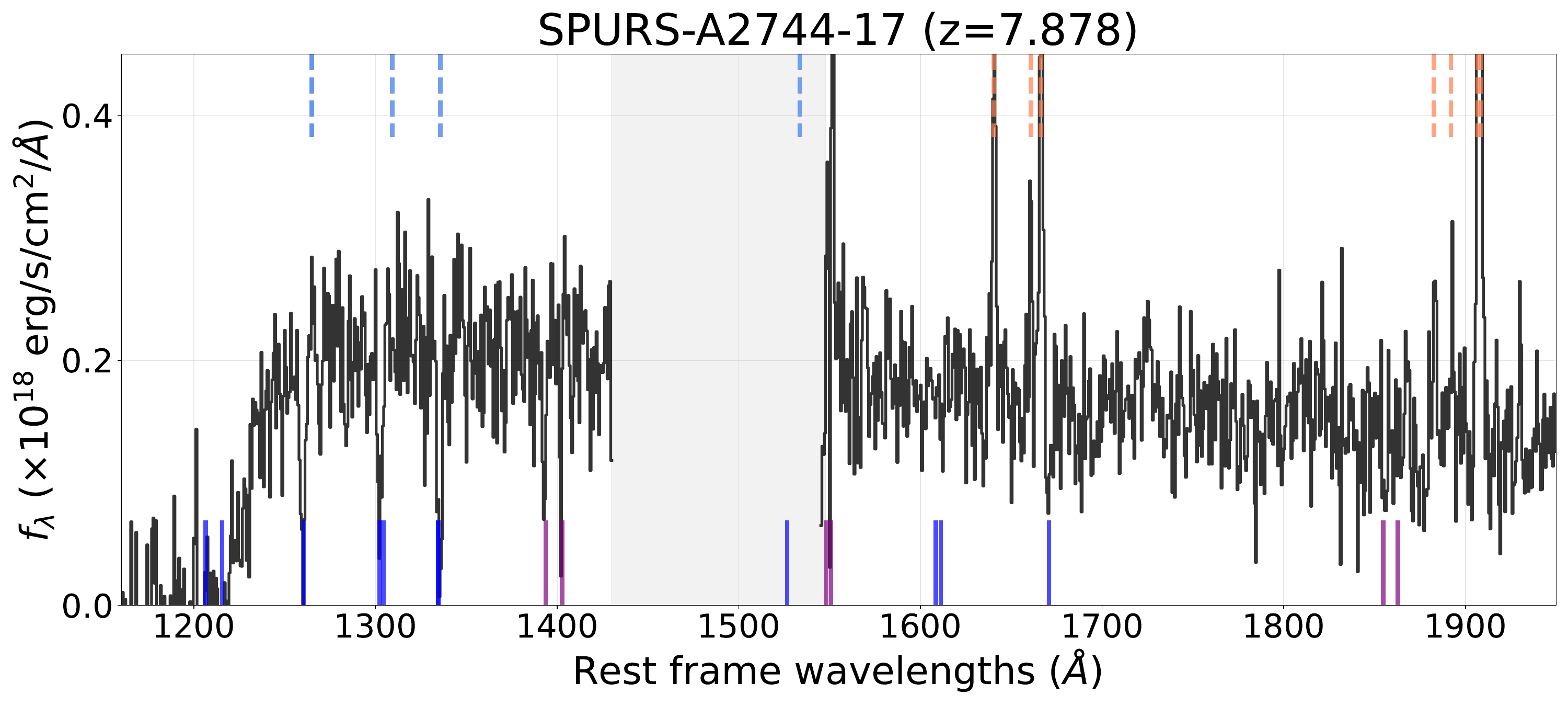} \\ 
       \includegraphics[width=0.485\linewidth]{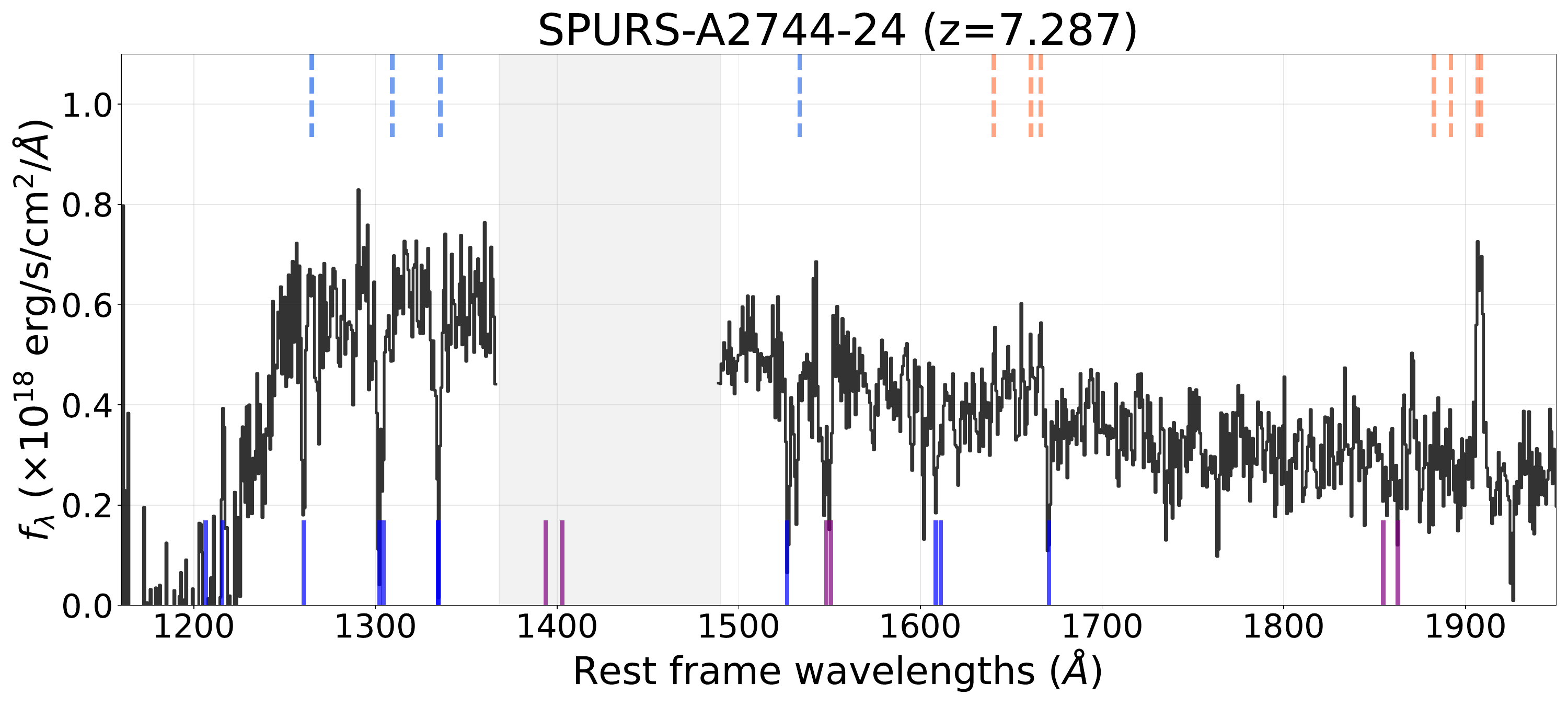}  & \includegraphics[width=0.485\linewidth]{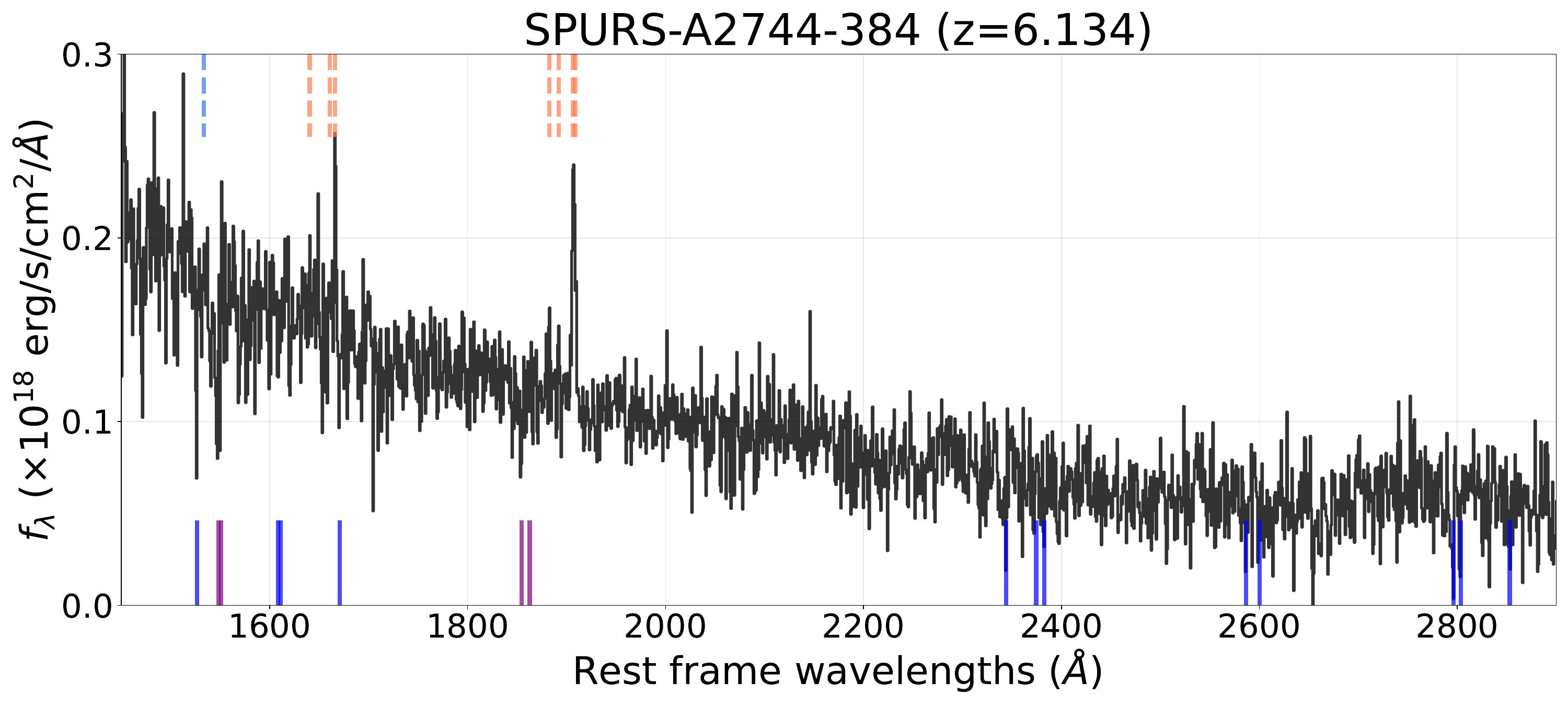} \\ 
       \includegraphics[width=0.485\linewidth]{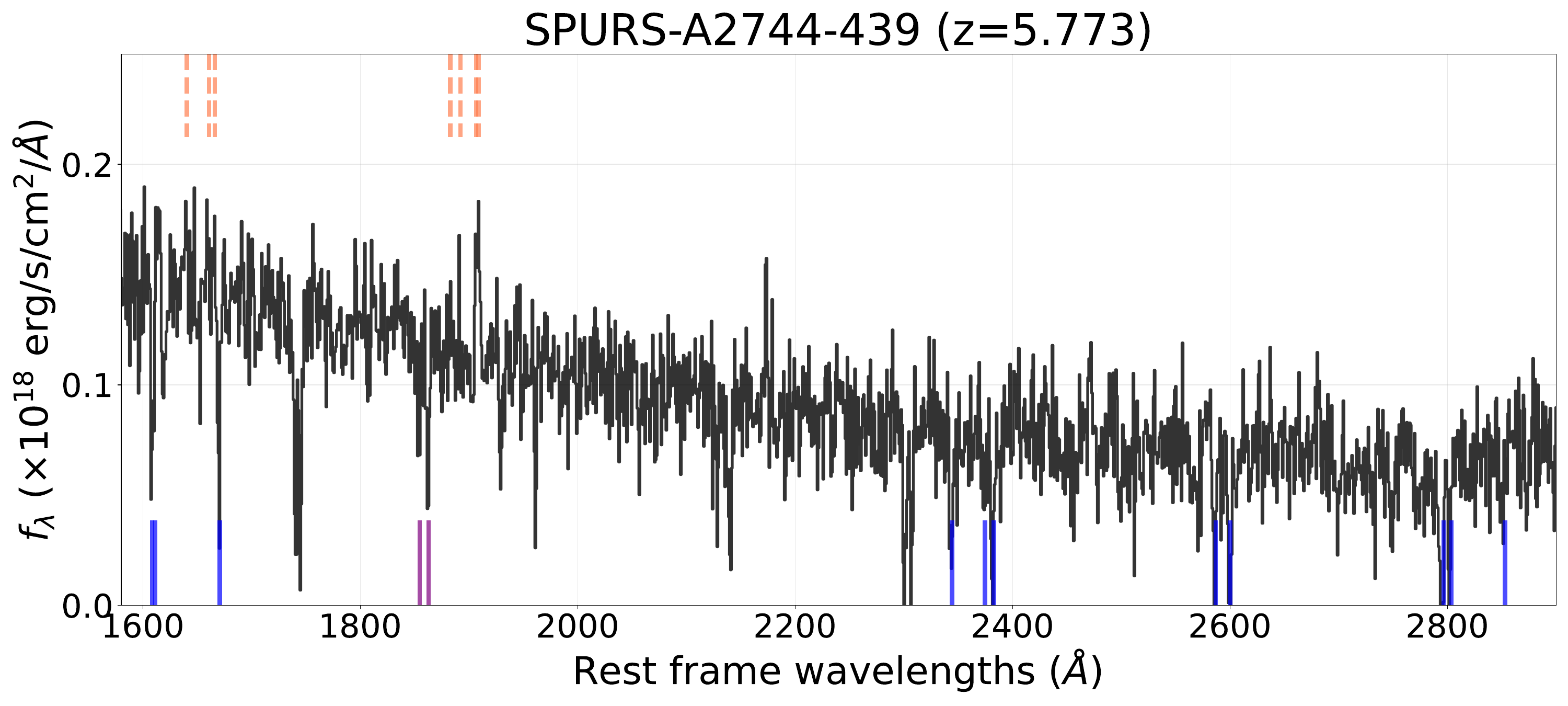}  & \includegraphics[width=0.485\linewidth]{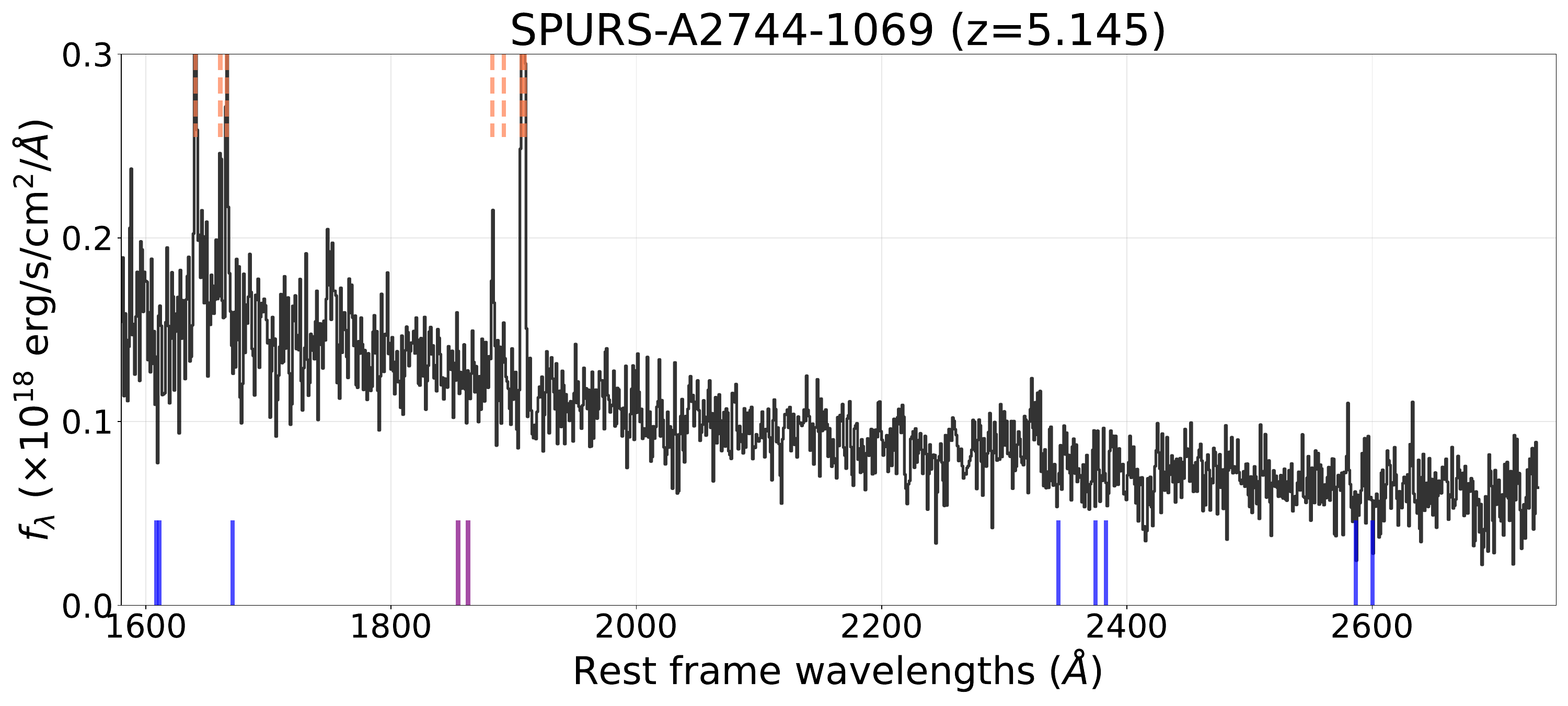} 
    \end{tabular}
    \includegraphics[width=0.7\linewidth]{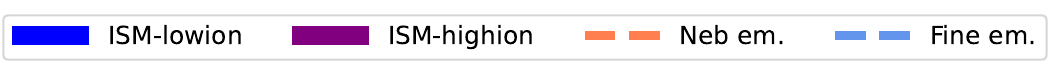} 
    \caption{\emph{Top panel}: $2\farcs\times2\farcs$ NIRCam color composite images probing the rest-frame UV wavelengths of all the galaxies in SPURS-A2744 pointing which have $z>5$ and continuum SNR$\geq3.5$. The NIRSpec shutter positions are overlaid in white. \emph{Bottom panels}: Deep G140M spectra obtained from SPURS. The gray regions in each panel denote regions of detector gap. Prominent low-ionization and high-ionization ISM absorption lines are marked in blue and purple respectively. Nebular and fine structure emission lines are marked as dashed orange and blue lines respectively. 
    }
    \label{fig:nircam-image}
\end{figure*}

\begin{deluxetable*}{cccccccccccc} \label{tab:galaxy-properties-basic}
\tablecaption{Table of galaxy properties. Stellar masses and SFR are derived from SED modeling assuming  a constant star formation history (Section~\ref{subsec:galprops}). All reported values have been corrected for the lensing magnification ($\mu$).}
\tablecolumns{12}
\tablehead{\colhead{objid} & \colhead{RA} & \colhead{Dec} & \colhead{z$_{\rm spec}$} &  \colhead{$M_{UV}$}  & \colhead{log(M$_*$)} & \colhead{log(SFR)$^a$} &  \colhead{sSFR}$^b$ &  \colhead{EW(H$\beta$)} & \colhead{EW(\ion{C}{3}])$^c$} & \colhead{$\mu^d$} & \colhead{Area$_{\rm eff}^e$}\\
 & ($\alpha$) & ($\delta$) & & & ($\Msun$) &  ($\Msun$ yr$^{-1}$) & (Gyr$^{-1}$) &  (\AA) & (\AA) &   &  (kpc$^2$) }
\startdata
7 & 3.61717 & -30.42555 & 9.31102 & -21.59 & $9.15_{-0.20}^{+0.20}$ & $1.09_{-0.04}^{+0.05}$ & $6_{-2}^{+5}$ & $27_{-4}^{+4}$ & 1.35 & 1.63 & 0.12 \\
17 & 3.60450 & -30.38046 & 7.87780 & -20.53 & $7.93_{-0.14}^{+0.05}$ & $0.94_{-0.14}^{+0.04}$ & $195_{-54}^{+35}$ & $567_{-94}^{+167}$ & 12.60 & 1.89 & 0.10 \\
24 & 3.60850 & -30.41853 & 7.28726 & -21.27 & $7.88_{-0.02}^{+0.14}$ & $0.88_{-0.02}^{+0.14}$ & $374_{-20}^{+38}$ & $90_{-2}^{+3}$ & 6.24 & 2.20 & 0.22 \\
384 & 3.62074 & -30.39612 & 6.13372 & -20.21 & $8.24_{-0.11}^{+0.08}$ & $0.70_{-0.06}^{+0.05}$ & $25_{-5}^{+7}$ & $75_{-3}^{+3}$ & 4.20 & 1.59 & 0.13 \\
439 & 3.61924 & -30.39446 & 5.77315 & -20.14 & $8.80_{-0.12}^{+0.11}$ & $1.12_{-0.04}^{+0.04}$ & $17_{-5}^{+8}$ & $45_{-1}^{+1}$ & 2.33 & 1.64 & 0.27 \\
1069 & 3.58381 & -30.37452 & 5.14519 & -19.32 & $7.45_{-0.05}^{+0.14}$ & $0.45_{-0.05}^{+0.14}$ & $262_{-107}^{+72}$ & $196_{-3}^{+3}$ & 11.47 & 2.72 & 0.13 \\
\enddata
\tablenotetext{}{(a) - Star formation rate averaged over the last 10 Myr, (b) - Instantaneous specific star formation rate at the observed epoch, (c)- Total equivalent width of the \ion{C}{3}]$\lambda\lambda$1907,09 doublet, (d) - Lensing magnification, (e) - Effective area. }
\end{deluxetable*}

\section{Data}\label{sec:data}

\subsection{Deep spectroscopic observations from SPURS}\label{subsec:observations}
This work is based on deep spectroscopic observations taken as part of the ongoing JWST Cycle 4 large program -- {SPURS} (GO-9214; PI: Mason \& Stark).
SPURS observed 75 sources in the Abell 2744 lensing cluster field  over two visits from November 6 to 9, 2025.
The spectra were obtained with the NIRSpec medium resolution ($R\sim1000$) gratings -- G140M/F100LP, G235M/F170LP, and G395M/F290LP -- which collectively span 0.97 -- 5.50 microns (rest-frame UV and optical wavelengths at $z\gtrsim5$).
The total exposure times were 29.47, 8.02 and 2.95 hours in the G140M, G235M and G395M gratings respectively. 
The data were reduced using {\tt msaexp} (v0.9.13), based on the official JWST pipeline version 1.16.1 and reference file mapping jwst 1303.pmap, following the 
procedures described in previous works \citep{Anna_Rubies, JWST_primal_Survey,Valentino2025}.
We adopt the extended wavelength extraction in {\tt msaexp} \citep{Valentino2025}, but note that all the absorption lines presented here are detected in the wavelength region free from contamination by higher order spectra. 
Because the G140M spectra were acquired during both visits, we combined them for each galaxy into a single spectrum using an inverse-variance weighted mean.
We refer the reader to \citet{Chen2026} for an overview of the observations and reduction, while a detailed description of the survey design and data reduction will be presented in a future paper.

We are interested in analyzing the ISM absorption lines in the rest-UV spectra probed by our deep G140M observations.  
This requires a robust spectroscopic redshift for the host galaxy and good continuum SNR.  
We measure spectroscopic redshifts for all the galaxies using optical emission lines covered by the G235M and G395M gratings (described in detail in Section~\ref{subsec:redshift-determination}) and then select galaxies at $z\geq5$ and with median continuum SNR$\geq3.5$ per wavelength element in the G140M spectra
We exclude one source which meets the SNR requirement (SPURS-A2744-415) from our selection due to potential continuum contamination from a nearby galaxy and arrive at a final sample of 6 galaxies.

The top panel of Figure~\ref{fig:nircam-image} shows the NIRCam color composite images of all the galaxies that satisfy our selection criteria, along with the location of the MSA shutters. 
We note that in all the galaxies, the MSA shutter adequately captures the brightest clump. 
The deep G140M spectra for all of the galaxies are shown in the bottom panels of Figure~\ref{fig:nircam-image}.
Key spectroscopic features such as damped \Lya\ absorption, numerous ISM absorption lines and rest-UV nebular emission lines, such as \ion{C}{3}], are prominently detected in \textit{individual galaxies}, highlighting the spectral resolution and depth of our observations and their suitability for the goals of this work.
{A detailed summary of our targets is presented in Table~\ref{tab:galaxy-properties-basic}.}

\subsection{Determining the systemic redshift}\label{subsec:redshift-determination}

\begin{figure}
    \centering
    \large{SPURS-A2744-7}\\
    \includegraphics[width=0.95\linewidth]{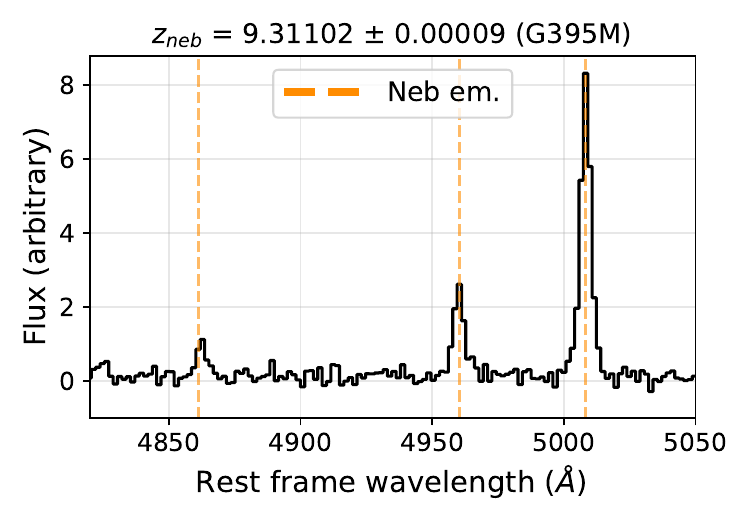}
    \includegraphics[width=\linewidth]{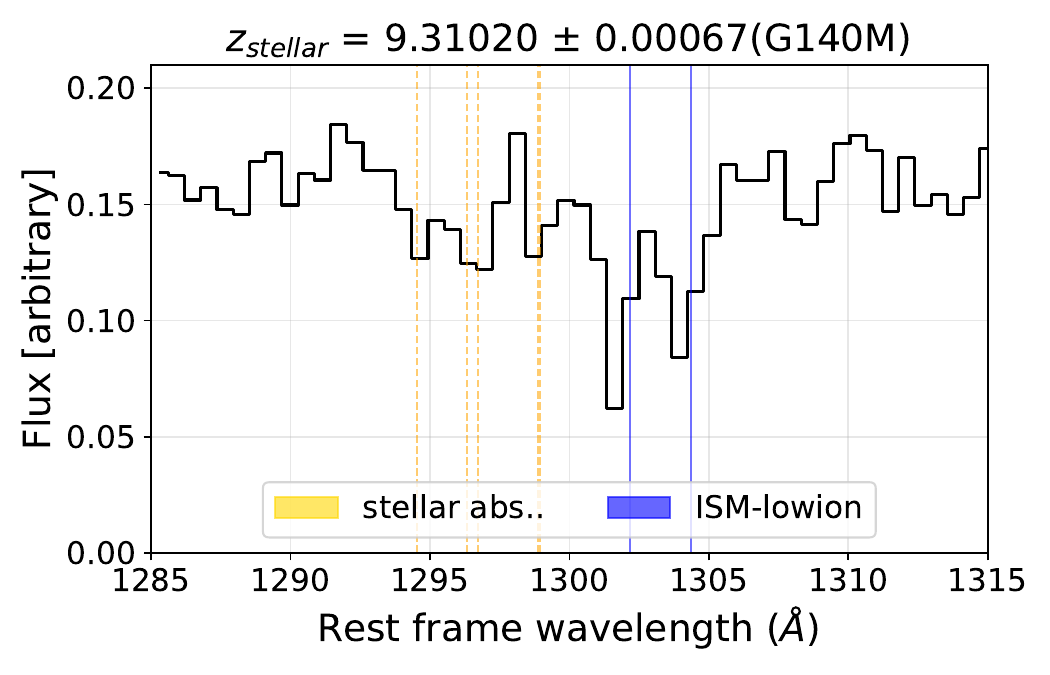}
    \caption{Determining the spectroscopic redshift using optical nebular emission and stellar absorption lines in SPURS-A2744-7.
    We obtain the systemic redshift for all our galaxies by fitting the [\ion{O}{3}]$\lambda\lambda$4959,5007 optical nebular emission line doublet prominently detected in the G235M and G395M gratings.
    The top panel shows the G395M spectrum around [\ion{O}{3}]+H$\beta$, the vertical orange lines denote the rest frame wavelengths of the nebular emission lines at the best fit redshift.
    The bottom panel shows part of the G140M spectrum for the same galaxy. 
    The \ion{Si}{3} stellar absorption lines (rest frame wavelengths marked by yellow dashed lines) agree well with the systemic redshift obtained from the nebular emission lines.
    By contrast, the \ion{O}{1}+\ion{Si}{2} $\lambda\lambda1302,1304$ ISM absorption lines (rest frame wavelengths marked by blue lines) are blueshifted in this source, providing clear evidence of outflowing gas. 
    }
    \label{fig:stel-abs}
\end{figure}

To determine the systemic redshifts for our sample, we use strong optical nebular emission lines, which trace the gas around star-forming regions, and are prominently detected in our G235M and G395M observations.
We perform a joint fit to the [\ion{O}{3}]$\lambda\lambda 4959,5007$ optical nebular emission line doublet which is clearly detected and spectrally resolved for every galaxy in our sample.
We model the doublet using two Gaussian components, fixing their amplitude ratio to the theoretical flux ratio of $2.98$ \citep{osterbrock-2006}.
The width of the lines, $\sigma_{\rm neb}$, is treated as a free parameter but is assumed to be identical for both lines.
We determine the best-fit values for the amplitude, width, and redshift by performing a least-squares fit. 
The uncertainties on the redshift are derived from the profile likelihood, with the $1\sigma$ confidence interval corresponding to the range in redshift for which $\Delta\chi^2=1$.
The high SNR of the [\ion{O}{3}] detections allows us to accurately determine the systemic redshift for all galaxies in our sample.
In Figure~\ref{fig:stel-abs} we show an example of the [\ion{O}{3}]$\lambda\lambda 4959,5007$ detection in SPURS-A2744-7, while the Appendix shows the profiles for the full sample.

In 4 out of the 6 galaxies, we find that the rest-UV \ion{C}{3}]-$\lambda\lambda$1907,09 nebular emission doublet is spectrally resolved in the G140M grating.
We find the systemic redshift obtained from the \ion{C}{3}] doublet is in agreement with those obtained from optical nebular lines. 
Additionally, in SPURS-A2744-7 \citep[discussed in detail in][]{Zuyi_SPURS_ID7}, we also detect multiple stellar photospheric absorption lines in our spectrum.
Figure~\ref{fig:stel-abs} shows the stellar absorption line complex \ion{Si}{3} around 1294 -- 1299\AA, as well as the \ion{O}{1}+\ion{Si}{2}-$\lambda\lambda1302,1304$ ISM absorption line.
We fit the stellar \ion{Si}{3} complex around 1294 -- 1299\,\AA\ using multiple Gaussian profiles \citep[following the approached described by][]{sunny_stellarkin}.
We measure the systemic redshift from the stellar lines as $z_{\rm stellar}= 9.31020 \pm 0.00067$, which is within $1.2\sigma$ of the systemic redshift obtained from the [\ion{O}{3}] nebular emission line ($z_{\rm neb}=9.31102 \pm 0.0009$).
Relative to this joint nebular and stellar systemic velocity, the \ion{O}{1}+\ion{Si}{2}-$\lambda\lambda1302,1304$ ISM absorption lines in this system are clearly blueshifted, providing clear evidence of outflowing gas.
Given the encouraging consistency between systemic redshifts in this exemplary case, we adopt the redshifts obtained from [\ion{O}{3}] as the systemic redshift for all of our galaxies. 

\begin{figure}
    \centering
   \includegraphics[width=0.95\linewidth]{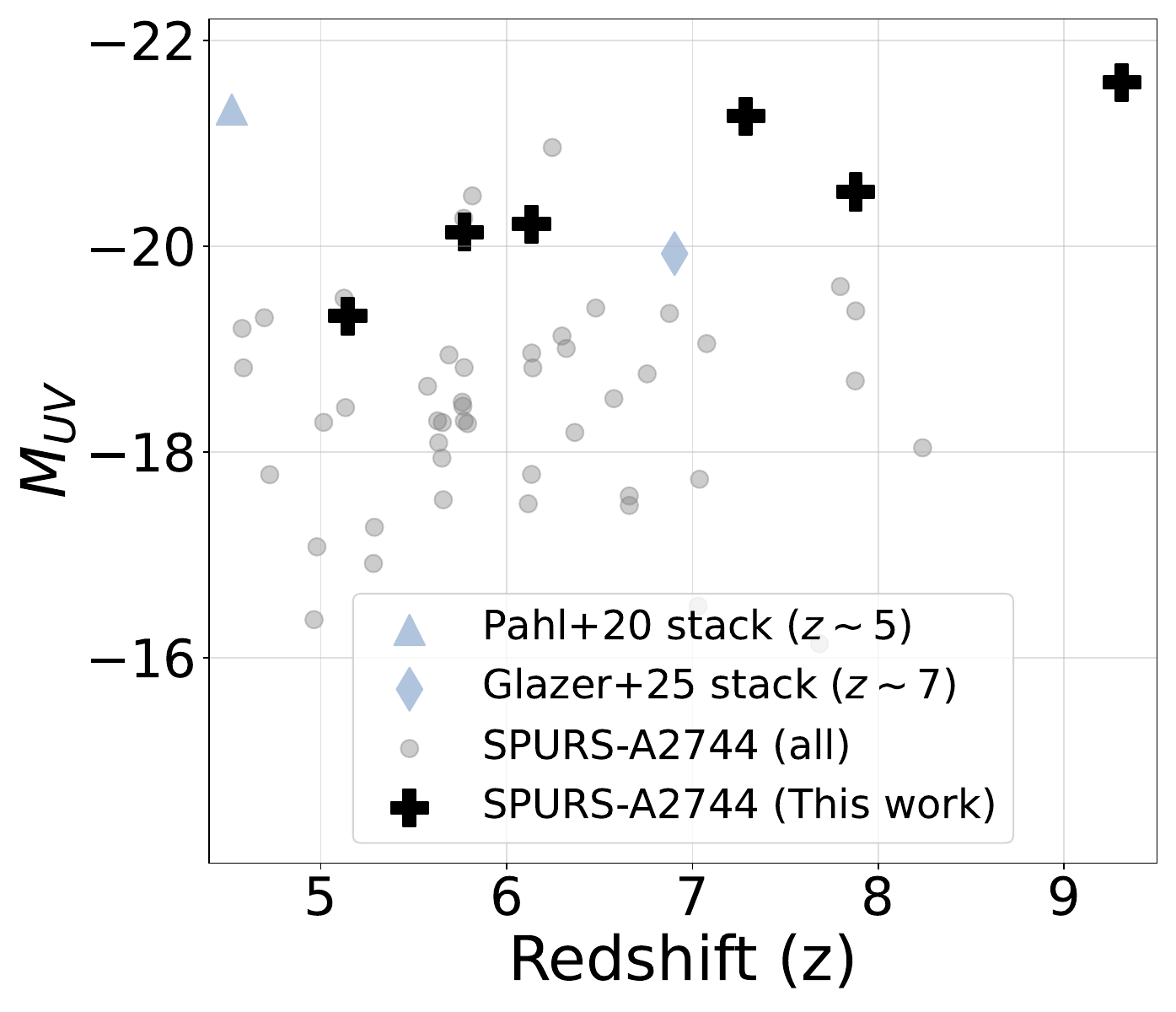}
    \caption{UV absolute magnitude ($M_{UV}$) as a function of redshift for all of the galaxies in the SPURS-A2744 pointing (gray) and the subset that meet our continuum selection criteria used in this work (Section~\ref{subsec:observations}, black crosses).
    We also show the mean $M_{UV}$ of the samples used in stacked studies in the literature \citep{Pahl_2020,pancakez_kelsey} which investigate UV ISM absorption lines in a similar redshift range to our sample.  
    The individual SPURS galaxies used in this work span comparable UV luminosity as the stacked studies. 
    }
    \label{fig:redshift-MUV}
\end{figure}

\subsection{Galaxy properties from ancillary data} \label{subsec:galprops}

We utilize the wealth of archival data over the Abell 2744 lensing cluster to derive physical properties for all of our galaxies.
Abell 2744 has deep archival imaging and spectroscopy from both HST and JWST, along with established lens models \citep[e.g.,][]{johan_richard_lensmodel_abel2744,Kawamata_lensmodel_abell2744,Castellano2016_astrodeep,Lotz2017_frontierfields,tomasso_glass,Bergamini2023_lensmodel,furtak_lensmodel,uncover_survey_first_release}.
Here we briefly summarize our methodology used to obtain the key galaxy properties ($M_{UV}$, stellar mass, sSFR, and $\Sigma_{SFR}$)  relevant for this work and refer the reader to \citet{Chen_overdensity} and references therein for a detailed description. 

We use the photometry measured from archival HST and JWST imaging presented in \citet{Endsley_2025_diversity_sfh}, with absolute UV magnitudes derived from the integrated Kron photometry described in the same work.
We correct for the effect of lensing magnification in all derived galaxy properties using the maps from \citet{furtak_lensmodel, uncover_survey_first_release}.

To infer physical properties, we use the SED modeling package BayEsian Analysis of GaLaxy sEds \citep[BEAGLE;][]{beagle_2016} assuming a Chabrier \citep{Chabrier2003} Initial Mass Function (IMF). We infer the stellar mass and star formation rate (SFR, averaged over the last 10 Myr) from the full NIRCam SED for each source, assuming a constant star formation history.
We measure galaxy sizes from the photometry to infer the star formation rate surface density ($\Sigma_{SFR}$): we model the F150W NIRCam images of each galaxy with multiple Sersic components using {\tt pysersic} \citep{pysersic_imad_pasha} and compute the effective area (A$_{\rm eff}$) as the area covered by the pixels having flux greater than half of the peak flux. 
This approach is similar to that used in \citet{topping_aurora_electrondensity}. 

Table~\ref{tab:galaxy-properties-basic} provides a summary of the measured galaxy properties. 
In Figure~\ref{fig:redshift-MUV} we show the lensing corrected UV absolute magnitude for our sample with clear detections of multiple ISM absorption lines, along with the full SPURS Abell 2744 sample, and those of composites used in previous stacked studies probing absorption lines in  the same redshift range \citep{Pahl_2020,pancakez_kelsey}. 
We find the SPURS galaxies used in this work have comparable UV brightness to galaxies in these existing stacked studies.
The SPURS galaxies also have small sizes (A$_{\rm eff}\sim0.2$ kpc$^2$) and high star formation surface densities ($\Sigma_{\rm SFR}\gtrsim50~{\rm M}_\odot{\rm yr}^{-1}{\rm kpc}^{-2}$), consistent with other luminous high-redshift galaxies during these epochs \citep[e.g.,][]{Morishita2024_sizeanalysis}.

In the following section, we measure the ISM gas kinematics and covering fraction from the absorption lines and then explore trends with galaxy properties in Section~\ref{sec:trends}.
\begin{figure*}
    \centering
    \includegraphics[width=\linewidth]{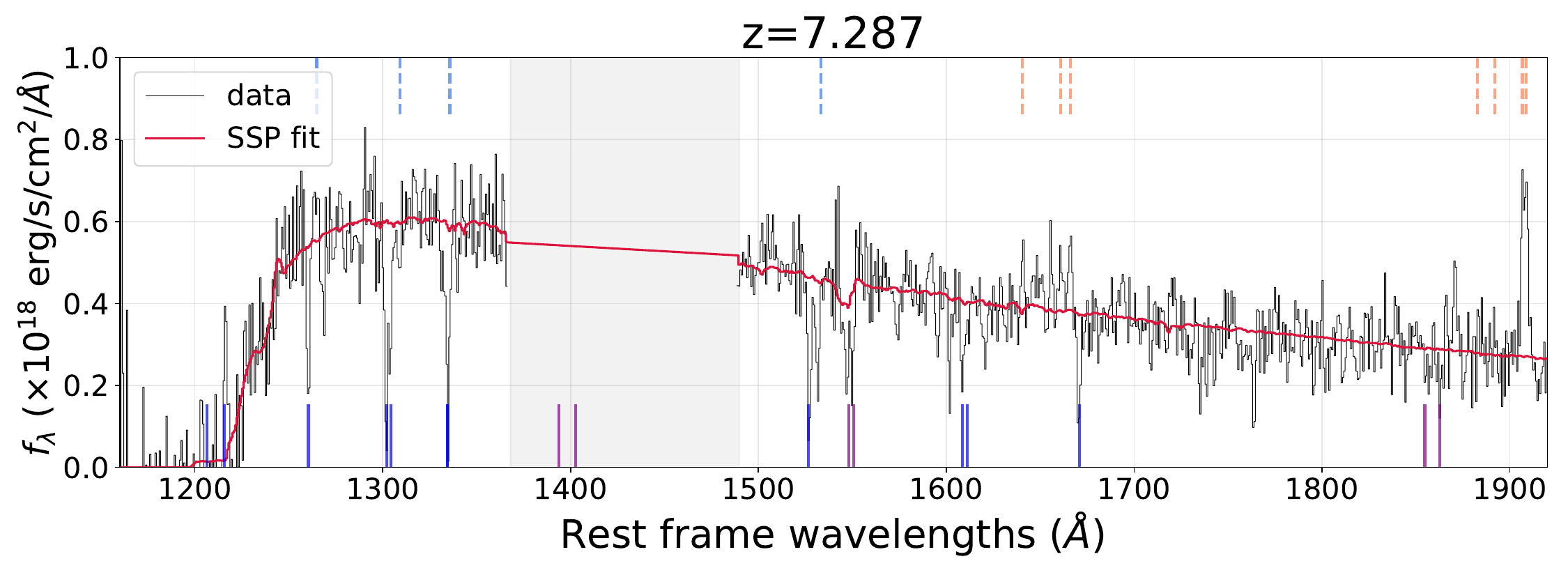}\\
        \includegraphics[width=0.7\linewidth]{figs/spectra_legend.pdf} \\

    \vspace{0.1in}
    \includegraphics[width=0.49\linewidth]{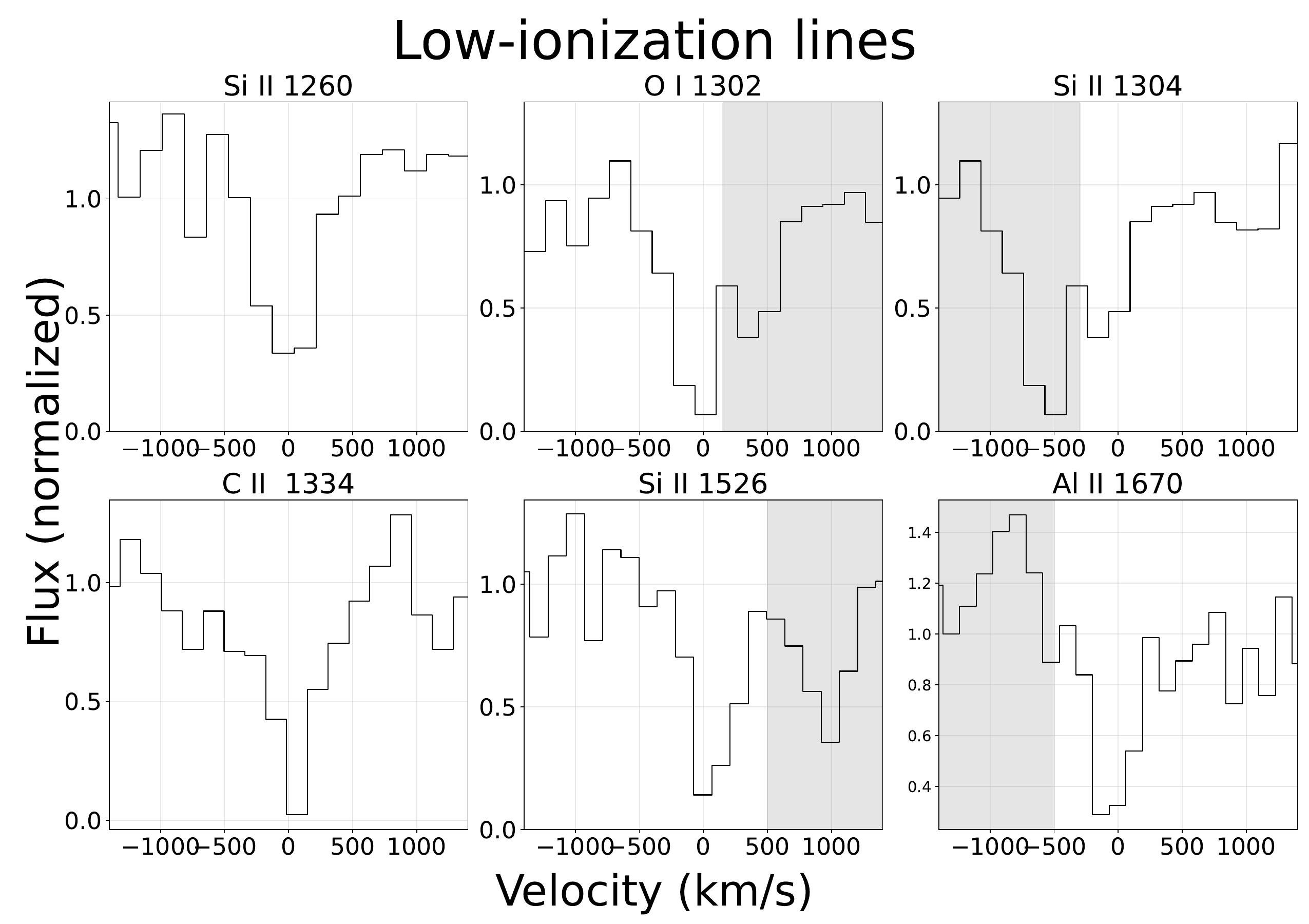}
    \raisebox{0.02in}{\includegraphics[width=0.45\linewidth]{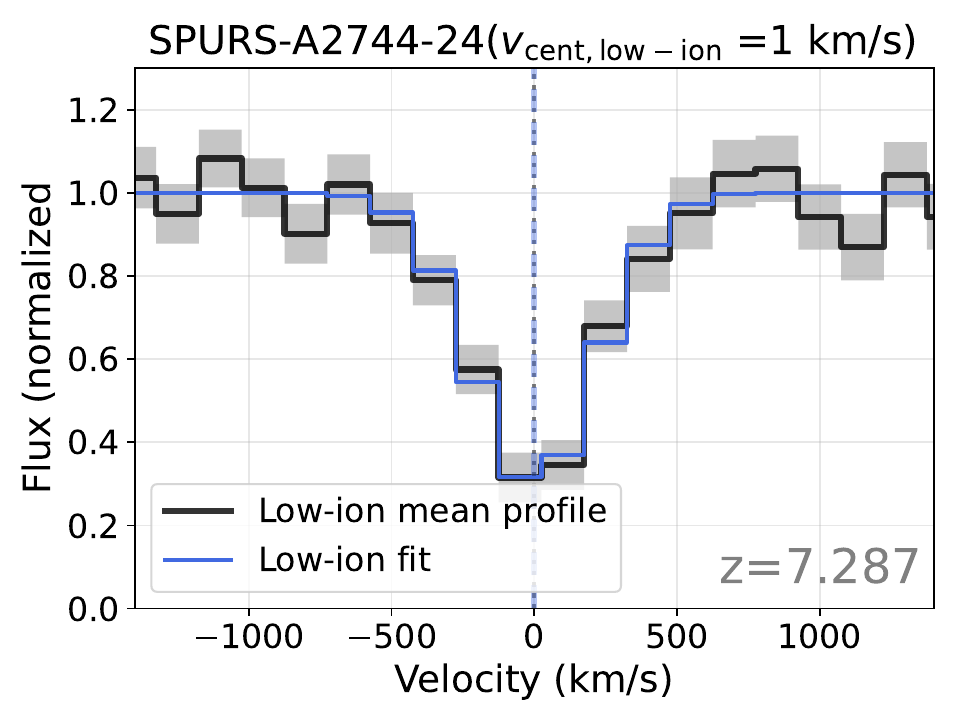}}

    \caption{ \emph{Top}: G140M spectra of SPURS-2744-24 at $z=7.287$. The low- and high-ionization absorption lines are marked in blue and purple respectively. Nebular emission lines such as \ion{C}{3}] and \ion{O}{3}] are shown as dashed orange lines. 
    The red line overlaid on the spectra is the median synthetic stellar population fit to the spectra (described in Section~\ref{subsec:continuum-norm}).  
    \emph{Bottom left}: A close-up of the  subset of the low-ionization lines used to generate our mean low-ionization absorption profile. The gray region in each panel shows the regions that are masked when constructing the mean profile. 
    \emph{Bottom right}: The mean ISM absorption lines from stacking the low-ionization lines is shown in black and a single Gaussian fit to the absorption profile is shown in blue. The gray regions denote the $1\sigma$ error. The velocity centroid ($v_{\rm cent, low-ion}$) is marked as blue dashed lines.  }
    \label{fig:demo-spectra}
\end{figure*}

\section{ISM gas kinematics and covering fraction}\label{sec:kinematics-and-coveringfraction}

The goal of this work is to probe the baryon cycle in individual galaxies during the epoch of reionization ($z\geq5$).
We will utilize optically thick ISM absorption lines in the rest-UV spectra which encode key information regarding the structure and dynamics of the ISM gas `in front' of the galaxy.
The relevant radiative transfer equation in the optically thick regime is as follows: 
\begin{equation}\label{eq:radtransfer}
    \frac{I(v)}{I_0}  = 1-C_f(v)\left( 1- e^{-\tau} \right) \approx 1-C_f(v)\ \text{when}\ \tau\gg1
\end{equation}
where $I$ is the observed intensity, $I_0$ is the stellar continuum intensity, $C_f$ is the covering fraction and $\tau$ is the optical depth.
The covering fraction inferred from the absorption lines traces the porosity of the ISM, while the absorption-line profile provides information about the kinematics of the ISM gas.

The individual low-ion absorption profiles in five out of the six galaxies in our sample exhibit similar absorption depths and equivalent widths (EW) such that Equation~\ref{eq:radtransfer} is a reasonable approximation, indicating that the absorption profiles likely trace the neutral gas covering fraction rather than variations in optical depth \citep{pettini-cb58-2002,jones2013,jones_dustinthewind, nicha2016-escape-fraction}. 
If the observed gas is not optically thick (for instance, if the accessible lines are relatively weak transitions, as is the case for SPURS-A2744-1069), the covering fraction values likely represent lower limits.

In this section, we outline our methodology for modeling the stellar continuum intensity ($I_0$, Section~\ref{subsec:continuum-norm}), including attenuation around Ly$\alpha$ from the ISM/CGM and IGM, characterizing the covering fraction $C_f(v)$ and determining the gas kinematics (Section~\ref{sec:ism_profiles}).

\subsection{Continuum modeling and normalization}\label{subsec:continuum-norm}

An important step in characterizing the interstellar absorption lines is first disentangling them from absorption arising in the stellar atmospheres that dominate the continuum.
To obtain a model for the continuum for each source we fit the full G140M spectrum of each galaxy in the sample using Charlot \& Bruzual (CB) stellar population synthesis models \citep{Charlot_Bruzual_2003}.
The CB2019 model version employed here includes evolutionary tracks and atmospheres for single stars, assuming a \citet{Chabrier2003} IMF with upper mass cutoff of 600~$M_\odot$, as described in more detail in \citet{vidal-garciaModellingUltravioletlineDiagnostics2017,platConstraintsProductionEscape2019,senchynaDirectConstraintsExtremely2022}.

We mask and do not include nebular emission lines in the fit, but include the nebular continuum predicted by \texttt{CLOUDY} as described in the aforementioned references at, as this can provide a substantial contribution to the UV continuum for young stellar populations.
We fix the ionization parameter to $\log U=-2.5$; while this choice has an impact on derived parameters, it has a negligible impact on the quality of the fit to the continuum over this spectral range \citep[competing mainly with the extinction; see e.g.][]{senchynaDirectConstraintsExtremely2022}.
We mask all lines with strong interstellar emission or absorption contribution to avoid biasing the stellar fits.
To account for dust attenuation, we adopt the \citet{gordonQuantitativeComparisonSmall2003} SMC Bar extinction curve, with $A_V$ allowed to vary from 0--2~magnitudes.

In order to reproduce the spectral region around \Lya\ in our $z>7$ spectra, where we cover up to $\approx 1200$\,\AA\ in the rest-frame, we account for Ly$\alpha$ damping wing attenuation from \HI in both the ISM/CGM and the IGM, similar to the approach of \citet{Mason2026}, adapted to grating resolution following \citet{Chen2026}.
We include both a Voigt absorption profile with variable column density ($18<\log N_\mathrm{HI}<24$), central wavelength ($\pm 500$\kms{} from systemic), and covering fraction ($0<C_f<1$) to model \Lya\ attenuation local to the galaxy; as well as \Lya\ emission \citep[with EW and velocity offset sampled using empirical priors from][]{Tang2024_Lyaprofile} and the IGM damping wing attenuation in a partially neutral IGM \citep[following][]{miralda-escudeReionizationIntergalacticMedium1998,barkanaDidUniverseReionize2002,Mesinger2008}, where we assume that the IGM is fully neutral beyond any potential ionized region each galaxy sits in and vary the distance of the source to the nearest neutral IGM region (0--100~cMpc). Using priors on the local \HI component that are informed by the low-ionization absorption features does not significantly impact our results. A detailed analysis of the damping wings in the context of reionization will be presented in an upcoming work.

For the purposes of this work, our fiducial continuum fits assume a simple single-age stellar population (SSP) for each galaxy.
We interpolate over a grid ranging in age over 1--40~Myr and in metallicity from $0.0001$--$0.03$ ($\sim 0.01$--$2\times Z_\odot$) spanning well beyond the range representative of our sample; and repeat each fit 100 times for spectra resampled according to the formal extraction uncertainties.
While this is a significant oversimplification of the star formation history of these galaxies, we find that this simple model provides a very reasonable fit to the UV continuum of these galaxies. We demonstrate this in Figure~\ref{fig:demo-spectra}, where we show the resulting SSP fit for SPURS-A2744-24 in the upper panel.
As a test, we also produce an alternative set of fits produced by modeling the spectrum as a linear combination of all SSPs in the grid \citep[following the approach of e.g.][]{chisholmConstrainingMetallicitiesAges2019}.
Adopting these alternative linear combination fits results in minimal shifts in derived ISM parameters (see Section~\ref{sec:ism_profiles}).
We find that the rest-UV spectra of these systems are well reproduced by SSPs with ages of $\sim 3$--5~Myr, metallicities of $Z=0.001$--$0.003$, and $A_V\sim 0$--$0.2$.
We will explore the stellar populations and physical conditions in these sources in more detail in a future work (e.g., Chen et al. in prep).

Here, we adopt the median model SSP spectrum fit as our best estimated continuum for each galaxy, and use the continuum normalized spectra for the rest of the analysis presented in this work.

\subsection{Characterizing the ISM absorption profiles}
\label{sec:ism_profiles}

\begin{figure*}[!t]
    \centering
    \Large{$v_{\text{cent,low-ion}} \simeq -100$\,\kms\ }\\
    \includegraphics[width=0.327\linewidth]{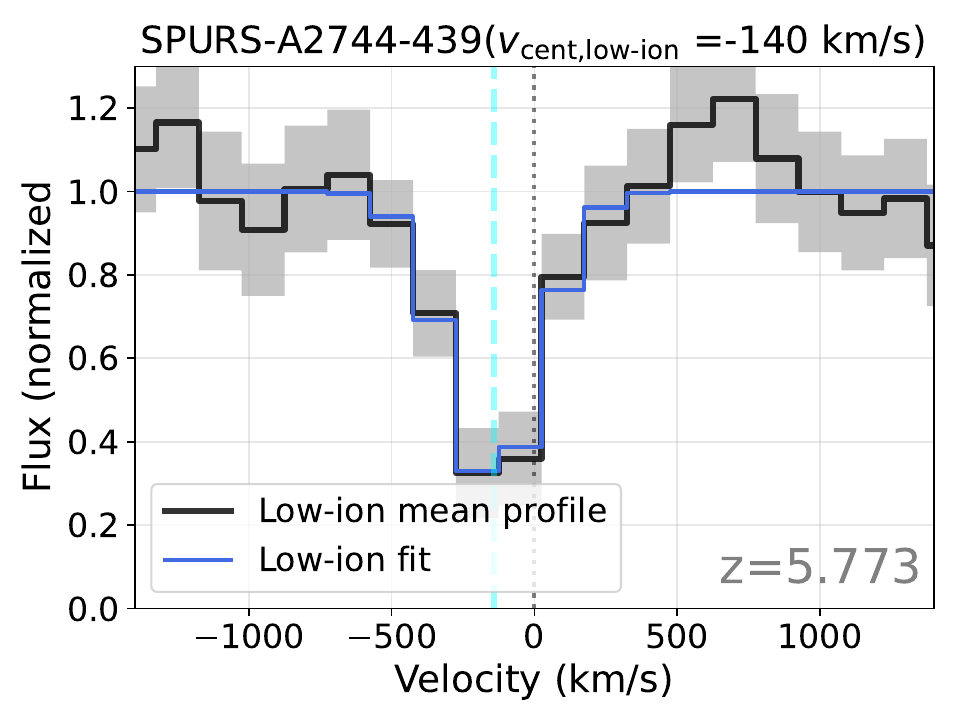}
    \includegraphics[width=0.327\linewidth]{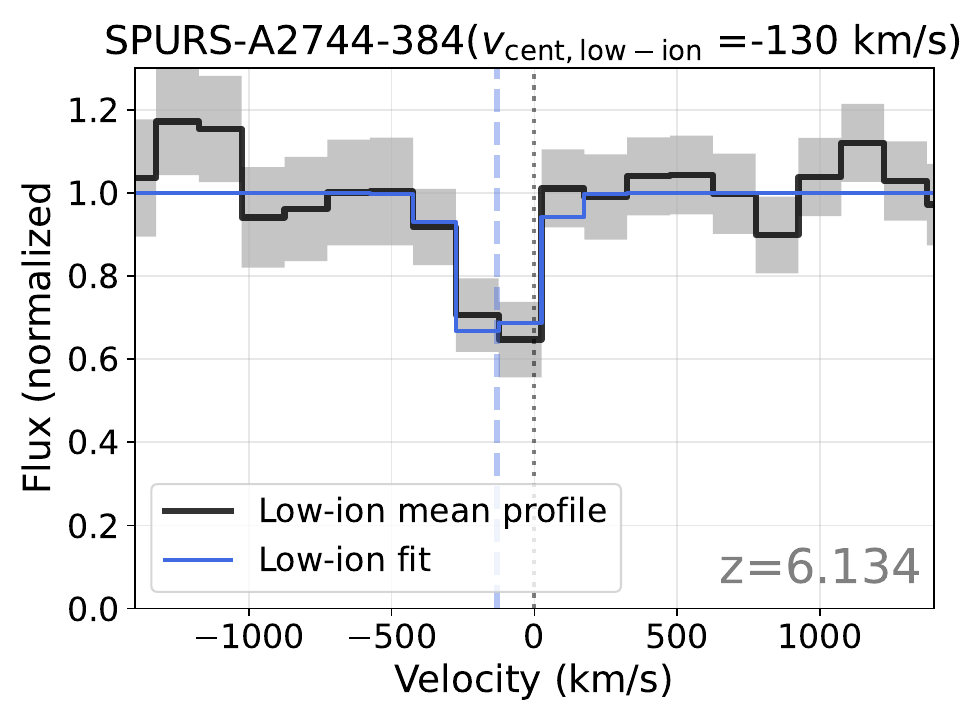}
    \includegraphics[width=0.327\linewidth]{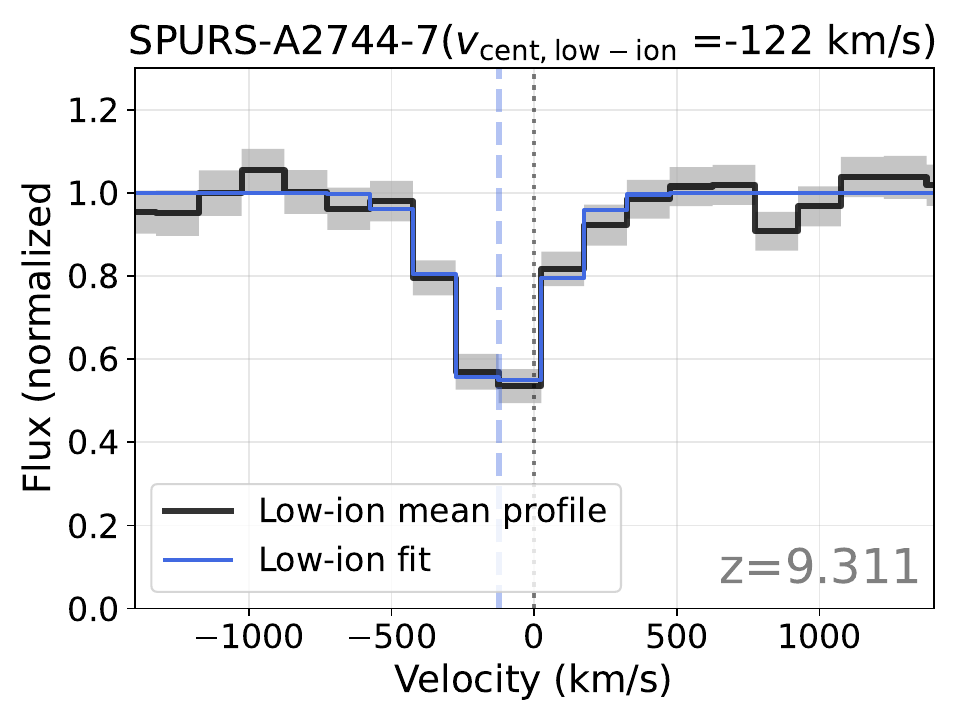}
\\ 
     \vspace{0.2in}

    \Large{$v_{\text{cent,low-ion}}\simeq 0$\,\kms} \\ 
    \includegraphics[width=0.327\linewidth]{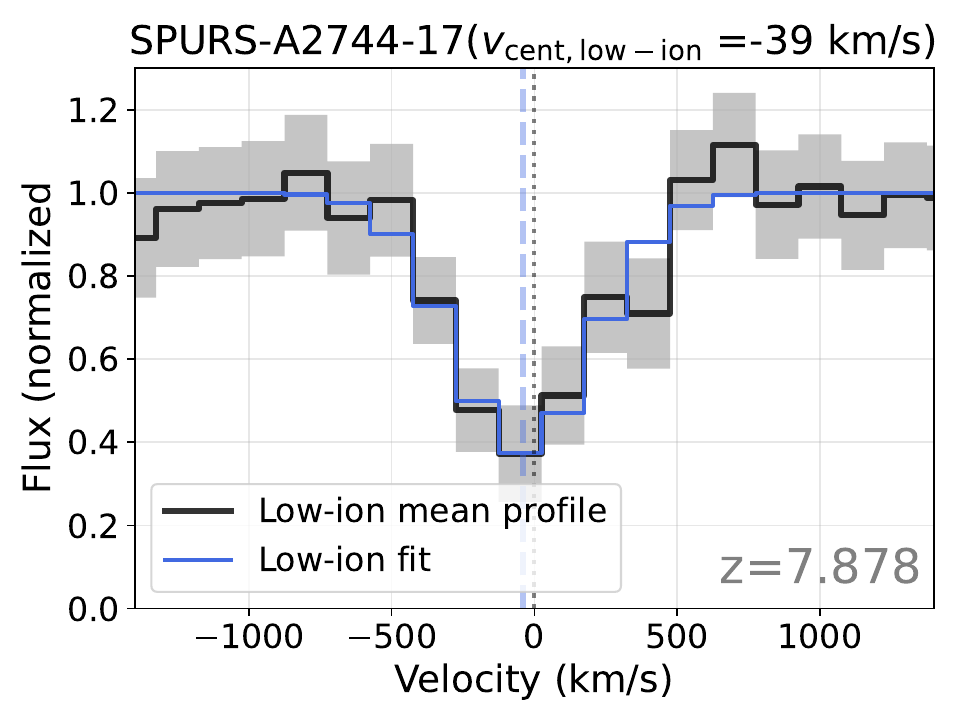}
    \includegraphics[width=0.327\linewidth]{figs/24_absprofile.pdf}
    \includegraphics[width=0.327\linewidth]{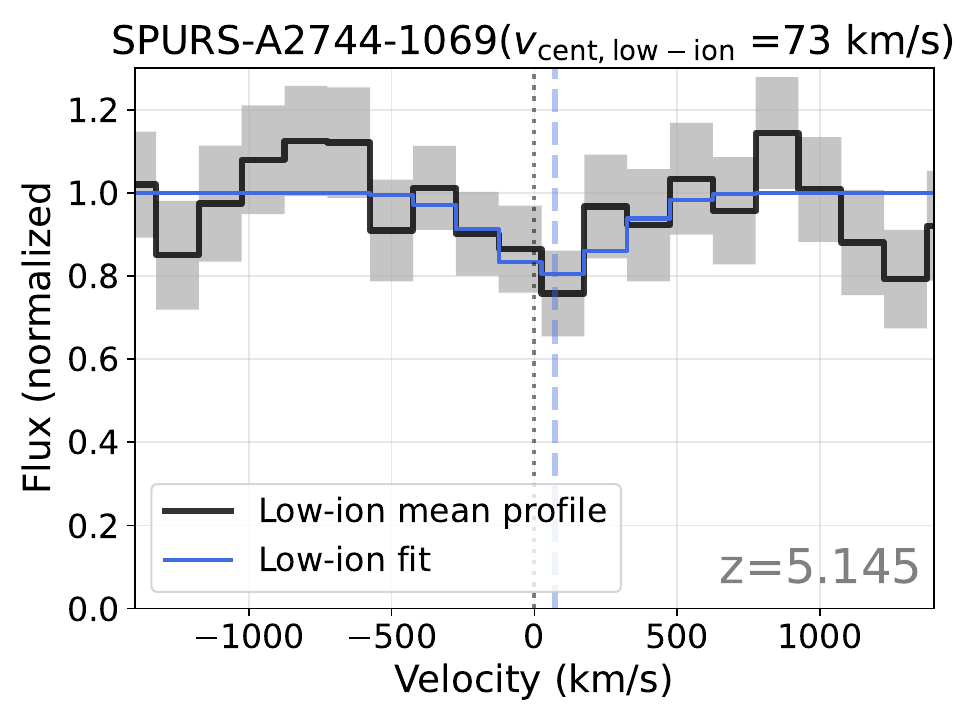}\\
   
    \caption{Mean low-ionization ISM absorption profiles for our sample. In each panel we show the normalized flux in black and the gray regions denote the $1\sigma$ error. The best fit Gaussian profile is shown in blue. The velocity centroid ($v_{\rm cent,low-ion}$)  is marked in each panel as blue dashed lines. We find that half of the galaxies in our sample exhibit clear signatures of outflowing gas (shown in top panels), with velocity centroids of $v_{\rm cent,low-ion} \simeq -100$\,\kms\ while the remaining systems are consistent with little to no net bulk motion, with $v_{\rm cent,low-ion} \simeq 0$\,\kms\ (shown in bottom panels).
    }
    \label{fig:ism-profiles}
\end{figure*}

The rest-frame UV spectra host several low-ionization metal transitions (e.g., \ion{Si}{2}, \ion{C}{2}) that predominantly trace the neutral \ion{H}{1} gas phase, alongside high-ionization metal transitions (e.g., \ion{Si}{4}, \ion{C}{4}) probing ionized gas.
Collectively, these transitions allow us to trace the $T \sim 10^4$~K multiphase gas structure and kinematics within the ISM. 
In this work, we are primarily interested in the covering fraction and kinematics of the low-ionization species which are unambiguously detected across our entire sample, allowing for consistent analysis. 
Where available, we use high-ionization species to probe the kinematics of the warmer gas phases.

For the low-ionization absorption lines, we obtain a mean ISM absorption profile for each galaxy by interpolating the detected strong low-ionization lines onto a common velocity grid of 150\,\kms\ and combining them using an inverse variance weighted mean.
At $z\gtrsim7$, we utilize \ion{O}{1}, \ion{Si}{2}, \ion{C}{2}, and \ion{Al}{2} while at lower redshifts, we utilize \ion{Si}{2}, \ion{Al}{2} \ion{Mg}{2} and \ion{Fe}{2} (see the Appendix for details). 
In the case of blended line transitions such as \ion{O}{1}+\ion{Si}{2} and \ion{Al}{2}, only the regions of interest that correspond to the transition are taken into account. 
We refer readers to \citet{kvgc_ESI2022} for a detailed discussion of our methodology.

Figure~\ref{fig:demo-spectra} demonstrates this procedure for one of the galaxies in our sample (SPURS-A2744-24). 
The top panel shows the rest-UV spectra in black, along with the stellar population fit used for continuum normalization in red. 
The bottom left panel shows the individual optically thick low-ionization ISM absorption lines used to produce our mean ISM absorption line profile. 
The bottom right panel shows the combined mean ISM absorption line obtained using an inverse variance weighted mean.
The mean low-ionization ISM absorption profile for each of our galaxies is presented in Figure~\ref{fig:ism-profiles}.

To derive the covering fraction and kinematics of the absorbing gas, we fit the mean absorption profiles using a single Gaussian profile of the following form, which adequately captures the shape of the absorption profiles: 
\begin{equation}
    C_f(v) = C_{\rm f, cent} \exp\left[\frac{-(v-v_{\rm cent})^2}{2\sigma_{\rm abs}^2}\right]
\end{equation}
The amplitude of the single Gaussian traces the peak covering fraction ($C_{\rm f, cent}$), the velocity centroid ($v_{\rm cent}$) traces the bulk motion of the ISM gas and the $\sigma_{\rm abs}$ measures the absorption line width. 
Each mean absorption profile is fitted using 250 realizations, and in each run, the spectra are perturbed by a random $1\sigma$ noise based on the error spectrum.
We obtain the best-fit and $1\sigma$ standard deviation of each quantity from the median and MAD/0.675 (where MAD is the median absolute deviation) from the 250 realizations, respectively.
In Figure~\ref{fig:ism-profiles} we show the best fit profile along with the mean ISM low-ionization absorption profiles for each galaxy.
We report the profile fit parameters for the low-ionization lines in Table~\ref{tab:fit-parameters} and for the high-ionization lines in Table~\ref{tab:fit-parameters-highions}.

We estimate the intrinsic (deconvolved) amplitude and width (FWHM) of ISM absorption lines from our best fit profile assuming a spectral resolution of $R=1300$ \citep[][which is $1.3\times$ higher than the pre-launch estimates quoted in the JWST User Documentation]{Anna_spectralresolution, anowar_spectralresolution}.
The velocity centroid is not impacted by the spectral resolution.
We use the deconvolved measurements of the amplitude and FWHM for the rest of the analysis presented in this paper.
We adopt the same methodology presented in this section to measure the covering fraction and velocity centroid for the low-ions from stacked spectra reported in the literature at $z\sim 2-7$ \citep[][]{du2018, Pahl_2020, pancakez_kelsey}, which we compare to in forthcoming sections.

\begin{figure*}[!t]
    \centering
    \Large{$v_{\text{cent, low-ion}}\simeq -100$\,\kms}, blueshifted $v_{\text{cent,high-ion}}$ \\ 
     \includegraphics[width=0.325\linewidth]{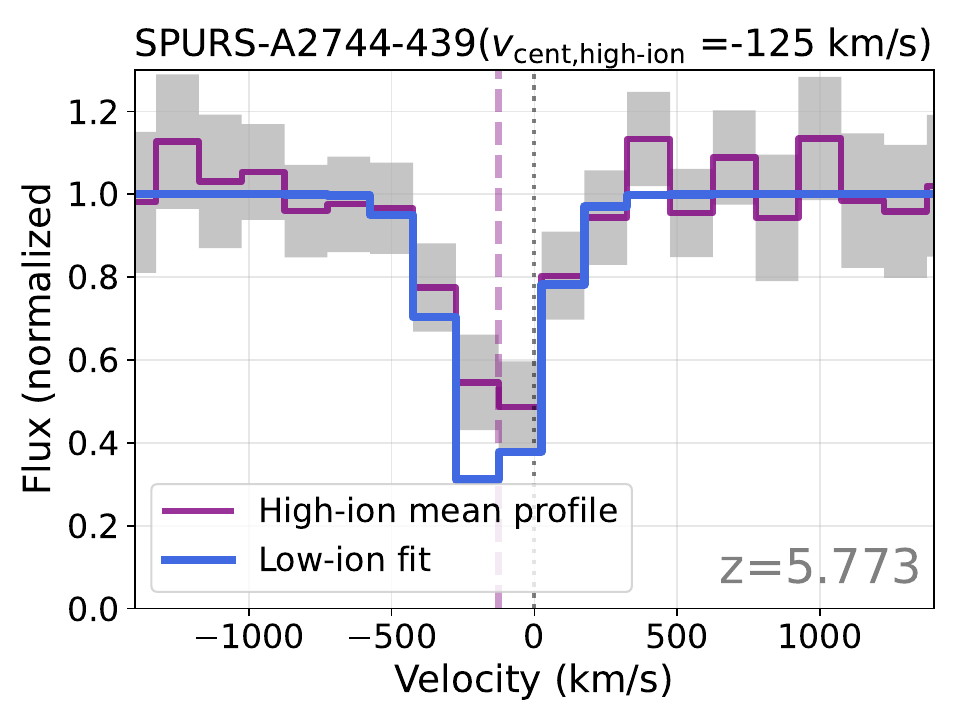}
     \includegraphics[width=0.325\linewidth]{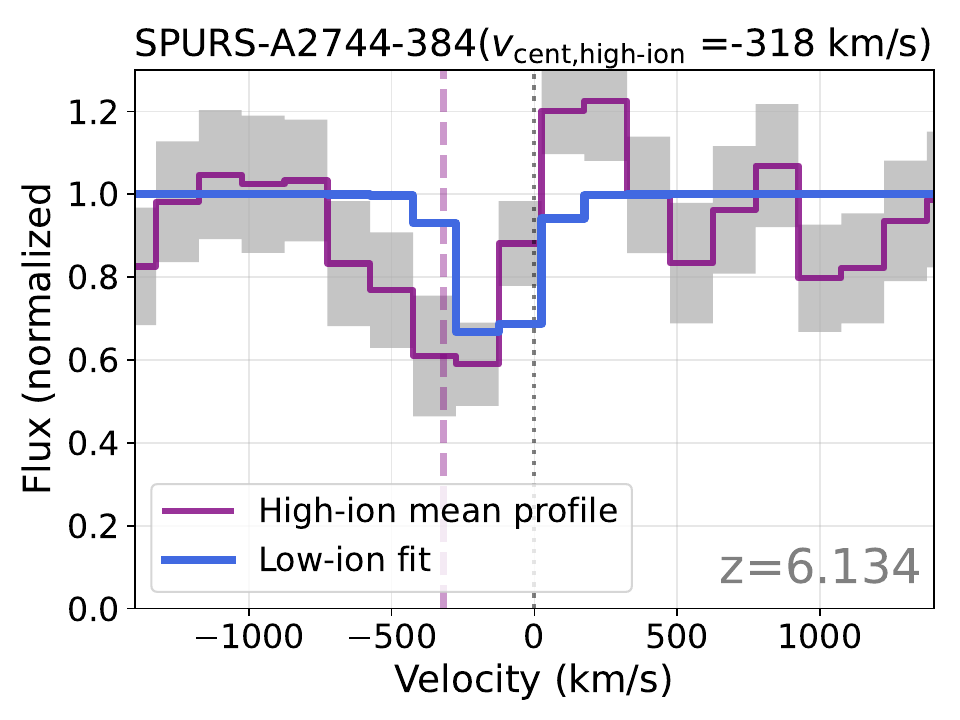}
     \includegraphics[width=0.325\linewidth]{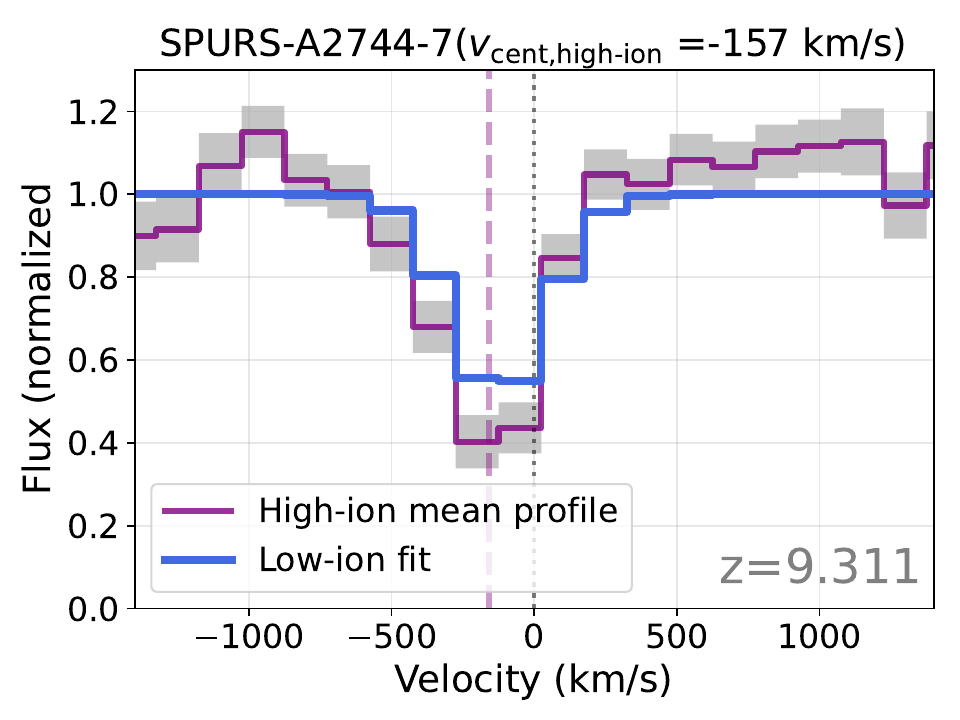}

     \vspace{0.2in}

    \Large{$v_{\text{cent,low-ion}}\simeq 0$\,\kms}, blueshifted $v_{\text{cent,high-ion}}$ \\ 
     \includegraphics[width=0.325\linewidth]{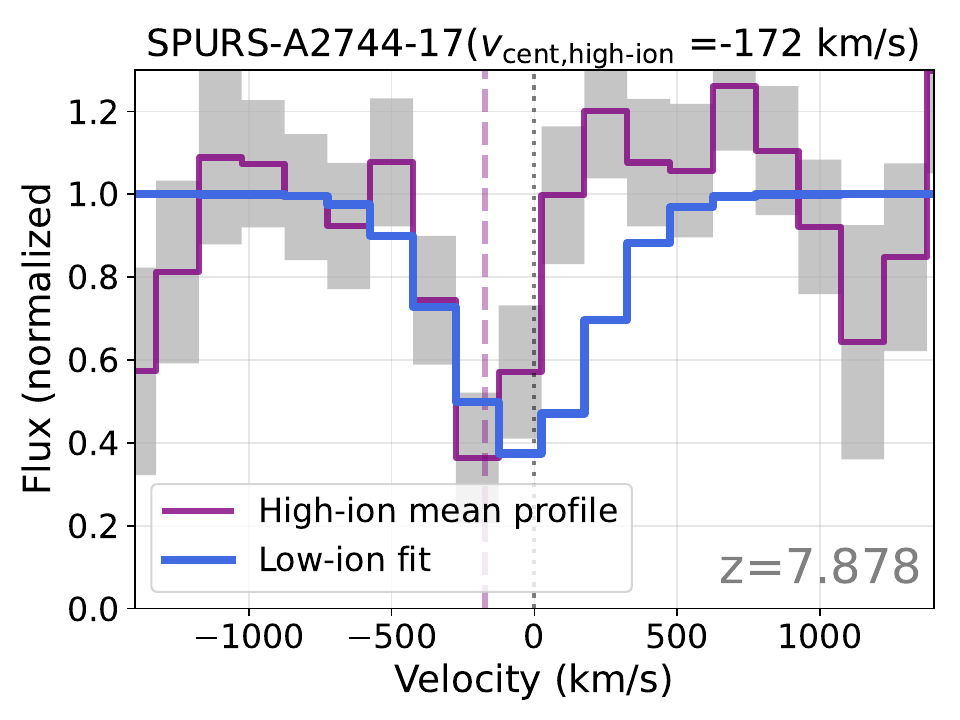}
     \includegraphics[width=0.325\linewidth]{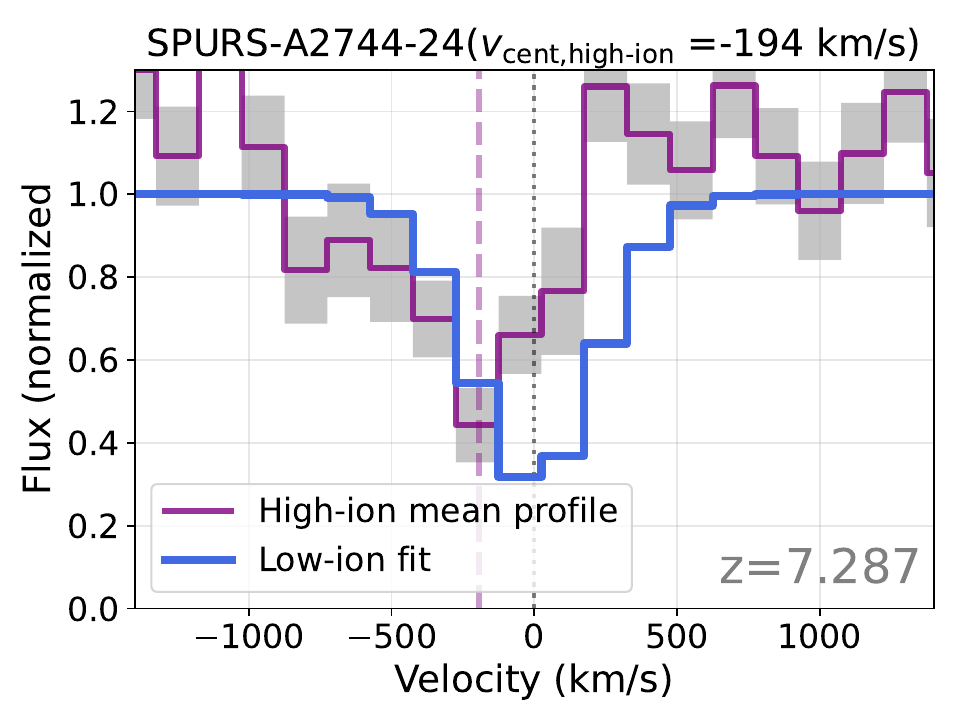}

    \caption{Comparison of the mean low-ionization profiles (probing the cool gas phase; blue) to the high-ionization profiles (warm phase; purple) for the five galaxies in our sample with requisite spectral coverage.
    The velocity centroid of the high-ions ($v_{\rm cent,high-ion}$) is marked in each panel as purple dashed lines.
    For the three sources with low-ionization velocity centroids of $v_{\rm cent,low-ion} \simeq -100$\,\kms\ we find the high-ionization profiles are similar to the low-ionization profiles, showing net blueshifted absorption (top panels).
    This is in contrast to the sources with little to no net bulk motion in their low-ionization gas (with $v_{\rm cent,low-ion} \simeq 0$\,\kms, bottom panels), which still show net blueshifted absorption in their high-ionization gas.
    We discuss the differences in cool and warm gas phase kinematics in detail in Section~\ref{sec:discussion}. 
    The individual high-ion absorption lines used to construct the mean profile for each galaxy are shown in the Appendix. 
   } \label{fig:low-ion-high-ion-comparison}
\end{figure*}

For 5 out of the 6 galaxies (SPURS-A2744-7, 17, 24, 384, and 439), we also probe the kinematics of the warmer gas phases using several high-ionization absorption lines (\ion{Si}{4}, \ion{C}{4}, \ion{Al}{3}). 
For each galaxy, we construct a mean absorption line profile from the high-ion transitions available within the observed spectral range (see the Appendix for details) and fit the resulting profile with a single Gaussian following the  same methodology used for the low-ionization lines described above. 
In SPURS-A2744-1069, while we have spectral coverage of \ion{Al}{3}, we do not include it in our analysis due to the lack of a statistically significant detection. 
We show the mean high-ionization absorption profiles for each galaxy in Figure~\ref{fig:low-ion-high-ion-comparison}, present our measured high-ionization absorption line fit parameters in Table~\ref{tab:fit-parameters-highions}.
We explore the kinematic differences between the high- and low-ionization lines as well as their physical implications in Sections~\ref{subsec:kinematics} and ~\ref{sec:discussion}.

We note that for SPURS-A2744-24 and 384, the mean high-ion absorption profile is obtained from \ion{C}{4} which is a resonant line and includes contributions from stellar features.
While our continuum fits (described in Section~\ref{subsec:continuum-norm}) account for these stellar features, residual redshifted \ion{C}{4} emission arising from scattering in an ionized outflow \citep{Jones2023_GLASS} or absorption associated with a poorly modeled stellar wind feature may bias the absorption profile. 
However, we note that using the more flexible linear combination fits (Section~\ref{subsec:continuum-norm}) to normalize the spectra instead results in the velocity centroids of the high-ion absorption profiles increasing only by $2\%$ on average. 
This uncertainty is within the quoted $1\sigma$ uncertainties of the velocity centroids and does not significantly affect the main results presented in this work. 
A detailed study quantifying the systemic uncertainities associated with stellar continuum fitting on low- and high-ionization absorption line measurements will be presented in a future study with an expanded sample (e.g., Chen et al. in prep).

\begin{deluxetable*}{ccccccc}\label{tab:fit-parameters}
\tablecaption{Table of derived ISM gas properties. 
Here we present our fitted measurements of the mean absorption line profile for the low-ionization gas (as described in detail in Section~\ref{sec:ism_profiles}).
The profile is of the form $C_f(v) = C_{\rm f,cent} \exp\left[\dfrac{-(v-v_{\rm cent})^2}{2\sigma_{\rm abs}^2}\right]$. 
We report the intrinsic (deconvolved) measurements, assuming a spectral resolution of $R=1300$, with the $1\sigma$ uncertainties.
The rows are sorted from top to bottom in increasing value of measured velocity centroid ($v_{\rm cent}$).
For a subset of the galaxies with spectral coverage of the \Lya\ break, we also quote our derived column density of the \HI gas.  
}
\tablehead{\colhead{objid} & \colhead{$z_{\rm spec}$} & \colhead{v$_{\rm cent}$} & \colhead{$\sigma_{\rm abs}\text{(deconv)}$} & \colhead{$C_{\rm f,cent}\text{(deconv)}$}  & \colhead{$\log(N_\mathrm{HI})$}   & \colhead{EW} \\
 &  &  (\kms) & (\kms) &  & (cm$^{-2}$) & (\kms) 
}
\startdata
439 & 5.77315 & -140 $\pm$ 14 & 128 $\pm$ 12 &  0.91 $\pm$ 0.08  & - & 290 $\pm$ 22 \\
384 & 6.13372 & -130 $\pm$ 17 & 66 $\pm$ 25 & 0.71 $\pm$ 0.24 & - & 118 $\pm$ 14 \\
7 & 9.31102 & -122 $\pm$ 9 & 136 $\pm$ 12 & 0.61 $\pm$ 0.04 & 19.9 $\pm$ 0.6 & 206 $\pm$ 9 \\
17 & 7.87780 & -39 $\pm$ 26 & 220 $\pm$ 25 & 0.69 $\pm$ 0.06 & 21.5 $\pm$  0.2 & 378 $\pm$ 29 \\
24 & 7.28726 & 1 $\pm$ 11 & 192 $\pm$ 12 & 0.79 $\pm$ 0.04 & 21.8 $\pm$ 0.1 & 381 $\pm$ 16 \\
1069 & 5.14519 & 73 $\pm$ 99 & 191 $\pm$ 115 & 0.23 $\pm$ 0.11 & - & 112 $\pm$ 38 \\
\enddata
\end{deluxetable*}

\begin{deluxetable*}{cccccc}\label{tab:fit-parameters-highions}
\tablecaption{Same as Table~\ref{tab:fit-parameters} but for the high-ion mean profiles.}
\tablehead{\colhead{objid} & \colhead{$z_{\rm spec}$} & \colhead{v$_{\rm cent}$}  & \colhead{$\sigma_{\rm abs}\text{(deconv)}$} & \colhead{$C_{\rm f,cent}\text{(deconv)}$} & \colhead{EW} \\
 &  & (\kms) & (\kms) & &  (\kms) 
}
\startdata
439 & 5.77315 & -125 $\pm$ 34 & 130 $\pm$ 23 & 0.67 $\pm$ 0.11 & 220 $\pm$ 22 \\
384 & 6.13372 & -318 $\pm$ 49 & 125 $\pm$ 41 & 0.61 $\pm$ 0.16 & 194 $\pm$ 34 \\
7 & 9.31102 & -157 $\pm$ 16 & 124 $\pm$ 11 & 0.84 $\pm$ 0.06 & 261 $\pm$ 12 \\
17 & 7.87780 & -172 $\pm$ 40 & 89 $\pm$ 17 & 0.92 $\pm$ 0.06 & 201 $\pm$ 30 \\
24 & 7.28726 & -194 $\pm$ 32 & 133 $\pm$ 31 & 0.67 $\pm$ 0.11 & 222 $\pm$ 27 \\
\enddata
\end{deluxetable*}

\section{Results}\label{sec:trends}

One of the striking visual aspects of the rest-UV spectra of our galaxies is the unambiguous detection of metal ion absorption across multiple species (Figure~\ref{fig:nircam-image}). 
In this work, we are interested in utilizing these ISM features to probe baryon cycle processes—namely inflows and outflows—in individual $z\sim$ 5 -- 9  galaxies and investigating how they scale with host galaxy properties such as stellar mass and sSFR. 
In the previous section, we have measured the ISM gas geometry and kinematics in the cool and warm gas phases as probed by low- and high-ionization absorption lines (Figures~\ref{fig:ism-profiles}, \ref{fig:low-ion-high-ion-comparison}).
Here, we present the absorption lines properties for each source (Section~\ref{subsec:notes}), the gas geometry (Section~\ref{subsec:ISM_porosity}) and kinematics of our sample (Section~\ref{subsec:kinematics}), and explore how the absorption lines properties correlate with host galaxy properties (Section~\ref{subsec:trends}).

\subsection{Notes on individual galaxies}\label{subsec:notes}
We begin by summarizing the key spectroscopic features we have measured for each galaxy in our sample. The individual absorption line profiles for each source are shown in the Appendix.

\textbf{SPURS-A2744-7} is a massive ($\log M_* (\Msun)$=$9.2$), extended galaxy at $z=9.3$ with a relatively low specific star formation rate ($\log \text{sSFR} =0.8\ {\rm Gyr}^{-1}$, assuming constant SFH, see Section~\ref{subsec:galprops}).
It was previously observed by JWST/NIRSpec as part of the GLASS program \citep[Gz9p3, DHZ1;][]{boyett-spurs7, tomasso_glass}, which first revealed it likely showed strong ISM absorption lines in 5\,hr G140H spectroscopy, and with the NIRSpec prism in UNCOVER \citep[UNCOVER-3686;][]{Fujimoto2024_UNCOVER,uncover_survey_first_release}. 
Recent ALMA observations \citep{Algera2025} detect [\ion{O}{3}] 88$\mu$m emission spatially offset from the rest-UV continuum, indicating an extended ionized gas reservoir within the galaxy.
In the SPURS G140M spectrum we clearly detect multiple ISM absorption lines with high SNR (\ion{Si}{2}-$\lambda$1260, \ion{O}{1}-$\lambda$1302, \ion{Si}{2}-$\lambda$1304, \ion{C}{2}-$\lambda$1334, \ion{Si}{2}-$\lambda$1526, \ion{Al}{2}-$\lambda$1670), along with damped \HI Ly$\alpha$ absorption.
We also detect a tentative weak \Lya\ emission with EW$\sim0.5\AA$ with velocity offset of $\sim$350\,\kms, consistent with expectations from UV bright galaxies during the early stages of reionization \citep{Dijkstra2011,Mason2018}.
A detailed study of this galaxy is presented in \citet{Zuyi_SPURS_ID7} and here we summarize the absorption lines.
The mean low-ionization ISM profile (Figure~\ref{fig:ism-profiles}) appears asymmetric, with an extended blueshifted tail and sharp ingress in the redshifted side. 
The velocity centroid derived from the mean ISM profile is  $v_{\rm cent, low-ion}=-122 \pm 9$\,\kms\ with an intrinsic covering fraction of $0.61\pm0.04$.
We detect high-ionization lines \ion{Si}{4}, \ion{C}{4} and \ion{Al}{3} with velocity centroid and FWHM similar to the low-ions (Figure~\ref{fig:low-ion-high-ion-comparison}).
The ISM/CGM component of the Ly$\alpha$ damping wing is best-fit by \HI column density of $\log(N_\mathrm{HI})=19.9\pm0.6$ cm$^{-2}$, indicating a cool gas reservoir, in addition to strong damping wing attenuation from the IGM \citep[see][]{Zuyi_SPURS_ID7}.

\textbf{SPURS-A2744-17} is a low mass ($\log M_* (\Msun)$=$7.9$) galaxy at $z=7.9$ with high sSFR ($\log \text{sSFR} =2.3\ {\rm Gyr}^{-1}$).
This galaxy was previously observed as part of GLASS and spectroscopically confirmed via the Lyman-break in the NIRISS/WFS spectra \citep[GLASS-100003, ZD2, GLASSZ8-1;][]{Borsani_lymanbreakgalaxies}.
It is also part of a known overdensity at $z=7.87$ \citep{Ishigaki_2016_ID17,Morishita_GLASS_protocluster,Witten_protocluster_GLASS}. 
Furthermore, this galaxy was also previously proposed as a candidate Lyman continuum leaker \citep{Mascia_2023_lyCLeaker_17, Jaskot_2024_ID17_LyCLeaker}, primarily based on its compact size and blue UV slope. 
However, we detect the same ISM absorption lines as SPURS-A2744-7, except \ion{Si}{2}-$\lambda1526$ which falls in the detector gap.
The intrinsic covering fraction of the low ions is $0.69\pm0.06$, and we infer a \ion{H}{1} column density of $\log(N_\mathrm{HI})=21.5\pm0.2$ cm$^{-2}$ from the damped Ly$\alpha$ profile. 
This indicates a substantial amount of neutral gas, and thus that SPURS-A2744-17 is unlikely to have significant ionizing photon escape along this line-of-sight \citep[e.g.,][]{Gazagnes2018,Saldana-Lopez2022}.
Unlike SPURS-A2744-7, the mean low-ion ISM profile is centered around 0\,\kms, and is broad, with an intrinsic FWHM of $518\pm59$\,\kms.
We detect \ion{Si}{4} in absorption, blueshifted with a velocity centroid of $-172\pm40$\,\kms\ and narrower than the low-ionization line profile, while \ion{C}{4} exhibits redshifted emission of $\sim100$\,\kms. The wavelength region immediately blueward of \ion{C}{4} is in a detector gap, limiting our ability to measure any absorption profile. However, as \ion{C}{4} is a resonant line, we note the redshifted emission also implies scattering in an ionized outflow \citep[e.g.,][]{Jones2023_GLASS}.
Thus, we find that the low- and high-ionization lines in this galaxy imply distinct kinematics.

\textbf{SPURS-A2744-24} is a low mass ($\log M_* (\Msun)$=$7.9$), high sSFR ($\log \text{sSFR} =2.6\ {\rm Gyr}^{-1}$) galaxy at $z=7.3$.
Morphologically, this galaxy has a large effective area (0.49 {\rm kpc}$^2$) compared to the rest of the sample, with several star-forming clumps in proximity (see Figure~\ref{fig:nircam-image}).
We detect the same ISM absorption lines as SPURS-A2744-7, with an intrinsic covering fraction of $0.79\pm0.04$.
Consistent with the high gas covering fraction, we infer a local \HI column density of $\log(N_\mathrm{HI})=21.8\pm0.1$ cm$^{-2}$ from the Ly$\alpha$ damping wing, indicating a large neutral gas reservoir along the line-of-sight to this galaxy.
We detect weak \Lya\ emission with EW$\sim1$ \AA\ redshifted from the systemic by $\sim150$\,\kms. This low \Lya\ velocity offset is expected for radiative transfer through lower column density gas \citep[$\log(N_\mathrm{HI})\approx20$ cm$^{-2}$,][]{Neufeld1990}, and thus consistent with a patchy or inhomogeneous \HI distribution in the ISM \citep[e.g.,][]{Hu2023_CLASSY,Almada2026}. 
Similar to SPURS-A2744-17, we find the mean ISM absorption line has a velocity centroid centered around systemic velocities ($v_{\rm cent}=1\pm11$\,\kms) and FWHM of $452\pm28$\,\kms.
For the high-ionization lines, we detect a blueshifted \ion{C}{4} in absorption with a velocity centroid of $-194\pm32$\,\kms\ (c.f. $1$\,\kms\ for the low ions), while \ion{Si}{4} falls in a detector gap.

We note that our velocity measurements of the $z>7$ sources are consistent with an independent analysis by \citet{Zhu2026} .

\textbf{SPURS-A2744-384} is a moderately massive galaxy ($\log M_* (\Msun)$=$8.2$) at $z=6.1$, with a moderate sSFR ($\log \text{sSFR} =1.4\ {\rm Gyr}^{-1}$).
We clearly detect \ion{Si}{2}-$\lambda1526$, \ion{Al}{2}-$\lambda1670$ and \ion{Mg}{2}-$\lambda\lambda2796,2802$.
The mean low-ionization absorption profile has an intrinsic covering fraction of $0.71\pm0.24$ and shows a clear blueshifted absorption profile with a velocity centroid of $-130\pm17$\,\kms.
We also detect weak \ion{Mg}{2} emission at redshifted velocities of $\sim150$\,\kms, which could arise from resonant scattering in an outflow \citep[e.g.,][]{Prochaska2011}. 
Similar to SPURS-A2744-7, we detect \ion{C}{4} in absorption with a blueshifted velocity centroid of  $-318\pm49$\,\kms.

\textbf{SPURS-A2744-439} is a massive galaxy ($\log M_* (\Msun)$=$8.8$) at $z=5.8$, with moderate sSFR ($\log \text{sSFR} =1.2\ {\rm Gyr}^{-1}$).
We clearly detect \ion{Al}{2}-$\lambda1670$, \ion{Mg}{2}-$\lambda\lambda2796,2802$ as well as numerous \ion{Fe}{2} lines.
The mean low-ionization ISM absorption profile has an intrinsic covering fraction of  $0.91\pm0.08$ and shows clear blueshifted absorption, with a velocity centroid of  $v_{\rm cent}= -140\pm14$\,\kms\ and FWHM of $301\pm28$\,\kms\ . 
The high-ionization absorption lines (\ion{Al}{3}-$\lambda\lambda1854,1862$) have similar velocity centroid as the low-ionization gas.  

\textbf{SPURS-A2744-1069} at $z=5.1$ has the lowest intrinsic UV luminosity in our sample ($M_\mathrm{UV}=-19.3$).
The inferred sSFR and stellar mass of the galaxy are $\log \text{sSFR} =2.4\ {\rm Gyr}^{-1}$ and $\log M_* (\Msun)$=$7.5$.
This galaxy also exhibits the lowest low-ion covering fraction of just $0.23\pm0.11$.
The wavelength region around \ion{Mg}{2} doublet falls in a detector gap.
The high-ionization absorption lines (\ion{Al}{3}-$\lambda\lambda1854,1862$) are similarly weak. 
Rest-UV nebular emission lines such as \ion{He}{2}, \ion{O}{3}] and \ion{C}{3}]  along with rest-optical nebular emission lines such as [\ion{O}{2}], [\ion{O}{3}], H$\beta$ and [\ion{Ne}{3}] are prominently detected in the G140M spectrum. 
The G395M spectrum reveals a broad H$\alpha$ line, indicating it may be a broad-line AGN.
The AGN nature of this source will be discussed in an upcoming paper.

These six $z>5$ galaxies represent a heterogeneous sample with a diversity of ISM kinematics and galaxy properties. We now seek to explore trends in their absorption line properties.

\begin{figure}
    \centering
     \includegraphics[width=\linewidth]{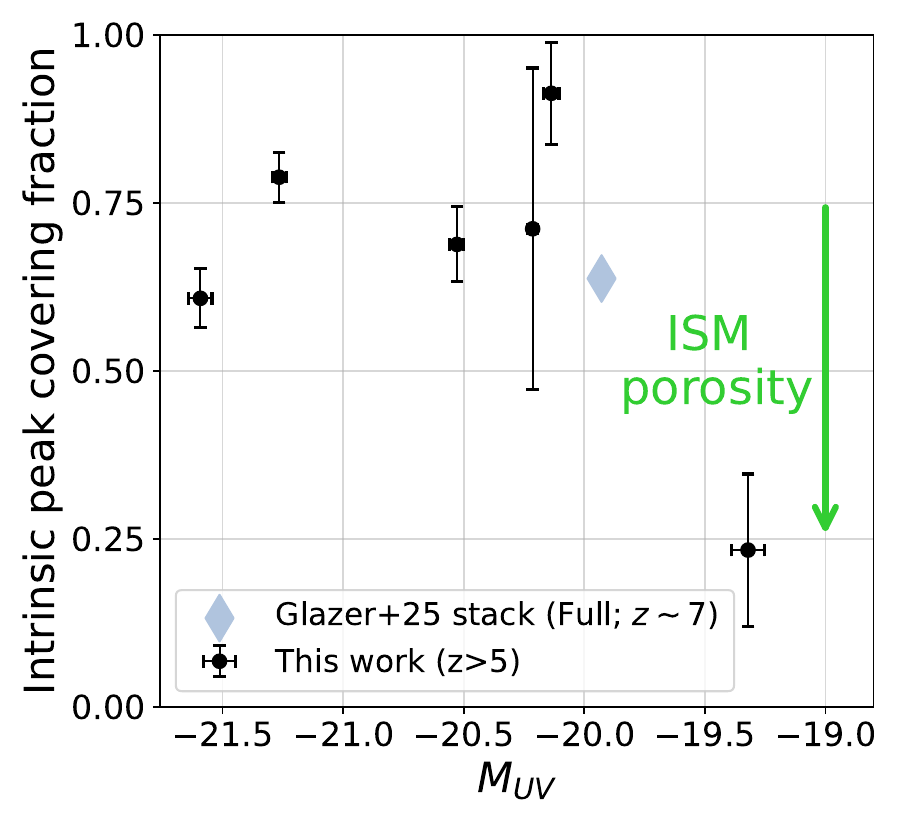}
    \caption{Intrinsic peak covering fraction ($C_f$) of the low-ionization gas versus UV absolute magnitude ($M_{UV}$) for our $z\sim5-9$ sample (black points) and stack of $z\sim7$ galaxies \citep[][]{pancakez_kelsey}. 
    We find that our sample exhibits a large diversity of ISM gas porosities with covering fractions ranging from 0.23 to 0.91.
    This implies that some galaxies are almost entirely surrounded by a neutral gas reservoir whereas the others possess more porous envelopes. 
    The covering fraction is comparable to those observed in the $z\sim7$ stack. 
    }
    \label{fig:Cf_MUV}
\end{figure}

\begin{figure}
    \centering
     \includegraphics[width=0.95\linewidth]{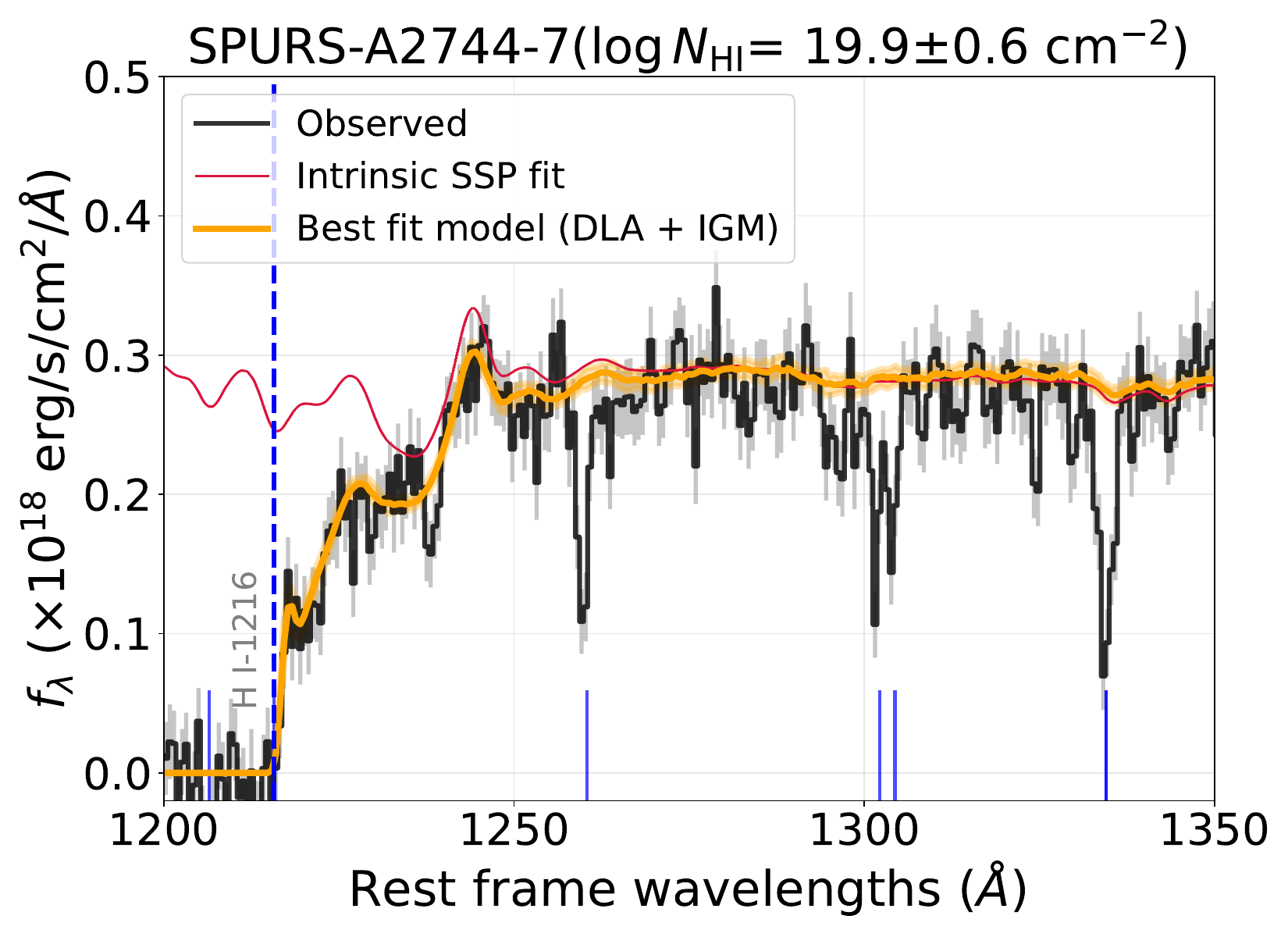}
    \includegraphics[width=0.95\linewidth]{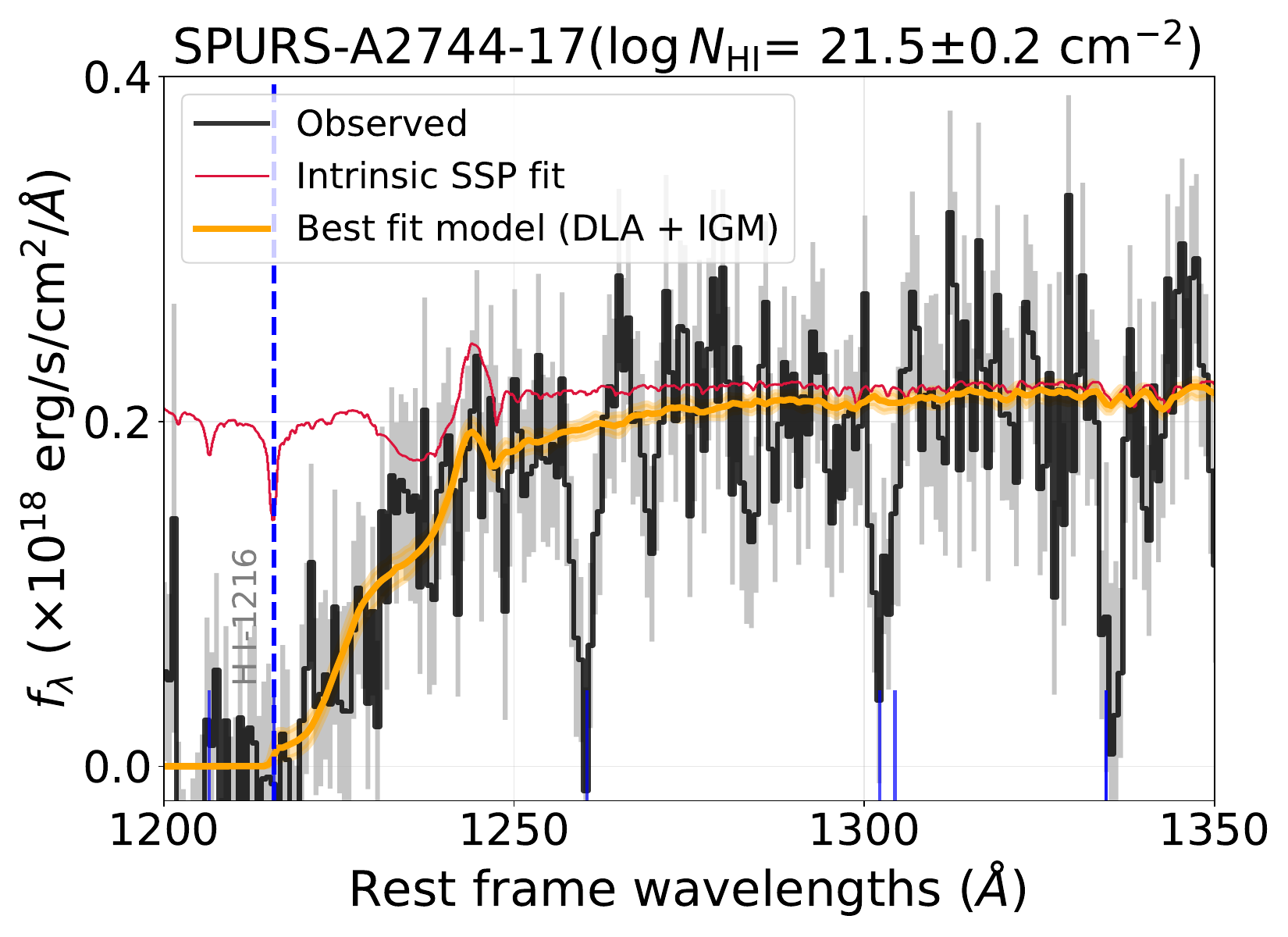}
     \includegraphics[width=0.95\linewidth]{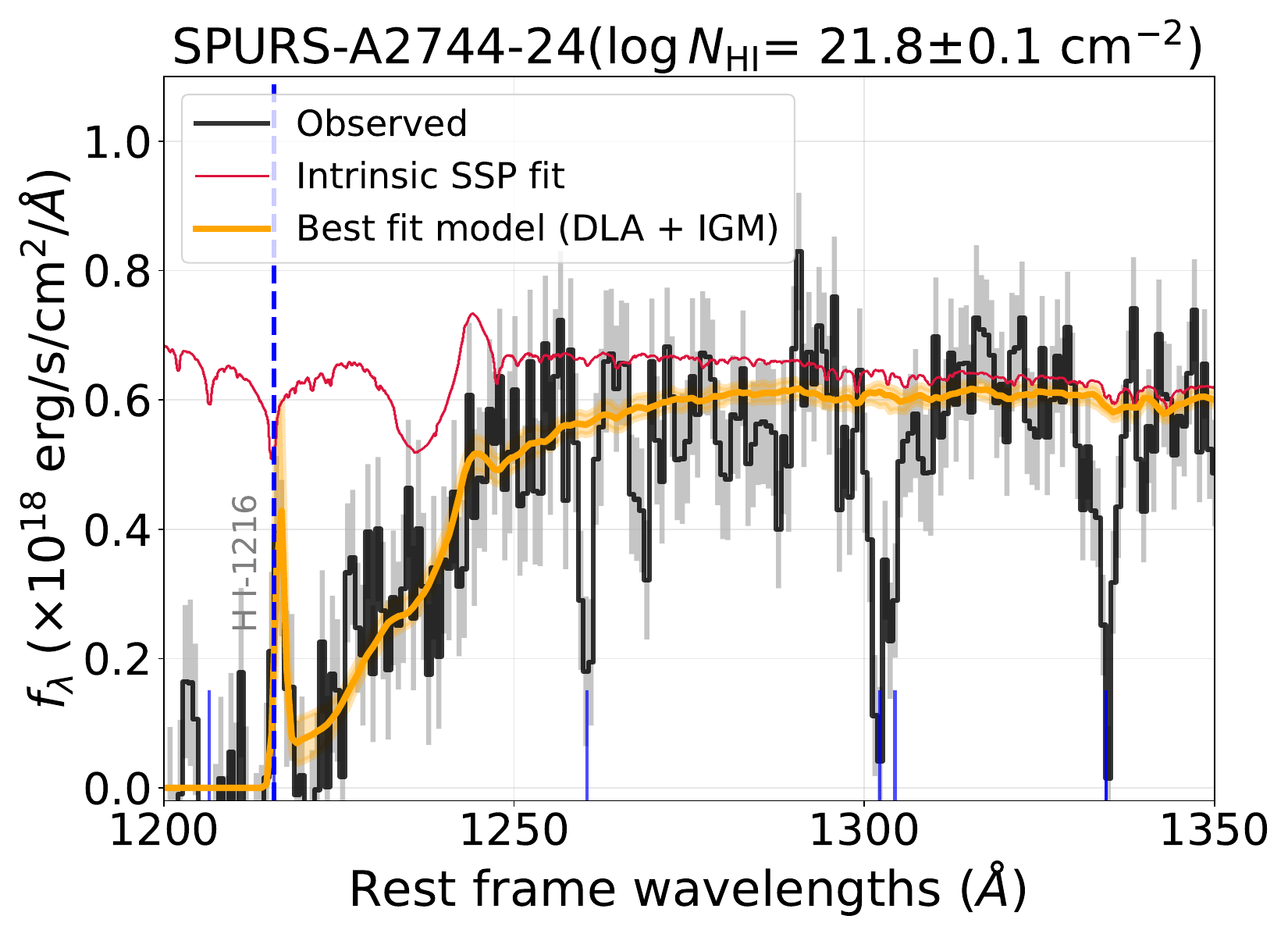}
    \caption{Fit to the \Lya\ profile in SPURS-A2744-7, 17 and 24. The black line shows the observed spectra with the error spectrum shown in gray. We show the best-fit absorption line model in orange and the intrinsic SSP fit in red. The low-ionization absorption lines are marked in blue. The profiles are sorted in increasing value of the column density ($N_{\HI}$).  
    }
    \label{fig:HI_fit}
\end{figure}
\subsection{ISM gas porosity}
\label{subsec:ISM_porosity}

The low-ionization absorption profiles (Figure~\ref{fig:ism-profiles}) reveal a striking diversity in absorption depths, painting a picture of a heterogeneous and patchy ISM. 
In Figure~\ref{fig:Cf_MUV} we show the intrinsic covering fractions for each source as a function of their UV magnitude.
We find that our galaxies have moderately high covering fractions and find no clear trend with UV luminosity.
However, we note that our least UV luminous galaxy also has the lowest covering fraction (SPURS-A2744-1069).
The intrinsic peak covering fraction of our sample has a mean and standard deviation of $0.66 \pm 0.23$.
The large scatter indicates that some galaxies are almost entirely surrounded by neutral gas while others possess more porous envelopes. 

We also show the covering fraction derived from the stacked profile of $z\sim7$ galaxies compiled by \citet{pancakez_kelsey}.
The covering fraction for the $z\sim7$ stack was determined with the same methodology employed in our sample, which is provided in Section~\ref{sec:ism_profiles}.
While our sample is slightly brighter than the mean UV luminosity of the stacked sample, we find our sources span a similar range in covering fraction.
To compare with other literature samples, which more commonly report absorption line EW, we measure the EW of \ion{Al}{2}-$\lambda$1670, which is detected in all sources in our sample.
We find the metal-enriched neutral gas reservoir is characterized by a mean absorption EW(\ion{Al}{2}-$\lambda$1670)=1.3$\pm$0.6 \,\AA, with a range from $0.6-2.1$\,\AA. 
Consistent with the derived covering fraction, this mean low-ion line EW is comparable to that reported by \citet{pancakez_kelsey} for the stack of $z\sim7$ galaxies  ($\approx 1.1$\,\AA). However, it is lower than those reported from stacked samples at $z\sim2-5$ \citep[$\approx1.5-2$\,\AA][]{shapley2003,jones2012,du2018,Pahl_2020,kvgc_ESI2022}.
This implies a potential decrease in covering fraction towards high redshifts, as discussed by \citet{pancakez_kelsey}. 

As noted in Section~\ref{subsec:notes}, for a subset of galaxies with spectral coverage of \Lya\ (SPURS-A2744-7, 17 and 24), we also measure the column density of the \HI gas reservoir surrounding these galaxies, finding a range of column densities  $\log(N_{\mathrm{HI}}) \approx 20 - 22 \,\mathrm{cm}^{-2}$. 
Figure~\ref{fig:HI_fit} shows the best-fit absorption line model to the \Lya\ profile.
The intrinsic peak low-ion covering fraction for this subset of galaxies is also high ($C_f\gtrsim60\%$).
While our sample is small, we find an increase in covering fraction with increasing \HI column density, consistent with findings at lower redshifts \citep[e.g.,][]{Reddy2016,jones_dustinthewind}, as would be expected for a clumpy ISM.
The \HI column densities we derive are in the same range as those inferred for $z\sim5-13$ galaxies from low resolution NIRSpec prism spectra \citep{Umeda2026,JWST_primal_Survey,Mason2026}, and the covering fractions and \HI column densities are also in the same range as commonly observed at higher resolution in star-forming galaxies at $z\sim2-5$ \citep[e.g.,][]{nicha2016-escape-fraction,jones_dustinthewind}, and metal-poor star-forming dwarf galaxies at $z\sim0$ \citep{McKinney2019_greenpeas,Hu2023_CLASSY}.

In summary, we find the neutral and low-ionization gas reservoirs surrounding our sample of galaxies are metal-enriched and inhomogeneous, with range of covering fractions.

\subsection{ISM gas kinematics}\label{subsec:kinematics}

We find the bulk of this cool neutral absorbing gas is moving at a wide range of velocities, as traced by the centroids of the absorption profiles ($v_{cent,\text{low-ion}}$).  
The measured velocity centroid values range from  $+$73 to $-$140\,\kms, with a sample mean and standard deviation of $-55 \pm 86$~\kms.
The intrinsic FWHM shows a moderate scatter, with a mean and standard deviation of $363 \pm 123$~\kms.
The low-ion FWHM is larger than those measured for the nebular lines ($83\pm 26$\,\kms) which implies that we are probing the kinematic structure of the ISM gas with the observed profiles.

\begin{figure}
    \centering
     \includegraphics[width=\linewidth]{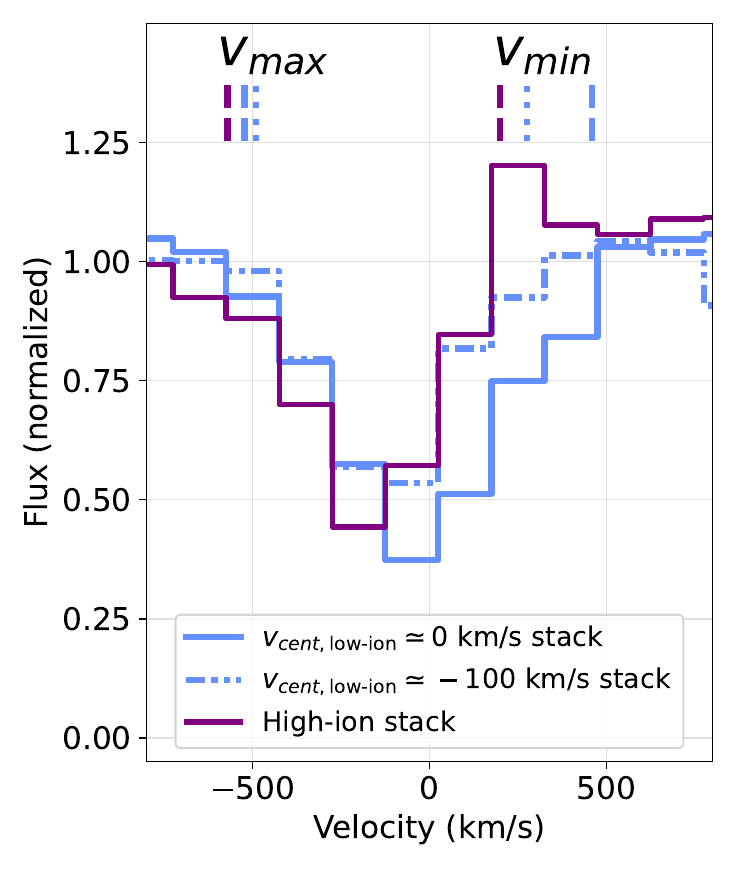}
    \caption{Comparison of the stacked absorption line profile for the low- and high-ionization lines. We divide the low-ion profiles into two groups based on their kinematics - $v_{\rm cent, low-ion}\simeq-100$~\kms\ (shown in dash-dotted blue) and $v_{\rm cent, low-ion}\simeq0$~\kms\ (shown in solid blue). The high-ion stack is shown in purple.
    We show the maximum and minimum velocities of absorption, computed as the 99\% and 1\% of absorption from best-fit profiles, with blue and purple vertical lines for the low- and high-ions respectively.
    The absorption profiles all have similar covering fractions at blueshifted velocities ($v<0$) and identical $v_{\rm max}$. However, they differ in $v_{\rm min}$, with the $v_{\rm cent} \simeq 0$\,\kms\ stack having a broader profile and showing excess absorption at systemic and redshifted velocities, resulting in a more positive (redshifted) $v_{\rm min}$ for this stack.
    }
    \label{fig:difference-lowions}
\end{figure}
\begin{figure*}[!t]
    \centering
    \includegraphics[width=0.93\linewidth]{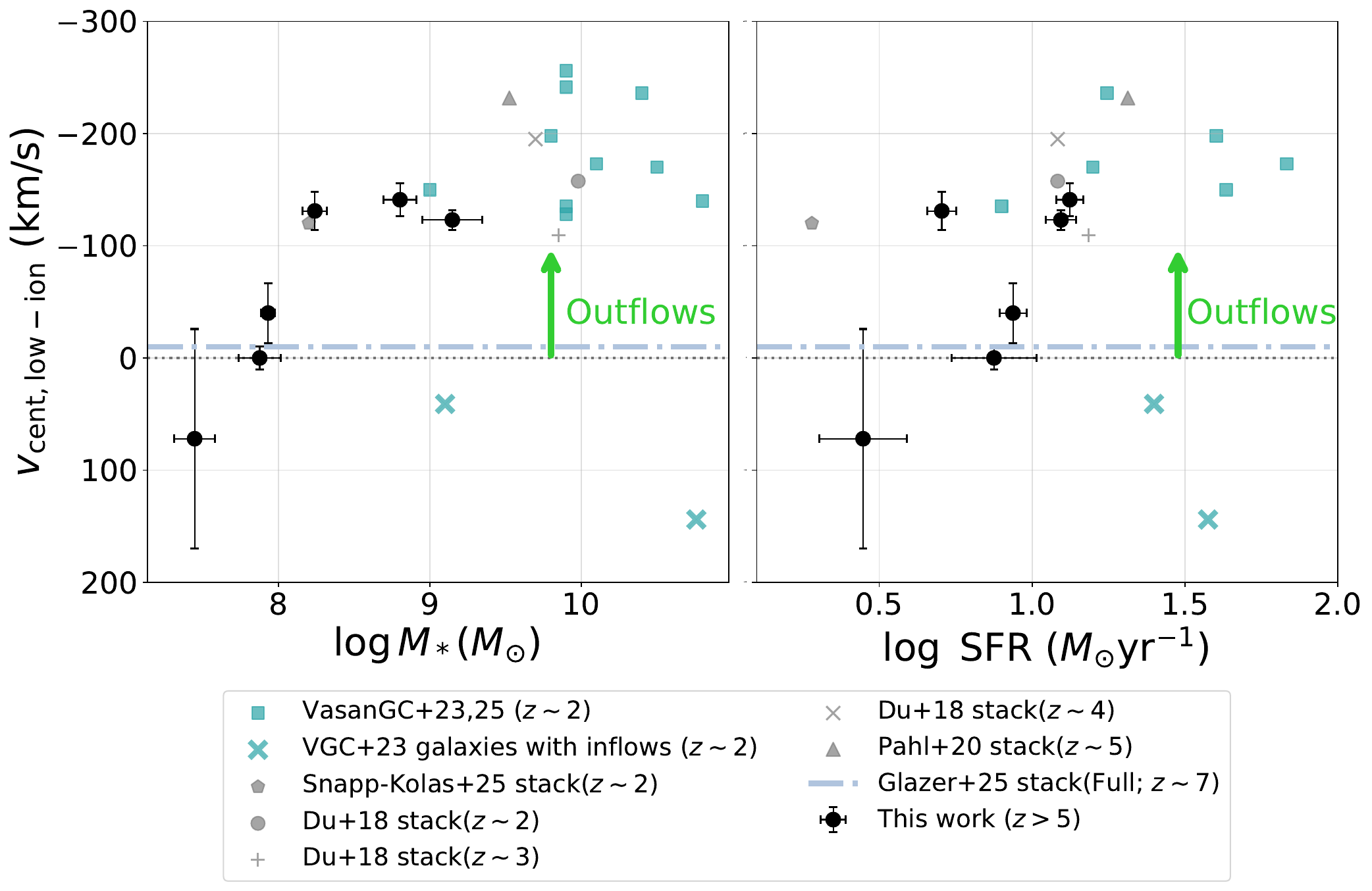}
    \caption{Low-ionization ISM absorption velocity centroid ($v_{\rm cent,low-ion}$) as a function of the stellar mass (left panel) and SFR (right panel).
    The galaxies from this work are denoted as black points. 
    The green points are centroid measurements from individual $z\sim2$ lensed galaxies \citep{kvgc_ESI2022,Vasan2025}. The gray points and the blue dash-dotted lines show the $v_{\rm cent}$ derived from spectral stacking studies at $z=2-7$ \citep[][Section~\ref{sec:ism_profiles}]{du2018, Pahl_2020, Snapp-Kolas_dwarfs_z2,pancakez_kelsey}. 
    The SPURS galaxies generally follow smooth trends in $v_{\rm cent}$, overlapping with measurements from lower-z galaxies.
     }
    \label{fig:galaxy-trends-stelmass-sfr}
\end{figure*}

Intriguingly, we find that our sample separates into two distinct kinematic groups. Half of the galaxies exhibit clear signatures of outflowing low ionization gas with blueshifted velocity centroids of $v_{\rm cent, low-ion} \simeq -100$\,\kms, while the remaining systems are consistent with little to no net bulk motion with $v_{\rm cent, low-ion} \simeq 0$\,\kms.
We demonstrate this in Figure~\ref{fig:ism-profiles} where we present the mean absorption profiles for each galaxy, and in Figure~\ref{fig:difference-lowions} where we show stacked absorption profiles from the two kinematic groups.
From the stacked profiles, we clearly see that galaxies with $v_{\rm cent, low-ion} \simeq 0$\,\kms\ are broader with higher covering fraction of gas present at systemic and redshifted velocities ($v\gtrsim0$~\kms). 
We explore differences between these groups in detail later in this paper.

\begin{figure*}
    \centering
    \begin{minipage}[t]{0.625\linewidth} 
        \centering
        \includegraphics[width=\linewidth]{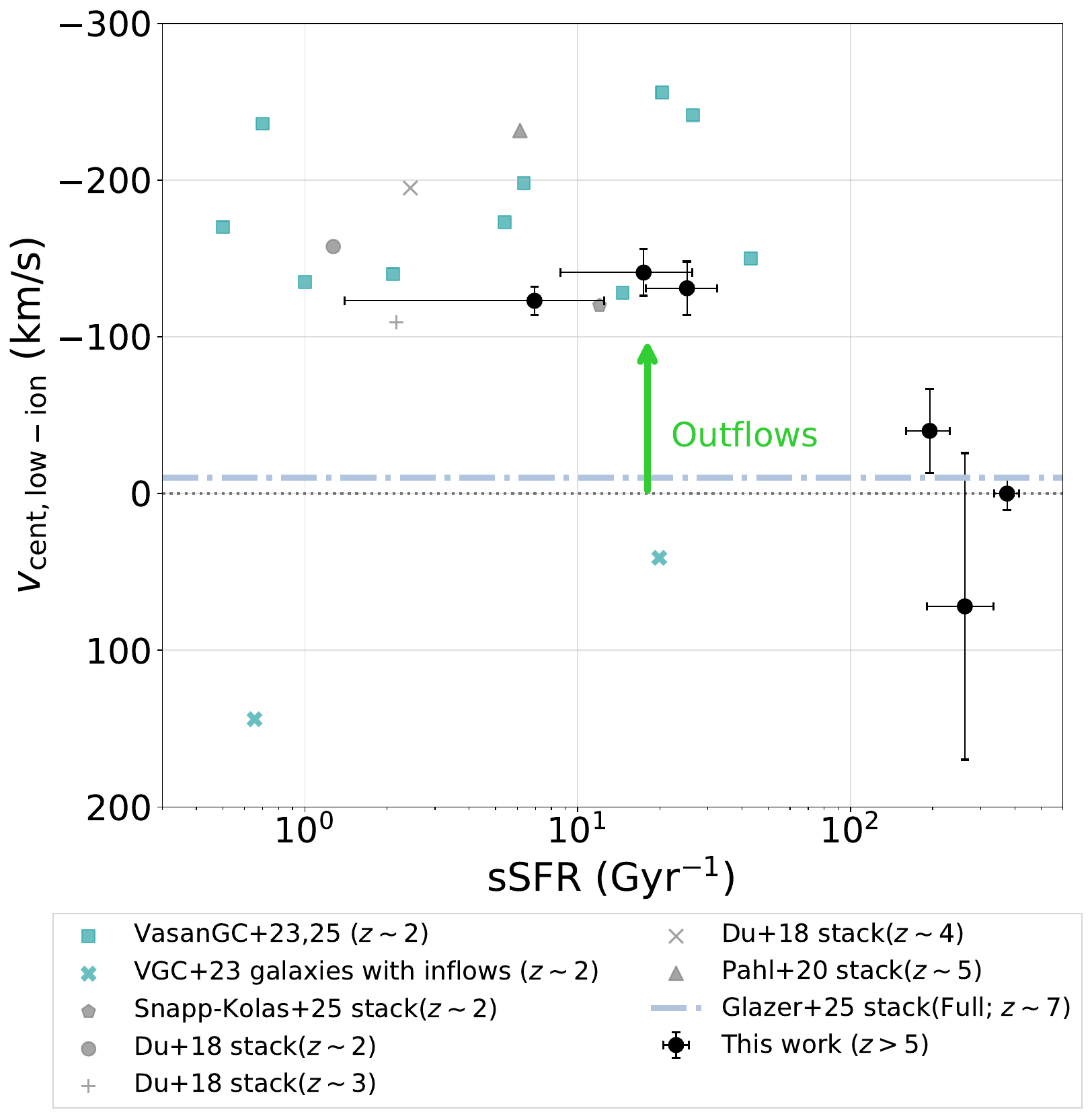}
    \end{minipage}
    \hspace{-0.2in}
    \raisebox{2.3in}{
        \begin{minipage}[t]{0.35\linewidth} 
            \centering
            \includegraphics[width=0.96\linewidth]{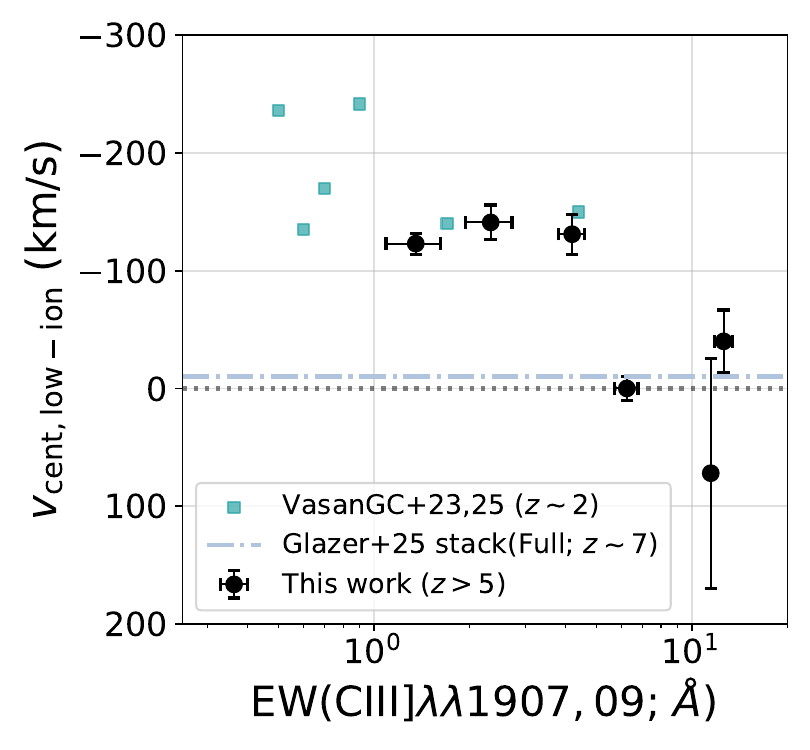}\\ 
           \includegraphics[width=\linewidth]{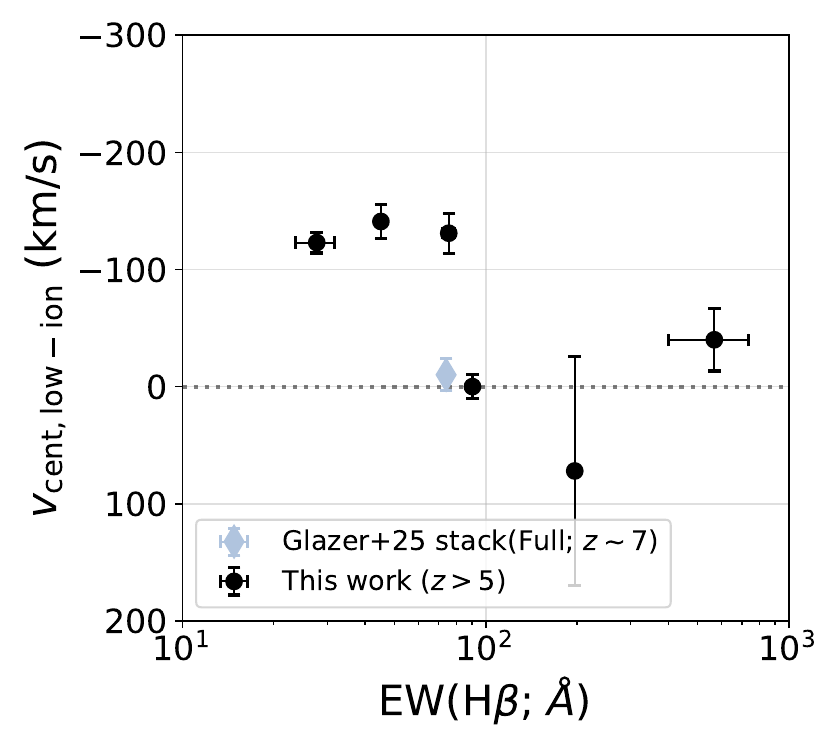}
        \end{minipage}
    }
    \caption{\emph{Left panel}: 
    Low-ionization ISM absorption velocity centroids ($v_{\rm cent,low-ion}$) as a function of sSFR. 
    The galaxies from this work are denoted in black.
    The green points are centroid measurements from individual $z\sim2$ lensed galaxies \citep{kvgc_ESI2022,Vasan2025}. 
    The gray points and the blue dash-dotted lines show the $v_{\rm cent}$ derived from spectral stacking studies at $z=2-7$ \citep[][Section~\ref{sec:ism_profiles}]{du2018, Pahl_2020, Snapp-Kolas_dwarfs_z2,pancakez_kelsey}.
    \emph{Right panels}: 
    Low-ionization ISM absorption velocity centroids ($v_{cent}$) as a function of EW(\ion{C}{3}]) and EW(H$\beta$). 
    We find that galaxies with high sSFR and high nebular line EW (which may be at the onset of a star formation burst cycle) typically have velocity centroids consistent with systemic velocities (no bulk outflows), while sources with lower sSFR and EW show net blueshifted absorption. 
    } 
    \label{fig:galaxy-trends-ssfr}
\end{figure*}

The low-ionization lines predominantly probe the neutral (\ion{H}{1}) gas phase of the baryon cycle. 
We now briefly discuss the high-ionization line profiles (Section~\ref{sec:ism_profiles}) which probe the warmer gas phases. 
Figure~\ref{fig:low-ion-high-ion-comparison} compares the mean absorption profile of the high-ionization gas to the low-ions in each source, and Figure~\ref{fig:difference-lowions} shows the stacked high- and low-ion absorption profiles for our entire sample, separating the sources with $v_{\rm cent, low-ion} \simeq -100$\,\kms and $v_{\rm cent, low-ion} \simeq 0$\,\kms.
Unlike the low-ions, we find that the high-ions exhibit blueshifted absorption in all galaxies in our sample.
We measure a mean velocity centroid for the high-ions of $-$211~\kms\ and a standard deviation of 63~\kms\ (c.f.  $-55 \pm 86$~\kms\ for the low-ions).
The differences in derived kinematics between the low- and high-ions seem to stem from the gas at $v\simeq0$~\kms. 
For instance, in SPURS-A2744-17, we measure a covering fraction at systemic velocities of $0.55$ for the low-ions compared to $\simeq0$ in the high-ions.
We also see this in the broader sample as shown in Figure~\ref{fig:difference-lowions}, where the covering fraction profile at blueshifted velocities of the low- and high-ion stacks is similar, whereas the covering fraction of the high-ions at $v\gtrsim0$~\kms is significantly lower.
The high-ion stack appears very similar to the $v_{\rm cent, low-ion} \simeq -100$\,\kms\ stack, 
which we discuss in Section~\ref{subsec:discussion-kinematics}.
Overall, this suggests that both the low- and high-ions trace an outflow component ($v<0$) with similar velocity structure.

Taken together, our observations point to a dynamic baryon cycle, characterized by a multiphase, metal-enriched gas reservoir with a large diversity in gas porosities and distinct kinematic structure.

\subsection{Trends with galaxy properties}\label{subsec:trends}

In this section, we seek to build a physical picture for the gas kinematics by assessing correlations between the kinematic profiles (Section~\ref{subsec:kinematics}) and host galaxy properties. 
Our galaxy sample spans $\simeq1-2$ orders of magnitude in stellar mass, SFR, and sSFR (Table~\ref{tab:galaxy-properties-basic}).
The stellar masses of our galaxies range from $\log M_*(\Msun)\simeq7-9$, with $\log(\mathrm{SFR})\simeq$~0.4 -- 1.1 $\Msun$ yr$^{-1}$ and  $\log(\mathrm{sSFR})\simeq$~0.8 -- 2.6 Gyr$^{-1}$ respectively.

Figure~\ref{fig:galaxy-trends-stelmass-sfr} shows the the velocity centroid of the low-ions as a function of stellar mass and SFR. 
We find that the velocity centroids correlate positively with both stellar mass and SFR. Naively, this would suggest that more massive and actively star forming galaxies are driving faster outflows in the cool gas phase, as found in cosmological simulations \citep[e.g.,][]{tng50, pandya2021}. 
However, Figure~\ref{fig:difference-lowions} shows that galaxies in our sample show similar blueshifted low- and high-ionization absorption wings, implying the presence of fast outflows, regardless of whether the average profile centroids are blueshifted or near the systemic redshift. 
Instead, as discussed above, the diversity of low-ionization absorption centroids in our sample appears to be driven by differences in the covering fraction of gas close to systemic velocities. 

We also investigate the trends between the velocity centroid and sSFR in Figure~\ref{fig:galaxy-trends-ssfr} (left panel) as well as empirical tracers of recent star formation, specifically the equivalent widths of \ion{C}{3}] and H$\beta$ nebular emission (right panel). 
We find that galaxies with high sSFR are characterized by $v_{\rm cent, low-ion} \simeq 0$\,\kms, while those with lower sSFR show velocities of $v_{\rm cent, low-ion} \simeq -100$\,\kms.
This trend is corroborated by an inverse relationship between $v_{\rm cent}$ and the nebular emission line equivalent widths.
We discuss the physical interpretations for this trend in Section~\ref{sec:discussion}.

In Figures~\ref{fig:galaxy-trends-stelmass-sfr} and \ref{fig:galaxy-trends-ssfr} we also plot measurements from lower-redshift star-forming galaxy samples at $z\sim2$--5 for comparison \footnote{We note that the galaxy properties for the lower-redshift studies used for comparison were derived assuming a constant formation history, consistent with this work (see Section~\ref{subsec:galprops}). In \citet{Pahl_2020}, the star formation history was assumed to be either constant or exponentially increasing.}.
These lower-redshift samples typically exhibit blueshifted ISM absorption velocities of $v_{\rm cent} \sim -150$~\kms, indicative of strong outflows, with a small minority having redshifted net velocities \citep[e.g.,][]{kvgc_ESI2022,weldon_inflow}. 
The subset of our $z=5$ -- 9 sample with blueshifted low-ionization absorption centroids agrees well with the trends found in other star forming galaxies at $z\sim2$--7 \citep{du2018, Pahl_2020,kvgc_ESI2022,Vasan2025, Snapp-Kolas_dwarfs_z2, pancakez_kelsey}.  
The subset of our sample with $v_{\rm cent, low-ion} \simeq 0$~\kms\ generally has lower stellar mass and higher sSFR than lower-redshift comparison samples. 
Collectively,  our sample appears to follow smooth trends in $v_{cent}$. 
We also find that the mean low-ionization velocity centroid measured from our sample ($-55 \pm 86$\,\kms) is consistent with that from stacked spectra at $z\sim7$ ($-20\pm50$\,\kms) complied by \citet{pancakez_kelsey} whose galaxies have similar mean $M_{UV}$ and $z$ (as shown in Figure~\ref{fig:redshift-MUV}). This suggests that our sample properties, including the large diversity of low ionization gas kinematics and covering fractions, are likely representative of the broader $z\sim5$-- 9 galaxy population. 

Overall, we find correlations between the kinematic measurements and the host galaxy properties in our sample. 
We discuss these results in the context of the baryon cycle and its implications for galaxy evolution in the next section.

\section{Discussion}\label{sec:discussion}
ISM absorption lines provide key insights into aspects of galaxy assembly that remain among the most enigmatic at high redshift: early chemical enrichment, feedback, and ionizing photon escape.
However, these absorption lines have so far been inaccessible for individual galaxies in the reionization era. 
The ultra-deep G140M spectroscopy presented here provides our first view of these interstellar lines in \emph{individual sources} at $z>5$, paving the way towards detailed constraints on outflows and interstellar gas, and their variation with galaxy properties.

In the sample of six galaxies assembled here with sufficient signal-to-noise in the UV, ISM absorption is ubiquitous and diverse.
The presence of such absorption provides
clear evidence of metal-enriched gas.
While absorption lines are  prominently detected in all six systems, we find a large diversity in gas porosity within the sample, and in kinematic structure for the low- and high-ions (Figures~\ref{fig:low-ion-high-ion-comparison}, \ref{fig:difference-lowions}).
{Here, we discuss the implications of our results in the context of the baryon cycle in Section~\ref{subsec:discussion-kinematics} and reionization in Section~\ref{subsec:discussion-reionization}}.

\subsection{Diversity of ISM gas kinematics in  $z>5$ galaxies and implications for the baryon cycle}\label{subsec:discussion-kinematics}

Our galaxies' ISM line profiles paint a picture of complex multiphase gas kinematics.
We find that the bulk of the neutral gas in the ISM, as traced by the low-ionization metals, has a large diversity of measured centroid velocities $v_{\rm cent, low-ion}$. 
We group our sample into two kinematic groups: half have $v_{\rm cent, low-ion}\approx-100$\,\kms, while the other half have $v_{\rm cent, low-ion}\approx0$\,\kms. 
This contrasts with our expectations based on lower-redshift studies, where nearly all star-forming galaxies with high star formation rate surface densities \citep[$\Sigma_{\rm SFR} \gtrsim 0.1 M_\odot{\rm yr}^{-1}{\rm kpc}^{-2}$; e.g.,][as is the case for every galaxy in our sample, see Section~\ref{subsec:galprops}]{heckman1990, Bron_duvet_outflows} show blueshifted velocity centroids in both low- and high-ionization lines at $\lesssim -100$\,\kms.
The stacked $z\sim7$ profile compiled by \citet{pancakez_kelsey} also demonstrated a reduced low-ionization gas centroid ($v_{\rm cent, low-ion}\approx-20\pm50$\,\kms) compared to lower redshift samples.
Here we discuss potential interpretations for this diversity of kinematics at $z>5$, based on our results.

\begin{figure*}
    \centering
   \includegraphics[width=0.75\linewidth]{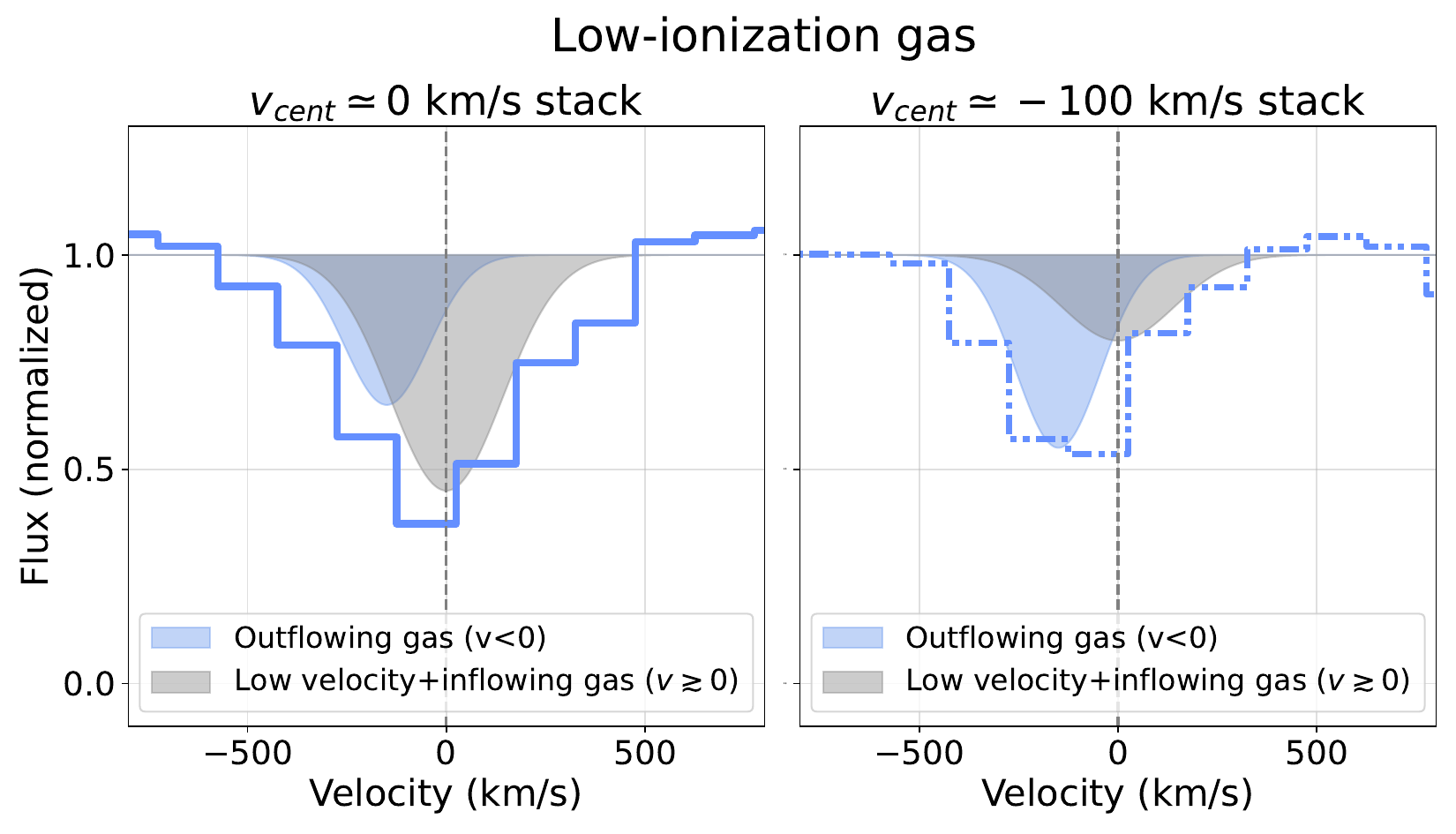}\\
        \vspace{0.15in}
      \includegraphics[width=0.94\linewidth]{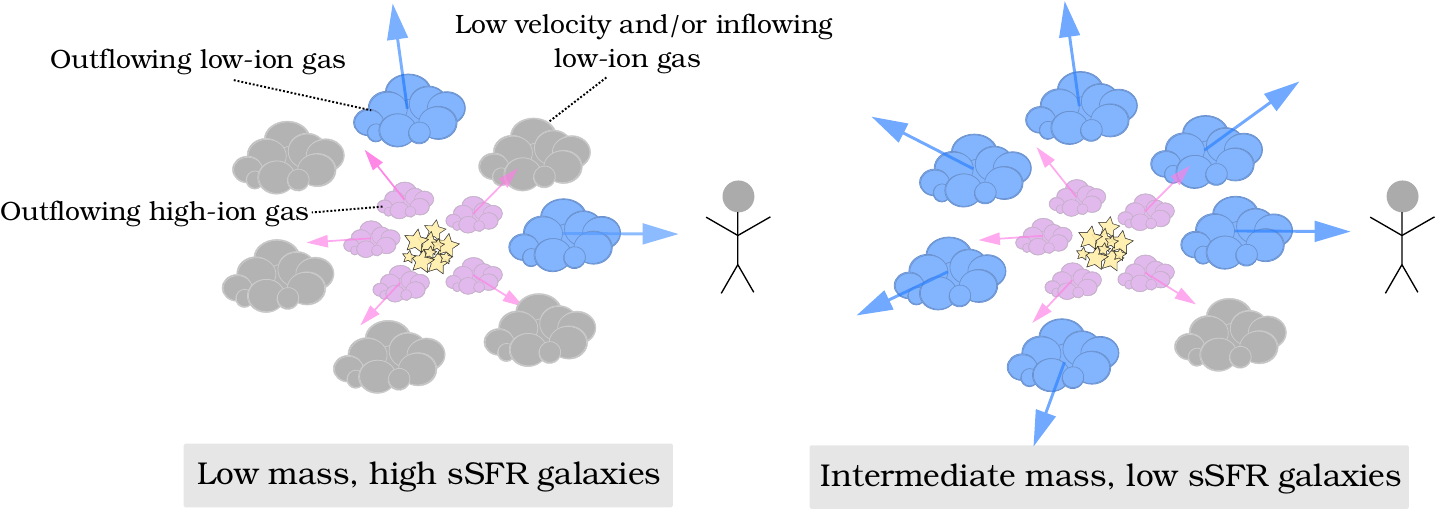}
    \caption{\emph{Top}: Stacked low-ionization absorption lines from galaxies with $v_{\rm cent,low-ion}\simeq0$\,\kms\ and $v_{\rm cen,low-iont}\simeq-100$ \kms plotted in blue and dash-dotted blue respectively. 
    The blue and gray Gaussian profiles in each panel illustrate the contributions of the ISM profile at systemic and blueshifted velocities. Systems with $v_{\rm cent,low-ion}\simeq0$\,\kms, which correspond to low mass, high sSFR in our sample, have significant absorption component at systemic velocities compared to the more massive, lower sSFR sample. 
    Both of the groups however have outflowing gas contributions in both the low-ions and high-ions (see Figure~\ref{fig:difference-lowions} for comparison). 
    A schematic illustrating these differences is shown in the bottom panel. }
    \label{fig:physical-scenario-B}
\end{figure*}

In our sample we find the two kinematic classes correlate with other galaxy properties: with the fast outflow-dominated ($v_{\rm cent,low-ion}\simeq -100$~\kms) galaxies having lower sSFR and higher stellar mass than those with velocity centroids close to systemic. 
Importantly, all of our galaxies do exhibit blueshifted absorption wings out to $\gtrsim 400$\,\kms\ in both low- and high-ionization lines (see Figure~\ref{fig:low-ion-high-ion-comparison}), indicating all galaxies have a fast component of outflowing gas with a range of temperatures and densities.
Galaxies with low‑ion centroids close to systemic ($v_{\rm cent, low-ion} \simeq 0$\,\kms) nevertheless show broader absorption profiles, implying a larger covering fraction of low-ionization gas close to systemic and/or at modestly redshifted velocities, rather than the absence of outflows in these sources. \citet{Trainor2015_faint_LAEs} discussed a similar picture for the absorption lines observed in a sample of \Lya-selected galaxies at $z\sim3$.
We illustrate this in Figure~\ref{fig:physical-scenario-B} with a schematic.
All galaxies exhibit an outflowing absorption component in the low- and high-ions, shown by the blue Gaussian in the top panels, as well as a component from low velocity, and potentially inflowing, low-ionization gas (gray Gaussian).
In this picture, galaxies with $v_{\rm cent, low-ion}\simeq0$\,\kms\ (left panel) have a larger fraction of neutral gas in their ISM close to systemic velocity within our sample.

We now consider possible physical origins for the differences in kinematics between the low- and high-ionization gas in the three sources in our sample with the bulk of their low-ionization gas close to systemic.
One possibility is that in these systems, with high sSFR and low stellar mass, the bulk of the low-ionization gas has not (yet) been efficiently accelerated to large velocities along with the hotter gas. 
Hydrodynamical simulations of multiphase winds show that entrainment of dense cold gas can be inefficient and slow, particularly when radiative cooling is weak
\citep[e.g.,][]{GronkeOh2018,HidalgoPineda2026}. 
Thus, near‑systemic low‑ion absorption may reflect an early stage of an outflow, or reduced cooling rates in these galaxies relative to typical lower redshift sources -- for example, due to lower metallicity or high densities where collisional de-excitation suppresses metal line cooling. These scenarios may be explored in larger samples.

Another possibility is a contribution from cold infalling gas at moderately redshifted velocities that is not resolved at our spectral resolution.
Such a high incidence of re-accretion of cool enriched gas (half of our small sample) would be surprising compared to lower redshift samples, where infall signatures are rare \citep[$\simeq 3$\% of galaxies at $z\sim 2$;][]{weldon_inflow,erin_inflows}, and as accretion is theoretically expected to be confined to narrow streams with relatively low covering fraction which may not be spatially coincident with outflows \citep{Dekel2009,Stewart2011_accretion}.
Nevertheless, simulations predict accretion is expected to be most prominent during the early stages of stellar mass assembly and in low-mass systems at the onset of a starburst \citep[e.g.,][]{muratov2015,sultan_coldaccretion}.

We also consider if the neutral gas component close to systemic may be related to environmental effects.
In particular, SPURS-A2744-17 is located in a known small-scale ($<100$\,pkpc) protocluster at $z\approx7.9$ \citep[][]{Ishigaki_2016_ID17,Morishita_GLASS_protocluster}, where several members show strong DLA signatures in their prism spectra, indicating local dense gas \citep{Chen2024_fesc,Witten_protocluster_GLASS,Mason2026,Terp2026}.
We infer a high \HI column density for SPURS-A2744-17 ($\log N_\mathrm{HI} =21.5\pm0.2$\,cm$^{-2}$) in our spectral fitting (see Section~\ref{subsec:continuum-norm}), providing additional evidence for dense neutral gas in its ISM/CGM. 
However, the other two sources are not in known strong overdensities, though we note SPURS-A2744-24 shows a very clumpy morphology (Figure~\ref{fig:nircam-image}) which may indicate a kinematically complex ISM.

The prospects for better understanding this kinematic diversity is promising. Growing samples of deep UV grating spectra from JWST will enable more detailed studies of multiphase outflows as a function of galaxy properties at high redshifts.
In particular, observations with higher spectral resolution will be an important next step to resolve low velocity blue and/or redshifted components in these sources to disentangle their absorption profiles.

\subsection{Neutral gas covering fractions during the Reionization era}\label{subsec:discussion-reionization}

Our results also reveal a diverse range of neutral gas covering fractions among $z > 5$ galaxies. We derive covering fractions ranging from $\approx0.23-0.91$, though the majority of our sample exhibits moderately high covering fractions ($C_\mathrm{f,cent} \gtrsim 60\%$; Figure~\ref{fig:Cf_MUV}). 
These results provide our first insights into the distribution of neutral gas covering fractions around galaxies during the epoch of reionization, one of our best indicators of ionizing photon escape \citep[e.g.,][]{Reddy2016,Gazagnes2018,Saldana-Lopez2022}.
Consistent with previous analyses of stacked $z\sim7$ spectra \citep{pancakez_kelsey}, we find a decrease in the median EW, or covering fraction, of neutral gas in our sample relative to lower redshift samples, indicating a more porous ISM.

The escape fraction of ionizing photons from galaxies is a key unknown in reionization  \citep[e.g.,][]{pahlUncontaminatedMeasurementEscaping2021,steidelKeckLymanContinuum2018}, though several hydrodynamical simulations predict some dependence on mass, with relatively low mass halos having the highest escape fractions due to their shallow potential wells \citep{Kimm2014,Paardekooper2015,Ma2020_FIRE2,Rosdahl2022_SPHINX}.
If UV luminosity generally traces halo mass, we may expect an anti-correlation between UV luminosity and ionizing photon escape fraction.
However, we find no clear trend between UV luminosity and neutral gas covering fraction in our sample (Figure~\ref{fig:Cf_MUV}). 
Nevertheless, we note the lowest UV luminosity source in our sample ($M_\mathrm{UV}=-19.3$) SPURS-A2744-1069 shows the lowest covering fraction: just $0.23\pm0.11$ (Table~\ref{tab:fit-parameters}).
As noted in Section~\ref{subsec:notes} its G395M spectrum reveals a broad H$\alpha$ line indicating it may be a broad-line AGN, and strong nebular emission lines suggesting a hard radiation field. If this source has a hard ionizing spectrum it may have ionized a significant fraction of its ISM, potentially facilitating Lyman continuum photon escape.

Building on work at lower redshifts, we also estimate the \HI gas covering fractions from our low-ionization lines.
Adopting the empirical relation between low-ionization and \HI covering fraction from \citet{Reddy2022}, the range of low-ionization covering fractions in our sample corresponds to \HI covering fractions\footnote{We note that \ion{H}{1} covering fractions estimated from low ionization absorption lines represent lower limits: at $R\sim1000$, narrow high covering fraction components may be unresolved, 
and if the gas is significantly metal-poor relative to the lower redshift samples where these empirical relations were established the \ion{H}{1} abundance may be underestimated.
However, detailed analysis of SPURS-A2744-7 suggests that at least this source is unlikely to have a very low ISM metallicity \citep[e.g.,][]{Zuyi_SPURS_ID7,Zhu2026}.} of {0.55-1}, and thus likely low ionizing photon escape fractions in the majority of our sample (assuming $f_\mathrm{esc} \approx 1-C_f(\rm HI)$).
Interestingly, we also note that SPURS-A2744-17, which was suggested to be a Lyman continuum leaker by \citet{Mascia_2023_lyCLeaker_17} and \citet{Jaskot_2024_ID17_LyCLeaker}, primarily based on its compact size and blue UV slope, has a high low-ionization covering fraction ($C_\mathrm{f,cent}=0.69\pm0.06$), indicating it may have a high covering fraction of \HI, and thus that it may not be leaking a large flux of ionizing photons along this line-of-sight. Larger samples of sources with continuum spectroscopy should enable us to establish the distribution of neutral gas covering fractions in the reionization era, and test how well other indirect indicators of Lyman continuum escape may predict the neutral gas covering fractions.

We also note that in the three $z>7$ sources where we have coverage of the Ly$\alpha$ break, the two sources with among the highest low-ionization covering fractions in our sample (SPURS-A2744-17 and SPURS-A2744-24) also show evidence for Damped Ly$\alpha$ absorption (DLA) troughs (Figure~\ref{fig:HI_fit}), with our inferred HI column densities $\log N_\mathrm{HI} \gtrsim {21}$\,cm$^{-2}$.
By contrast, SPURS-A2744-7 shows weaker low-ionization lines and correspondingly weaker DLA absorption \citep[see Figure~\ref{fig:HI_fit}, also][]{Zuyi_SPURS_ID7}.
This adds additional evidence that sources with a high covering fraction of low-ionization gas have a high covering fraction of \HI, consistent with expectations from low-$z$ samples \citep[e.g.,][]{shapley2003,Reddy2016,Tanvir2019,Reddy2022}. 
While larger samples will be needed to establish trends between \HI column densities and neutral gas covering fractions in the reionization era, these observations are consistent with a picture where the strongest DLAs in galaxy spectra correspond to sightlines with a high covering fraction of metal-enriched neutral gas clouds in the ISM, whereas sources with weaker Ly$\alpha$ absorption trace a more porous ISM that is more conducive to Lyman continuum escape.

While our current sample is still small, ongoing efforts to obtain deep UV medium- and high-resolution grating spectroscopy from SPURS and other Cycle 4 and 5 surveys will continue to grow the sample of $z>5$ sources where these measurements are possible. In particular, upcoming observations in more lensed fields (e.g. GO-12435 PI: Senchyna, GO-9645, PI: Treu, Fontana, Roberts-Borsani) will be particularly interesting for increasing the sample of intrinsically UV faint sources, which have long been posited as the primary drivers of reionization.

\section{Conclusions}\label{sec:conclusion}
In this paper, we present ultra-deep JWST/NIRSpec G140M observations of six individual UV-luminous $z\sim5-9$ galaxies in the Abell 2744 field, observed as part of the SPURS Cycle 4 Large Program.
We unambiguously detect a suite of metal absorption lines in these sources.
We use the absorption lines to characterize the gas covering fraction and ISM gas kinematics in our sample, and explore how the ISM geometry and kinematics may be linked to galaxy properties.
Our main findings are summarized below: 

\begin{enumerate}
    \item We detect ISM absorption from metal ion transitions in all galaxies in our continuum SNR-selected sample, from both low and high ionization states.
    The presence of metals implies that the ISM/CGM of these luminous $z\sim5-9$ galaxies has already been significantly enriched by previous episodes of star formation, consistent with the emerging picture from JWST observations of rapid galaxy enrichment in the first billion years \citep[e.g.][]{morishitaDiverseOxygenAbundance2024,curtiJADESInsightsLowmass2024,liInsightsMetalEnrichment2025}.

    \item The low-ionization absorption lines reveal a large diversity of neutral gas porosity, with peak covering fractions ranging from 0.23 to 0.91. This suggests that some of our galaxies have patchy `picket-fence' like ISM distributions, while others are largely engulfed by neutral gas -- likely inhibiting ionizing photon escape along the line-of-sight.
    The mean low-ion EW is comparable to that reported for stacked $z\sim7$ galaxies, but smaller than those reported at lower redshifts, suggesting an increase in ISM porosity with increasing redshift in our sample.
    For the three $z>7$ galaxies where our spectra cover the Ly$\alpha$ break, we infer \HI gas reservoirs with a range of column densities $\log(N_{\HI}) \sim {20-22}$\,cm$^{-2}$, within the range inferred in lower redshift samples.
    {While our sample is small, we find an increase in low-ion covering fraction with increasing \HI column density, as expected for a patchy, metal-enriched ISM.}
    
    \item We find that the bulk motion of the neutral gas reservoirs, as traced by the low-ionization lines, shows a large diversity in our sample, with the measured velocity centroid ($v_{\rm cent, \text{low-ion}}$) ranging from $+$73 to $-$140\,\kms\ and intrinsic FWHM ranging from 155 to 518\,\kms.
    By contrast, all galaxies exhibit outflow-dominated kinematics  in the high-ionization lines, with a mean and scatter of $v_{\rm cent, \text{high-ion}}=-211 \pm 63$\,\kms, even when the low-ionization gas centroids are consistent with the systemic velocity.
    The blueshifted absorption wings from the low- and high-ions are broadly consistent in our sample, suggesting that they trace multi-phase outflows with a range of temperatures and densities.

    \item  We find our sample falls into two kinematic groups based on their low-ionization absorption line profiles. Half of our sample exhibits clear signatures of outflowing low-ionization gas with blueshifted velocity centroids of $v_{\rm cent, \text{low-ion}}\simeq-100$\,\kms, while the remaining systems are consistent with little to no net bulk motion with $v_{\rm cent, \text{low-ion}}\simeq0$\,\kms.
    The variation in the low-ionization absorption centroids appears to be driven by differences in gas close to systemic velocities. 
    
    \item We find correlations between the low-ion velocity centroid and the host galaxy properties, which span $\simeq1-2$ orders of magnitude. Galaxies with $v_{\rm cent, \text{low-ion}}\simeq0$\,\kms\ have higher sSFR and lower stellar mass than those with outflow-dominated kinematics ($v_{\rm cent, \text{low-ion}}\simeq-100$\,\kms). 
    Given that the low- and high-ions share similar blueshifted absorption wings, the presence of cool gas reservoirs close to systemic velocities around low-mass and high sSFR galaxies does not reflect an absence of feedback.
    Instead, it may indicate that the bulk of the cool gas has not been efficiently entrained in fast multiphase outflows in these sources, and/or reflect a dense CGM/IGM environment, potentially including contributions from infalling gas at moderate redshifted velocities that are unresolved at our spectral resolution.
\end{enumerate}

This work provides a first glimpse of interstellar absorption-line properties in individual galaxies at $z>5$, highlighting broad diversity in ISM structure and kinematics. 
Upcoming and ongoing deep rest-UV spectroscopic surveys will significantly increase the available sample for absorption line studies, enabling more detailed population-level analysis.
It will be particularly promising to extend these studies to higher spectral resolution and spatially resolved rest-UV observations,
to map how outflowing and inflowing gas are distributed in galaxies.

\section*{Acknowledgments}
KVGC was supported by NASA through the STScI grants  JWST-GO-03777, JWST-GO-04265, and JWST-GO-05974.
We would like to thank Tony Pahl, Gwen Rudie and Max Gronke for insightful discussions.
CAM acknowledges support by the European Union ERC grant RISES (101163035), Carlsberg Foundation (CF22-1322), and VILLUM FONDEN (37459). Views and opinions expressed are those of the author(s) only and do not necessarily reflect those of the European Union or the European Research Council. Neither the European Union nor the granting authority can be held responsible for them.
The Cosmic Dawn Center (DAWN) is funded by the Danish National Research Foundation under grant DNRF140.
ZC acknowledges support from the VILLUM FONDEN (37459).
VG acknowledges support from the Carlsberg Foundation under grant CF22-1322. 
MA is supported by FONDECYT grant number 1252054, and gratefully acknowledges support from ANID Basal Project FB210003,  ANID MILENIO NCN2024\_112 and ANID + Vinculaci\'on Internacional + FOVI250261.
LW acknowledges support from the Gavin Boyle Fellowship at the Kavli Institute for Cosmology, Cambridge and from the Kavli Foundation.
J. G-L. is supported by FONDECYT grant No. 1252054, and gratefully acknowledges support from ANID Basal Project FB210003 and ANID MILENIO NCN2024\_112

This work is based in part on observations made with the NASA/ESA/CSA JWST. The data were obtained from the Mikulski Archive for Space Telescopes at the Space Telescope Science Institute, which is operated by the Association of Universities for Research in As- tronomy, Inc., under NASA contract NAS 5-03127 for JWST. These observations are associated with program GO 9214. We thank our program coordinator, Christian Soto, and our NIRSpec reviewer, Diane Karakla. The Tycho supercomputer hosted at the SCIENCE HPC center at the University of Copenhagen was used for supporting this work.

\bibliography{refs}{}

@ARTICLE{Mesinger2008,
       author = {{Mesinger}, Andrei and {Furlanetto}, Steven R.},
        title = "{Ly{\ensuremath{\alpha}} damping wing constraints on inhomogeneous reionization}",
      journal = {\mnras},
         year = 2008,
        month = apr,
       volume = {385},
       number = {3},
        pages = {1348-1358},
          doi = {10.1111/j.1365-2966.2007.12836.x},
archivePrefix = {arXiv},
       eprint = {0710.0371},
 primaryClass = {astro-ph},
       adsurl = {https://ui.adsabs.harvard.edu/abs/2008MNRAS.385.1348M}
}

@ARTICLE{Neufeld1990,
       author = {{Neufeld}, David A.},
        title = "{The Transfer of Resonance-Line Radiation in Static Astrophysical Media}",
      journal = {\apj},
         year = 1990,
        month = feb,
       volume = {350},
        pages = {216},
          doi = {10.1086/168375},
       adsurl = {https://ui.adsabs.harvard.edu/abs/1990ApJ...350..216N}
}

@ARTICLE{Almada2026,
       author = {{Almada Monter}, Silvia and {Gronke}, Max and {Chang}, Seok-Jun},
        title = "{Lyman-{\ensuremath{\alpha}} escape through anisotropic media}",
      journal = {\mnras},
         year = 2026,
        month = apr,
       volume = {547},
       number = {2},
          eid = {stag330},
        pages = {stag330},
          doi = {10.1093/mnras/stag330},
archivePrefix = {arXiv},
       eprint = {2509.19184},
 primaryClass = {astro-ph.GA},
       adsurl = {https://ui.adsabs.harvard.edu/abs/2026MNRAS.547ag330A}
}

@ARTICLE{Chen2024_fesc,
       author = {{Chen}, Zuyi and {Stark}, Daniel P. and {Mason}, Charlotte and {Topping}, Michael W. and {Whitler}, Lily and {Tang}, Mengtao and {Endsley}, Ryan and {Charlot}, St{\'e}phane},
        title = "{JWST spectroscopy of z   5-8 UV-selected galaxies: new constraints on the evolution of the Ly {\ensuremath{\alpha}} escape fraction in the reionization era}",
      journal = {\mnras},
         year = 2024,
        month = mar,
       volume = {528},
       number = {4},
        pages = {7052-7075},
          doi = {10.1093/mnras/stae455},
archivePrefix = {arXiv},
       eprint = {2311.13683},
 primaryClass = {astro-ph.GA},
       adsurl = {https://ui.adsabs.harvard.edu/abs/2024MNRAS.528.7052C}
}

@ARTICLE{Tanvir2019,
       author = {{Tanvir}, N.~R. and {Fynbo}, J.~P.~U. and {de Ugarte Postigo}, A. and {Japelj}, J. and {Wiersema}, K. and {Malesani}, D. and {Perley}, D.~A. and {Levan}, A.~J. and {Selsing}, J. and {Cenko}, S.~B. and {Kann}, D.~A. and {Milvang-Jensen}, B. and {Berger}, E. and {Cano}, Z. and {Chornock}, R. and {Covino}, S. and {Cucchiara}, A. and {D'Elia}, V. and {Gargiulo}, A. and {Goldoni}, P. and {Gomboc}, A. and {Heintz}, K.~E. and {Hjorth}, J. and {Izzo}, L. and {Jakobsson}, P. and {Kaper}, L. and {Kr{\"u}hler}, T. and {Laskar}, T. and {Myers}, M. and {Piranomonte}, S. and {Pugliese}, G. and {Rossi}, A. and {S{\'a}nchez-Ram{\'\i}rez}, R. and {Schulze}, S. and {Sparre}, M. and {Stanway}, E.~R. and {Tagliaferri}, G. and {Th{\"o}ne}, C.~C. and {Vergani}, S. and {Vreeswijk}, P.~M. and {Wijers}, R.~A.~M.~J. and {Watson}, D. and {Xu}, D.},
        title = "{The fraction of ionizing radiation from massive stars that escapes to the intergalactic medium}",
      journal = {\mnras},
         year = 2019,
        month = mar,
       volume = {483},
       number = {4},
        pages = {5380-5408},
          doi = {10.1093/mnras/sty3460},
archivePrefix = {arXiv},
       eprint = {1805.07318},
 primaryClass = {astro-ph.GA},
       adsurl = {https://ui.adsabs.harvard.edu/abs/2019MNRAS.483.5380T}
}

@ARTICLE{Umeda2026,
       author = {{Umeda}, Hiroya and {Ouchi}, Masami and {Nakajima}, Kimihiko and {Harikane}, Yuichi and {Ono}, Yoshiaki and {Xu}, Yi and {Isobe}, Yuki and {Zhang}, Yechi},
        title = "{JWST Measurements of Neutral Hydrogen Fractions and Ionized Bubble Sizes at z = 7─12 Obtained with Ly{\ensuremath{\alpha}} Damping Wing Absorptions in 27 Bright Continuum Galaxies}",
      journal = {\apj},
         year = 2024,
        month = aug,
       volume = {971},
       number = {2},
          eid = {124},
        pages = {124},
          doi = {10.3847/1538-4357/ad554e},
archivePrefix = {arXiv},
       eprint = {2306.00487},
 primaryClass = {astro-ph.GA},
       adsurl = {https://ui.adsabs.harvard.edu/abs/2024ApJ...971..124U}
}

@ARTICLE{Terp2026,
       author = {{Terp}, Chamilla and {Heintz}, Kasper E. and {Matthee}, Jorryt and {Naidu}, Rohan P. and {Oesch}, Pascal A. and {Witten}, Callum and {Kashino}, Daichi and {Pollock}, Clara L. and {Di Cesare}, Claudia and {Torralba}, Alberto},
        title = "{All the Massive Galaxy Overdensities during Reionization: JWST Rest-Frame Optical Selection Reveals Young, Chemically Evolved Galaxies Embedded in Dense, Neutral Gas at z > 5}",
      journal = {arXiv e-prints},
         year = 2026,
        month = feb,
          eid = {arXiv:2602.09091},
        pages = {arXiv:2602.09091},
          doi = {10.48550/arXiv.2602.09091},
archivePrefix = {arXiv},
       eprint = {2602.09091},
 primaryClass = {astro-ph.GA},
       adsurl = {https://ui.adsabs.harvard.edu/abs/2026arXiv260209091T}
}

@ARTICLE{HidalgoPineda2026,
       author = {{Hidalgo-Pineda}, Fernando and {Gronke}, Max and {Grete}, Philipp},
        title = "{The launching of galactic winds from a multiphase ISM}",
      journal = {\mnras},
         year = 2026,
        month = may,
       volume = {548},
       number = {1},
          eid = {stag539},
        pages = {stag539},
          doi = {10.1093/mnras/stag539},
archivePrefix = {arXiv},
       eprint = {2510.14829},
 primaryClass = {astro-ph.GA},
       adsurl = {https://ui.adsabs.harvard.edu/abs/2026MNRAS.548ag539H}
}

@ARTICLE{GronkeOh2018,
       author = {{Gronke}, Max and {Oh}, S. Peng},
        title = "{The growth and entrainment of cold gas in a hot wind}",
      journal = {\mnras},
         year = 2018,
        month = oct,
       volume = {480},
       number = {1},
        pages = {L111-L115},
          doi = {10.1093/mnrasl/sly131},
archivePrefix = {arXiv},
       eprint = {1806.02728},
 primaryClass = {astro-ph.GA},
       adsurl = {https://ui.adsabs.harvard.edu/abs/2018MNRAS.480L.111G}
}

@ARTICLE{Mason2018,
       author = {{Mason}, Charlotte A. and {Treu}, Tommaso and {de Barros}, Stephane and {Dijkstra}, Mark and {Fontana}, Adriano and {Mesinger}, Andrei and {Pentericci}, Laura and {Trenti}, Michele and {Vanzella}, Eros},
        title = "{Beacons into the Cosmic Dark Ages: Boosted Transmission of Ly{\ensuremath{\alpha}} from UV Bright Galaxies at z {\ensuremath{\gtrsim}} 7}",
      journal = {\apjl},
         year = 2018,
        month = apr,
       volume = {857},
       number = {2},
          eid = {L11},
        pages = {L11},
          doi = {10.3847/2041-8213/aabbab},
archivePrefix = {arXiv},
       eprint = {1801.01891},
 primaryClass = {astro-ph.CO},
       adsurl = {https://ui.adsabs.harvard.edu/abs/2018ApJ...857L..11M}
}

@ARTICLE{Dijkstra2011,
       author = {{Dijkstra}, Mark and {Mesinger}, Andrei and {Wyithe}, J. Stuart B.},
        title = "{The detectability of Ly{\ensuremath{\alpha}} emission from galaxies during the epoch of reionization}",
      journal = {\mnras},
         year = 2011,
        month = jul,
       volume = {414},
       number = {3},
        pages = {2139-2147},
          doi = {10.1111/j.1365-2966.2011.18530.x},
archivePrefix = {arXiv},
       eprint = {1101.5160},
 primaryClass = {astro-ph.CO},
       adsurl = {https://ui.adsabs.harvard.edu/abs/2011MNRAS.414.2139D}
}

@ARTICLE{Algera2025,
       author = {{Algera}, Hiddo S.~B. and {Weaver}, John R. and {Bakx}, Tom J.~L.~C. and {Aravena}, Manuel and {Bouwens}, Rychard J. and {Cescon}, Karin and {Chen}, Chian-Chou and {da Cunha}, Elisabete and {Dayal}, Pratika and {Faisst}, Andreas and {Ferrara}, Andrea and {Fujimoto}, Seiji and {Hashimoto}, Takuya and {Heintz}, Kasper and {Herrera-Camus}, Rodrigo and {Hodge}, Jacqueline and {Inami}, Hanae and {Inoue}, Akio K. and {Matthee}, Jorryt and {Meyer}, Romain and {Mizukoshi}, Shoichiro and {Mondal}, Chayan and {Nanayakkara}, Themiya and {Oesch}, Pascal A. and {Pallottini}, Andrea and {R{\"o}ttgering}, Huub and {Rowland}, Lucie E. and {Schouws}, Sander and {Smit}, Renske and {Sommovigo}, Laura and {Stark}, Daniel P. and {Sugahara}, Yuma and {Vallini}, Livia and {Vijarnwannaluk}, Bovornpratch and {van der Werf}, Paul and {Werner}, Norbert and {Witstok}, Joris and {Xiao}, Mengyuan},
        title = "{A first systematic study of [OIII] 88$μ$m at $z>8$: two luminous oxygen lines and a powerful ionized outflow in the first 600 million years}",
      journal = {arXiv e-prints},
         year = 2025,
        month = dec,
          eid = {arXiv:2512.14486},
        pages = {arXiv:2512.14486},
          doi = {10.48550/arXiv.2512.14486},
archivePrefix = {arXiv},
       eprint = {2512.14486},
 primaryClass = {astro-ph.GA},
       adsurl = {https://ui.adsabs.harvard.edu/abs/2025arXiv251214486A}
}

@ARTICLE{Trainor2015_faint_LAEs,
       author = {{Trainor}, Ryan F. and {Steidel}, Charles C. and {Strom}, Allison L. and {Rudie}, Gwen C.},
        title = "{The Spectroscopic Properties of Ly{\ensuremath{\alpha}}-Emitters at z {\ensuremath{\sim}}2.7: Escaping Gas and Photons from Faint Galaxies}",
      journal = {\apj},
         year = 2015,
        month = aug,
       volume = {809},
       number = {1},
          eid = {89},
        pages = {89},
          doi = {10.1088/0004-637X/809/1/89},
archivePrefix = {arXiv},
       eprint = {1506.08205},
 primaryClass = {astro-ph.GA},
       adsurl = {https://ui.adsabs.harvard.edu/abs/2015ApJ...809...89T}
}

@ARTICLE{Tang2024_Lyaprofile,
       author = {{Tang}, Mengtao and {Stark}, Daniel P. and {Ellis}, Richard S. and {Sun}, Fengwu and {Topping}, Michael and {Robertson}, Brant and {Tacchella}, Sandro and {Arribas}, Santiago and {Baker}, William M. and {Bhatawdekar}, Rachana and {Boyett}, Kristan and {Bunker}, Andrew J. and {Charlot}, St{\'e}phane and {Chen}, Zuyi and {Chevallard}, Jacopo and {Jones}, Gareth C. and {Kumari}, Nimisha and {Lyu}, Jianwei and {Maiolino}, Roberto and {Maseda}, Michael V. and {Saxena}, Aayush and {Whitler}, Lily and {Williams}, Christina C. and {Willott}, Chris and {Witstok}, Joris},
        title = "{Ly{\ensuremath{\alpha}} emission in galaxies at z ≃ 5-6: new insight from JWST into the statistical distributions of Ly{\ensuremath{\alpha}} properties at the end of reionization}",
      journal = {\mnras},
         year = 2024,
        month = jun,
       volume = {531},
       number = {2},
        pages = {2701-2730},
          doi = {10.1093/mnras/stae1338},
archivePrefix = {arXiv},
       eprint = {2402.06070},
 primaryClass = {astro-ph.GA},
       adsurl = {https://ui.adsabs.harvard.edu/abs/2024MNRAS.531.2701T}
}

@ARTICLE{Bergamini2023_lensmodel,
       author = {{Bergamini}, P. and {Acebron}, A. and {Grillo}, C. and {Rosati}, P. and {Caminha}, G.~B. and {Mercurio}, A. and {Vanzella}, E. and {Angora}, G. and {Brammer}, G. and {Meneghetti}, M. and {Nonino}, M.},
        title = "{New high-precision strong lensing modeling of Abell 2744. Preparing for JWST observations}",
      journal = {\aap},
         year = 2023,
        month = feb,
       volume = {670},
          eid = {A60},
        pages = {A60},
          doi = {10.1051/0004-6361/202244575},
archivePrefix = {arXiv},
       eprint = {2207.09416},
 primaryClass = {astro-ph.CO},
       adsurl = {https://ui.adsabs.harvard.edu/abs/2023A&A...670A..60B}
}

@ARTICLE{Lotz2017_frontierfields,
       author = {{Lotz}, J.~M. and {Koekemoer}, A. and {Coe}, D. and {Grogin}, N. and {Capak}, P. and {Mack}, J. and {Anderson}, J. and {Avila}, R. and {Barker}, E.~A. and {Borncamp}, D. and {Brammer}, G. and {Durbin}, M. and {Gunning}, H. and {Hilbert}, B. and {Jenkner}, H. and {Khandrika}, H. and {Levay}, Z. and {Lucas}, R.~A. and {MacKenty}, J. and {Ogaz}, S. and {Porterfield}, B. and {Reid}, N. and {Robberto}, M. and {Royle}, P. and {Smith}, L.~J. and {Storrie-Lombardi}, L.~J. and {Sunnquist}, B. and {Surace}, J. and {Taylor}, D.~C. and {Williams}, R. and {Bullock}, J. and {Dickinson}, M. and {Finkelstein}, S. and {Natarajan}, P. and {Richard}, J. and {Robertson}, B. and {Tumlinson}, J. and {Zitrin}, A. and {Flanagan}, K. and {Sembach}, K. and {Soifer}, B.~T. and {Mountain}, M.},
        title = "{The Frontier Fields: Survey Design and Initial Results}",
      journal = {\apj},
         year = 2017,
        month = mar,
       volume = {837},
       number = {1},
          eid = {97},
        pages = {97},
          doi = {10.3847/1538-4357/837/1/97},
archivePrefix = {arXiv},
       eprint = {1605.06567},
 primaryClass = {astro-ph.GA},
       adsurl = {https://ui.adsabs.harvard.edu/abs/2017ApJ...837...97L}
}

@ARTICLE{Dekel2009,
       author = {{Dekel}, A. and {Birnboim}, Y. and {Engel}, G. and {Freundlich}, J. and {Goerdt}, T. and {Mumcuoglu}, M. and {Neistein}, E. and {Pichon}, C. and {Teyssier}, R. and {Zinger}, E.},
        title = "{Cold streams in early massive hot haloes as the main mode of galaxy formation}",
      journal = {\nat},
         year = 2009,
        month = jan,
       volume = {457},
       number = {7228},
        pages = {451-454},
          doi = {10.1038/nature07648},
archivePrefix = {arXiv},
       eprint = {0808.0553},
 primaryClass = {astro-ph},
       adsurl = {https://ui.adsabs.harvard.edu/abs/2009Natur.457..451D}
}

@ARTICLE{Anna_Rubies,
       author = {{de Graaff}, Anna and {Brammer}, Gabriel and {Weibel}, Andrea and {Lewis}, Zach and {Maseda}, Michael V. and {Oesch}, Pascal A. and {Bezanson}, Rachel and {Boogaard}, Leindert A. and {Cleri}, Nikko J. and {Cooper}, Olivia R. and {Gottumukkala}, Rashmi and {Greene}, Jenny E. and {Hirschmann}, Michaela and {Hviding}, Raphael E. and {Katz}, Harley and {Labb{\'e}}, Ivo and {Leja}, Joel and {Matthee}, Jorryt and {McConachie}, Ian and {Miller}, Tim B. and {Naidu}, Rohan P. and {Price}, Sedona H. and {Rix}, Hans-Walter and {Setton}, David J. and {Suess}, Katherine A. and {Wang}, Bingjie and {Whitaker}, Katherine E. and {Williams}, Christina C.},
        title = "{RUBIES: A complete census of the bright and red distant Universe with JWST/NIRSpec}",
      journal = {\aap},
         year = 2025,
        month = may,
       volume = {697},
          eid = {A189},
        pages = {A189},
          doi = {10.1051/0004-6361/202452186},
archivePrefix = {arXiv},
       eprint = {2409.05948},
 primaryClass = {astro-ph.GA},
       adsurl = {https://ui.adsabs.harvard.edu/abs/2025A&A...697A.189D}
}

@ARTICLE{Stewart2011_accretion,
       author = {{Stewart}, Kyle R. and {Kaufmann}, Tobias and {Bullock}, James S. and {Barton}, Elizabeth J. and {Maller}, Ariyeh H. and {Diemand}, J{\"u}rg and {Wadsley}, James},
        title = "{Observing the End of Cold Flow Accretion Using Halo Absorption Systems}",
      journal = {\apjl},
         year = 2011,
        month = jul,
       volume = {735},
       number = {1},
          eid = {L1},
        pages = {L1},
          doi = {10.1088/2041-8205/735/1/L1},
archivePrefix = {arXiv},
       eprint = {1012.2128},
 primaryClass = {astro-ph.CO},
       adsurl = {https://ui.adsabs.harvard.edu/abs/2011ApJ...735L...1S}
}

@ARTICLE{Rosdahl2022_SPHINX,
       author = {{Rosdahl}, Joakim and {Blaizot}, J{\'e}r{\'e}my and {Katz}, Harley and {Kimm}, Taysun and {Garel}, Thibault and {Haehnelt}, Martin and {Keating}, Laura C. and {Martin-Alvarez}, Sergio and {Michel-Dansac}, L{\'e}o and {Ocvirk}, Pierre},
        title = "{LyC escape from SPHINX galaxies in the Epoch of Reionization}",
      journal = {\mnras},
         year = 2022,
        month = sep,
       volume = {515},
       number = {2},
        pages = {2386-2414},
          doi = {10.1093/mnras/stac1942},
archivePrefix = {arXiv},
       eprint = {2207.03232},
 primaryClass = {astro-ph.GA},
       adsurl = {https://ui.adsabs.harvard.edu/abs/2022MNRAS.515.2386R}
}

@ARTICLE{Robertson2022_ARA,
       author = {{Robertson}, Brant E.},
        title = "{Galaxy Formation and Reionization: Key Unknowns and Expected Breakthroughs by the James Webb Space Telescope}",
      journal = {\araa},
         year = 2022,
        month = aug,
       volume = {60},
        pages = {121-158},
          doi = {10.1146/annurev-astro-120221-044656},
archivePrefix = {arXiv},
       eprint = {2110.13160},
 primaryClass = {astro-ph.CO},
       adsurl = {https://ui.adsabs.harvard.edu/abs/2022ARA&A..60..121R}
}

@ARTICLE{Hu2023_CLASSY,
       author = {{Hu}, Weida and {Martin}, Crystal L. and {Gronke}, Max and {Gazagnes}, Simon and {Hayes}, Matthew and {Chisholm}, John and {Heckman}, Timothy and {Mingozzi}, Matilde and {Roy}, Namrata and {Senchyna}, Peter and {Xu}, Xinfeng and {Berg}, Danielle A. and {James}, Bethan L. and {Stark}, Daniel P. and {Arellano-C{\'o}rdova}, Karla Z. and {Henry}, Alaina and {Jaskot}, Anne E. and {Kumari}, Nimisha and {Parker}, Kaelee S. and {Scarlata}, Claudia and {Wofford}, Aida and {Amor{\'\i}n}, Ricardo O. and {Leonhardes-Barboza}, Naunet and {Brinchmann}, Jarle and {Carr}, Cody and {Aloisi}, Alessandra},
        title = "{CLASSY VII Ly{\ensuremath{\alpha}} Profiles: The Structure and Kinematics of Neutral Gas and Implications for LyC Escape in Reionization-era Analogs}",
      journal = {\apj},
         year = 2023,
        month = oct,
       volume = {956},
       number = {1},
          eid = {39},
        pages = {39},
          doi = {10.3847/1538-4357/aceefd},
archivePrefix = {arXiv},
       eprint = {2307.04911},
 primaryClass = {astro-ph.GA},
       adsurl = {https://ui.adsabs.harvard.edu/abs/2023ApJ...956...39H}
}

@ARTICLE{McKinney2019_greenpeas,
       author = {{McKinney}, Jed H. and {Jaskot}, Anne E. and {Oey}, M.~S. and {Yun}, Min S. and {Dowd}, Tara and {Lowenthal}, James D.},
        title = "{Neutral Gas Properties and Ly{\ensuremath{\alpha}} Escape in Extreme Green Pea Galaxies}",
      journal = {\apj},
         year = 2019,
        month = mar,
       volume = {874},
       number = {1},
          eid = {52},
        pages = {52},
          doi = {10.3847/1538-4357/ab08eb},
archivePrefix = {arXiv},
       eprint = {1902.08204},
 primaryClass = {astro-ph.GA},
       adsurl = {https://ui.adsabs.harvard.edu/abs/2019ApJ...874...52M}
}

@ARTICLE{Mason2026,
       author = {{Mason}, Charlotte A. and {Chen}, Zuyi and {Stark}, Daniel P. and {Yi Lu}, Ting and {Topping}, Michael and {Tang}, Mengtao},
        title = "{Constraints on the z {\ensuremath{\sim}} 6{\ensuremath{-}}13 intergalactic medium from JWST spectroscopy of Lyman-alpha damping wings in galaxies}",
      journal = {\aap},
         year = 2026,
        month = jan,
       volume = {705},
          eid = {A114},
        pages = {A114},
          doi = {10.1051/0004-6361/202553820},
archivePrefix = {arXiv},
       eprint = {2501.11702},
 primaryClass = {astro-ph.GA},
       adsurl = {https://ui.adsabs.harvard.edu/abs/2026A&A...705A.114M}
}

@ARTICLE{Chen2026,
       author = {{Chen}, Zuyi and {Stark}, Daniel P. and {Mason}, Charlotte A. and {Plat}, Adele and {Gelli}, Viola and {Senchyna}, Peter and {Keerthi Vasan G.}, C. and {Endsley}, Ryan and {Tang}, Mengtao and {Topping}, Michael W. and {Whitler}, Lily},
        title = "{SPURS: Bursty Star Formation in an Extremely Luminous Weak Emission Line Galaxy at $z=9.3$}",
      journal = {arXiv e-prints},
         year = 2026,
        month = apr,
          eid = {arXiv:2604.21516},
        pages = {arXiv:2604.21516},
          doi = {10.48550/arXiv.2604.21516},
archivePrefix = {arXiv},
       eprint = {2604.21516},
 primaryClass = {astro-ph.GA},
       adsurl = {https://ui.adsabs.harvard.edu/abs/2026arXiv260421516C}
}

@ARTICLE{Borsani_lymanbreakgalaxies,
       author = {{Roberts-Borsani}, Guido and {Morishita}, Takahiro and {Treu}, Tommaso and {Brammer}, Gabriel and {Strait}, Victoria and {Wang}, Xin and {Bradac}, Marusa and {Acebron}, Ana and {Bergamini}, Pietro and {Boyett}, Kristan and {Calabr{\'o}}, Antonello and {Castellano}, Marco and {Fontana}, Adriano and {Glazebrook}, Karl and {Grillo}, Claudio and {Henry}, Alaina and {Jones}, Tucker and {Malkan}, Matthew and {Marchesini}, Danilo and {Mascia}, Sara and {Mason}, Charlotte and {Mercurio}, Amata and {Merlin}, Emiliano and {Nanayakkara}, Themiya and {Pentericci}, Laura and {Rosati}, Piero and {Santini}, Paola and {Scarlata}, Claudia and {Trenti}, Michele and {Vanzella}, Eros and {Vulcani}, Benedetta and {Willott}, Chris},
        title = "{Early Results from GLASS-JWST. I: Confirmation of Lensed z {\ensuremath{\geq}} 7 Lyman-break Galaxies behind the Abell 2744 Cluster with NIRISS}",
      journal = {\apjl},
         year = 2022,
        month = oct,
       volume = {938},
       number = {2},
          eid = {L13},
        pages = {L13},
          doi = {10.3847/2041-8213/ac8e6e},
archivePrefix = {arXiv},
       eprint = {2207.11387},
 primaryClass = {astro-ph.GA},
       adsurl = {https://ui.adsabs.harvard.edu/abs/2022ApJ...938L..13R}
}

@ARTICLE{Zhu2026,
       author = {{Zhu}, Yongda and {Ji}, Zhiyuan and {Becker}, George D. and {Ding}, Jiani and {Egami}, Eiichi and {Fan}, Xiaohui and {Jin}, Xiangyu and {Liu}, Weizhe and {Lyu}, Jianwei and {Ma}, Zheng and {Narisetty}, Suprabhas and {Rieke}, George H. and {Wu}, Yunjing and {Yue}, Minghao and {Zhang}, Junyu and {Rieke}, Marcia J.},
        title = "{Early metal-enriched baryon cycling before the midpoint of cosmic reionization}",
      journal = {arXiv e-prints},
         year = 2026,
        month = apr,
          eid = {arXiv:2604.21218},
        pages = {arXiv:2604.21218},
          doi = {10.48550/arXiv.2604.21218},
archivePrefix = {arXiv},
       eprint = {2604.21218},
 primaryClass = {astro-ph.GA},
       adsurl = {https://ui.adsabs.harvard.edu/abs/2026arXiv260421218Z}
}

@ARTICLE{Saldana-Lopez2022,
       author = {{Saldana-Lopez}, Alberto and {Schaerer}, Daniel and {Chisholm}, John and {Flury}, Sophia R. and {Jaskot}, Anne E. and {Worseck}, G{\'a}bor and {Makan}, Kirill and {Gazagnes}, Simon and {Mauerhofer}, Valentin and {Verhamme}, Anne and {Amor{\'\i}n}, Ricardo O. and {Ferguson}, Harry C. and {Giavalisco}, Mauro and {Grazian}, Andrea and {Hayes}, Matthew J. and {Heckman}, Timothy M. and {Henry}, Alaina and {Ji}, Zhiyuan and {Marques-Chaves}, Rui and {McCandliss}, Stephan R. and {Oey}, M. Sally and {{\"O}stlin}, G{\"o}ran and {Pentericci}, Laura and {Thuan}, Trinh X. and {Trebitsch}, Maxime and {Vanzella}, Eros and {Xu}, Xinfeng},
        title = "{The Low-Redshift Lyman Continuum Survey. Unveiling the ISM properties of low-z Lyman-continuum emitters}",
      journal = {\aap},
         year = 2022,
        month = jul,
       volume = {663},
          eid = {A59},
        pages = {A59},
          doi = {10.1051/0004-6361/202141864},
archivePrefix = {arXiv},
       eprint = {2201.11800},
 primaryClass = {astro-ph.GA},
       adsurl = {https://ui.adsabs.harvard.edu/abs/2022A&A...663A..59S}
}

@ARTICLE{Reddy2022,
       author = {{Reddy}, Naveen A. and {Topping}, Michael W. and {Shapley}, Alice E. and {Steidel}, Charles C. and {Sanders}, Ryan L. and {Du}, Xinnan and {Coil}, Alison L. and {Mobasher}, Bahram and {Price}, Sedona H. and {Shivaei}, Irene},
        title = "{The Effects of Stellar Population and Gas Covering Fraction on the Emergent Ly{\ensuremath{\alpha}} Emission of High-redshift Galaxies}",
      journal = {\apj},
         year = 2022,
        month = feb,
       volume = {926},
       number = {1},
          eid = {31},
        pages = {31},
          doi = {10.3847/1538-4357/ac3b4c},
archivePrefix = {arXiv},
       eprint = {2108.05363},
 primaryClass = {astro-ph.GA},
       adsurl = {https://ui.adsabs.harvard.edu/abs/2022ApJ...926...31R}
}

@ARTICLE{Gazagnes2018,
       author = {{Gazagnes}, S. and {Chisholm}, J. and {Schaerer}, D. and {Verhamme}, A. and {Rigby}, J.~R. and {Bayliss}, M.},
        title = "{Neutral gas properties of Lyman continuum emitting galaxies: Column densities and covering fractions from UV absorption lines}",
      journal = {\aap},
         year = 2018,
        month = aug,
       volume = {616},
          eid = {A29},
        pages = {A29},
          doi = {10.1051/0004-6361/201832759},
archivePrefix = {arXiv},
       eprint = {1802.06378},
 primaryClass = {astro-ph.GA},
       adsurl = {https://ui.adsabs.harvard.edu/abs/2018A&A...616A..29G}
}

@ARTICLE{Reddy2016,
       author = {{Reddy}, Naveen A. and {Steidel}, Charles C. and {Pettini}, Max and {Bogosavljevi{\'c}}, Milan and {Shapley}, Alice E.},
        title = "{The Connection Between Reddening, Gas Covering Fraction, and the Escape of Ionizing Radiation at High Redshift}",
      journal = {\apj},
         year = 2016,
        month = sep,
       volume = {828},
       number = {2},
          eid = {108},
        pages = {108},
          doi = {10.3847/0004-637X/828/2/108},
archivePrefix = {arXiv},
       eprint = {1606.03452},
 primaryClass = {astro-ph.GA},
       adsurl = {https://ui.adsabs.harvard.edu/abs/2016ApJ...828..108R}
}

@ARTICLE{sultan_coldaccretion,
       author = {{Sultan}, Imran and {Faucher-Gigu{\`e}re}, Claude-Andr{\'e} and {Stern}, Jonathan and {Sun}, Guochao},
        title = "{Cold vs. Hot Gas Accretion and Angular Momentum in FIRE Simulations: From Halo to Galaxy Scales}",
      journal = {arXiv e-prints},
         year = 2026,
        month = apr,
          eid = {arXiv:2604.14273},
        pages = {arXiv:2604.14273},
archivePrefix = {arXiv},
       eprint = {2604.14273},
 primaryClass = {astro-ph.GA},
       adsurl = {https://ui.adsabs.harvard.edu/abs/2026arXiv260414273S}
}

@ARTICLE{Mascia_2023_lyCLeaker_17,
       author = {{Mascia}, S. and {Pentericci}, L. and {Calabr{\`o}}, A. and {Treu}, T. and {Santini}, P. and {Yang}, L. and {Napolitano}, L. and {Roberts-Borsani}, G. and {Bergamini}, P. and {Grillo}, C. and {Rosati}, P. and {Vulcani}, B. and {Castellano}, M. and {Boyett}, K. and {Fontana}, A. and {Glazebrook}, K. and {Henry}, A. and {Mason}, C. and {Merlin}, E. and {Morishita}, T. and {Nanayakkara}, T. and {Paris}, D. and {Roy}, N. and {Williams}, H. and {Wang}, X. and {Brammer}, G. and {Brada{\v{c}}}, M. and {Chen}, W. and {Kelly}, P.~L. and {Koekemoer}, A.~M. and {Trenti}, M. and {Windhorst}, R.~A.},
        title = "{Closing in on the sources of cosmic reionization: First results from the GLASS-JWST program}",
      journal = {\aap},
         year = 2023,
        month = apr,
       volume = {672},
          eid = {A155},
        pages = {A155},
          doi = {10.1051/0004-6361/202345866},
archivePrefix = {arXiv},
       eprint = {2301.02816},
 primaryClass = {astro-ph.GA},
       adsurl = {https://ui.adsabs.harvard.edu/abs/2023A&A...672A.155M}
}

@ARTICLE{Jaskot_2024_ID17_LyCLeaker,
       author = {{Jaskot}, Anne E. and {Silveyra}, Anneliese C. and {Plantinga}, Anna and {Flury}, Sophia R. and {Hayes}, Matthew and {Chisholm}, John and {Heckman}, Timothy and {Pentericci}, Laura and {Schaerer}, Daniel and {Trebitsch}, Maxime and {Verhamme}, Anne and {Carr}, Cody and {Ferguson}, Henry C. and {Ji}, Zhiyuan and {Giavalisco}, Mauro and {Henry}, Alaina and {Marques-Chaves}, Rui and {{\"O}stlin}, G{\"o}ran and {Saldana-Lopez}, Alberto and {Scarlata}, Claudia and {Worseck}, G{\'a}bor and {Xu}, Xinfeng},
        title = "{Multivariate Predictors of Lyman Continuum Escape. II. Predicting Lyman Continuum Escape Fractions for High-redshift Galaxies}",
      journal = {\apj},
         year = 2024,
        month = oct,
       volume = {973},
       number = {2},
          eid = {111},
        pages = {111},
          doi = {10.3847/1538-4357/ad5557},
archivePrefix = {arXiv},
       eprint = {2406.10179},
 primaryClass = {astro-ph.GA},
       adsurl = {https://ui.adsabs.harvard.edu/abs/2024ApJ...973..111J}
}

@ARTICLE{Morishita_GLASS_protocluster,
       author = {{Morishita}, Takahiro and {Roberts-Borsani}, Guido and {Treu}, Tommaso and {Brammer}, Gabriel and {Mason}, Charlotte A. and {Trenti}, Michele and {Vulcani}, Benedetta and {Wang}, Xin and {Acebron}, Ana and {Bah{\'e}}, Yannick and {Bergamini}, Pietro and {Boyett}, Kristan and {Bradac}, Marusa and {Calabr{\`o}}, Antonello and {Castellano}, Marco and {Chen}, Wenlei and {De Lucia}, Gabriella and {Filippenko}, Alexei V. and {Fontana}, Adriano and {Glazebrook}, Karl and {Grillo}, Claudio and {Henry}, Alaina and {Jones}, Tucker and {Kelly}, Patrick L. and {Koekemoer}, Anton M. and {Leethochawalit}, Nicha and {Lu}, Ting-Yi and {Marchesini}, Danilo and {Mascia}, Sara and {Mercurio}, Amata and {Merlin}, Emiliano and {Metha}, Benjamin and {Nanayakkara}, Themiya and {Nonino}, Mario and {Paris}, Diego and {Pentericci}, Laura and {Rosati}, Piero and {Santini}, Paola and {Strait}, Victoria and {Vanzella}, Eros and {Windhorst}, Rogier A. and {Xie}, Lizhi},
        title = "{Early Results from GLASS-JWST. XIV. A Spectroscopically Confirmed Protocluster 650 Million Years after the Big Bang}",
      journal = {\apjl},
         year = 2023,
        month = apr,
       volume = {947},
       number = {2},
          eid = {L24},
        pages = {L24},
          doi = {10.3847/2041-8213/acb99e},
archivePrefix = {arXiv},
       eprint = {2211.09097},
 primaryClass = {astro-ph.GA},
       adsurl = {https://ui.adsabs.harvard.edu/abs/2023ApJ...947L..24M}
}

@ARTICLE{Zuyi_SPURS_ID7,
       author = {{Chen}, Zuyi and {Stark}, Daniel P. and {Mason}, Charlotte A. and {Plat}, Adele and {Gelli}, Viola and {Senchyna}, Peter and {Keerthi Vasan G.}, C. and {Endsley}, Ryan and {Tang}, Mengtao and {Topping}, Michael W. and {Whitler}, Lily},
        title = "{SPURS: Bursty Star Formation in an Extremely Luminous Weak Emission Line Galaxy at $z=9.3$}",
      journal = {arXiv e-prints},
         year = 2026,
        month = apr,
          eid = {arXiv:2604.21516},
        pages = {arXiv:2604.21516},
          doi = {10.48550/arXiv.2604.21516},
archivePrefix = {arXiv},
       eprint = {2604.21516},
 primaryClass = {astro-ph.GA},
       adsurl = {https://ui.adsabs.harvard.edu/abs/2026arXiv260421516C}
}

@ARTICLE{tomasso_glass,
       author = {{Treu}, T. and {Roberts-Borsani}, G. and {Bradac}, M. and {Brammer}, G. and {Fontana}, A. and {Henry}, A. and {Mason}, C. and {Morishita}, T. and {Pentericci}, L. and {Wang}, X. and {Acebron}, A. and {Bagley}, M. and {Bergamini}, P. and {Belfiori}, D. and {Bonchi}, A. and {Boyett}, K. and {Boutsia}, K. and {Calabr{\'o}}, A. and {Caminha}, G.~B. and {Castellano}, M. and {Dressler}, A. and {Glazebrook}, K. and {Grillo}, C. and {Jacobs}, C. and {Jones}, T. and {Kelly}, P.~L. and {Leethochawalit}, N. and {Malkan}, M.~A. and {Marchesini}, D. and {Mascia}, S. and {Mercurio}, A. and {Merlin}, E. and {Nanayakkara}, T. and {Nonino}, M. and {Paris}, D. and {Poggianti}, B. and {Rosati}, P. and {Santini}, P. and {Scarlata}, C. and {Shipley}, H.~V. and {Strait}, V. and {Trenti}, M. and {Tubthong}, C. and {Vanzella}, E. and {Vulcani}, B. and {Yang}, L.},
        title = "{The GLASS-JWST Early Release Science Program. I. Survey Design and Release Plans}",
      journal = {\apj},
         year = 2022,
        month = aug,
       volume = {935},
       number = {2},
          eid = {110},
        pages = {110},
          doi = {10.3847/1538-4357/ac8158},
archivePrefix = {arXiv},
       eprint = {2206.07978},
 primaryClass = {astro-ph.GA},
       adsurl = {https://ui.adsabs.harvard.edu/abs/2022ApJ...935..110T}
}

@ARTICLE{Anna_spectralresolution,
       author = {{de Graaff}, Anna and {Rix}, Hans-Walter and {Carniani}, Stefano and {Suess}, Katherine A. and {Charlot}, St{\'e}phane and {Curtis-Lake}, Emma and {Arribas}, Santiago and {Baker}, William M. and {Boyett}, Kristan and {Bunker}, Andrew J. and {Cameron}, Alex J. and {Chevallard}, Jacopo and {Curti}, Mirko and {Eisenstein}, Daniel J. and {Franx}, Marijn and {Hainline}, Kevin and {Hausen}, Ryan and {Ji}, Zhiyuan and {Johnson}, Benjamin D. and {Jones}, Gareth C. and {Maiolino}, Roberto and {Maseda}, Michael V. and {Nelson}, Erica and {Parlanti}, Eleonora and {Rawle}, Tim and {Robertson}, Brant and {Tacchella}, Sandro and {{\"U}bler}, Hannah and {Williams}, Christina C. and {Willmer}, Christopher N.~A. and {Willott}, Chris},
        title = "{Ionised gas kinematics and dynamical masses of z {\ensuremath{\gtrsim}} 6 galaxies from JADES/NIRSpec high-resolution spectroscopy}",
      journal = {\aap},
         year = 2024,
        month = apr,
       volume = {684},
          eid = {A87},
        pages = {A87},
          doi = {10.1051/0004-6361/202347755},
archivePrefix = {arXiv},
       eprint = {2308.09742},
 primaryClass = {astro-ph.GA},
       adsurl = {https://ui.adsabs.harvard.edu/abs/2024A&A...684A..87D}
}

@ARTICLE{anowar_spectralresolution,
       author = {{Shajib}, Anowar J. and {Treu}, Tommaso and {Melo}, Alejandra and {Roberts-Borsani}, Guido and {Knabel}, Shawn and {Cappellari}, Michele and {Frieman}, Joshua A.},
        title = "{An accurate measurement of the spectral resolution of the JWST Near Infrared Spectrograph}",
      journal = {\aap},
         year = 2025,
        month = oct,
       volume = {702},
          eid = {L12},
        pages = {L12},
          doi = {10.1051/0004-6361/202556281},
archivePrefix = {arXiv},
       eprint = {2507.03746},
 primaryClass = {astro-ph.IM},
       adsurl = {https://ui.adsabs.harvard.edu/abs/2025A&A...702L..12S}
}

@ARTICLE{NIRSpec_boker,
       author = {{B{\"o}ker}, T. and {Beck}, T.~L. and {Birkmann}, S.~M. and {Giardino}, G. and {Keyes}, C. and {Kumari}, N. and {Muzerolle}, J. and {Rawle}, T. and {Zeidler}, P. and {Abul-Huda}, Y. and {Alves de Oliveira}, C. and {Arribas}, S. and {Bechtold}, K. and {Bhatawdekar}, R. and {Bonaventura}, N. and {Bunker}, A.~J. and {Cameron}, A.~J. and {Carniani}, S. and {Charlot}, S. and {Curti}, M. and {Espinoza}, N. and {Ferruit}, P. and {Franx}, M. and {Jakobsen}, P. and {Karakla}, D. and {L{\'o}pez-Caniego}, M. and {L{\"u}tzgendorf}, N. and {Maiolino}, R. and {Manjavacas}, E. and {Marston}, A.~P. and {Moseley}, S.~H. and {Ogle}, P. and {Perna}, M. and {Pe{\~n}a-Guerrero}, M. and {Pirzkal}, N. and {Plesha}, R. and {Proffitt}, C.~R. and {Rauscher}, B.~J. and {Rix}, H.-W. and {Rodr{\'\i}guez del Pino}, B. and {Rustamkulov}, Z. and {Sabbi}, E. and {Sing}, D.~K. and {Sirianni}, M. and {te Plate}, M. and {{\'U}beda}, L. and {Wahlgren}, G.~M. and {Wislowski}, E. and {Wu}, R. and {Willott}, Chris J.},
        title = "{In-orbit Performance of the Near-infrared Spectrograph NIRSpec on the James Webb Space Telescope}",
      journal = {\pasp},
         year = 2023,
        month = mar,
       volume = {135},
       number = {1045},
          eid = {038001},
        pages = {038001},
          doi = {10.1088/1538-3873/acb846},
archivePrefix = {arXiv},
       eprint = {2301.13766},
 primaryClass = {astro-ph.IM},
       adsurl = {https://ui.adsabs.harvard.edu/abs/2023PASP..135c8001B}
}

@ARTICLE{NIRSpec_Jakobsen,
       author = {{Jakobsen}, P. and {Ferruit}, P. and {Alves de Oliveira}, C. and {Arribas}, S. and {Bagnasco}, G. and {Barho}, R. and {Beck}, T.~L. and {Birkmann}, S. and {B{\"o}ker}, T. and {Bunker}, A.~J. and {Charlot}, S. and {de Jong}, P. and {de Marchi}, G. and {Ehrenwinkler}, R. and {Falcolini}, M. and {Fels}, R. and {Franx}, M. and {Franz}, D. and {Funke}, M. and {Giardino}, G. and {Gnata}, X. and {Holota}, W. and {Honnen}, K. and {Jensen}, P.~L. and {Jentsch}, M. and {Johnson}, T. and {Jollet}, D. and {Karl}, H. and {Kling}, G. and {K{\"o}hler}, J. and {Kolm}, M.-G. and {Kumari}, N. and {Lander}, M.~E. and {Lemke}, R. and {L{\'o}pez-Caniego}, M. and {L{\"u}tzgendorf}, N. and {Maiolino}, R. and {Manjavacas}, E. and {Marston}, A. and {Maschmann}, M. and {Maurer}, R. and {Messerschmidt}, B. and {Moseley}, S.~H. and {Mosner}, P. and {Mott}, D.~B. and {Muzerolle}, J. and {Pirzkal}, N. and {Pittet}, J.-F. and {Plitzke}, A. and {Posselt}, W. and {Rapp}, B. and {Rauscher}, B.~J. and {Rawle}, T. and {Rix}, H.-W. and {R{\"o}del}, A. and {Rumler}, P. and {Sabbi}, E. and {Salvignol}, J.-C. and {Schmid}, T. and {Sirianni}, M. and {Smith}, C. and {Strada}, P. and {te Plate}, M. and {Valenti}, J. and {Wettemann}, T. and {Wiehe}, T. and {Wiesmayer}, M. and {Willott}, C.~J. and {Wright}, R. and {Zeidler}, P. and {Zincke}, C.},
        title = "{The Near-Infrared Spectrograph (NIRSpec) on the James Webb Space Telescope. I. Overview of the instrument and its capabilities}",
      journal = {\aap},
         year = 2022,
        month = may,
       volume = {661},
          eid = {A80},
        pages = {A80},
          doi = {10.1051/0004-6361/202142663},
archivePrefix = {arXiv},
       eprint = {2202.03305},
 primaryClass = {astro-ph.IM},
       adsurl = {https://ui.adsabs.harvard.edu/abs/2022A&A...661A..80J}
}

@BOOK{osterbrock-2006,
       author = {{Osterbrock}, Donald E. and {Ferland}, Gary J.},
        title = "{Astrophysics of gaseous nebulae and active galactic nuclei}",
         year = 2006,
       adsurl = {https://ui.adsabs.harvard.edu/abs/2006agna.book.....O}
}

@ARTICLE{boyett-spurs7,
       author = {{Boyett}, Kristan and {Trenti}, Michele and {Leethochawalit}, Nicha and {Calabr{\'o}}, Antonello and {Metha}, Benjamin and {Roberts-Borsani}, Guido and {Dalmasso}, Nicol{\'o} and {Yang}, Lilan and {Santini}, Paola and {Treu}, Tommaso and {Jones}, Tucker and {Henry}, Alaina and {Mason}, Charlotte A. and {Morishita}, Takahiro and {Nanayakkara}, Themiya and {Roy}, Namrata and {Wang}, Xin and {Fontana}, Adriano and {Merlin}, Emiliano and {Castellano}, Marco and {Paris}, Diego and {Brada{\v{c}}}, Maru{\v{s}}a and {Malkan}, Matt and {Marchesini}, Danilo and {Mascia}, Sara and {Glazebrook}, Karl and {Pentericci}, Laura and {Vanzella}, Eros and {Vulcani}, Benedetta},
        title = "{A massive interacting galaxy 510 million years after the Big Bang}",
      journal = {Nature Astronomy},
         year = 2024,
        month = may,
       volume = {8},
        pages = {657-672},
          doi = {10.1038/s41550-024-02218-7},
archivePrefix = {arXiv},
       eprint = {2303.00306},
 primaryClass = {astro-ph.GA},
       adsurl = {https://ui.adsabs.harvard.edu/abs/2024NatAs...8..657B}
}

@ARTICLE{erin_inflows,
       author = {{Coleman}, Erin and {Keerthi}, Vasan G.~C. and {Chen}, Yuguang and {Jones}, Tucker and {Rhoades}, Sunny and {Ellis}, Richard and {Stark}, Dan and {Leethochawalit}, Nicha and {Sanders}, Ryan and {Mortensen}, Kris and {Glazebrook}, Karl and {Kacprzak}, Glenn G.},
        title = "{Detection of Gas Inflow during the Onset of a Starburst in a Low-mass Galaxy at z = 2.45}",
      journal = {\apjl},
         year = 2024,
        month = dec,
       volume = {977},
       number = {1},
          eid = {L23},
        pages = {L23},
          doi = {10.3847/2041-8213/ad93d0},
archivePrefix = {arXiv},
       eprint = {2409.00308},
 primaryClass = {astro-ph.GA},
       adsurl = {https://ui.adsabs.harvard.edu/abs/2024ApJ...977L..23C}
}

@ARTICLE{Prochaska2011,
       author = {{Prochaska}, J. Xavier and {Kasen}, Daniel and {Rubin}, Kate},
        title = "{Simple Models of Metal-line Absorption and Emission from Cool Gas Outflows}",
      journal = {\apj},
         year = 2011,
        month = jun,
       volume = {734},
       number = {1},
          eid = {24},
        pages = {24},
          doi = {10.1088/0004-637X/734/1/24},
archivePrefix = {arXiv},
       eprint = {1102.3444},
 primaryClass = {astro-ph.GA},
       adsurl = {https://ui.adsabs.harvard.edu/abs/2011ApJ...734...24P}
}

@article{kvgc_ESI2022,
doi = {10.3847/1538-4357/acf462},
url = {https://dx.doi.org/10.3847/1538-4357/acf462},
year = {2023},
month = {dec},
publisher = {The American Astronomical Society},
volume = {959},
number = {2},
pages = {124},
author = {Keerthi {Vasan G. C.} and Tucker Jones and Ryan L. Sanders and Richard S. Ellis and Daniel P. Stark and Glenn G. Kacprzak and Tania M. Barone and Kim-Vy H. Tran and Karl Glazebrook and Colin Jacobs},
title = {Resolved Velocity Profiles of Galactic Winds at Cosmic Noon},
journal = {The Astrophysical Journal}
}

@ARTICLE{Chabrier2003,
       author = {{Chabrier}, Gilles},
        title = "{Galactic Stellar and Substellar Initial Mass Function}",
      journal = {\pasp},
         year = 2003,
        month = jul,
       volume = {115},
       number = {809},
        pages = {763-795},
          doi = {10.1086/376392},
archivePrefix = {arXiv},
       eprint = {astro-ph/0304382},
 primaryClass = {astro-ph},
       adsurl = {https://ui.adsabs.harvard.edu/abs/2003PASP..115..763C}
}

@ARTICLE{weldon_inflow,
       author = {{Weldon}, Andrew and {Reddy}, Naveen A. and {Topping}, Michael W. and {Shapley}, Alice E. and {Du}, Xinnan and {Price}, Sedona H. and {Sanders}, Ryan L. and {Coil}, Alison L. and {Mobasher}, Bahram and {Kriek}, Mariska and {Siana}, Brian and {Rezaee}, Saeed},
        title = "{The MOSDEF-LRIS survey: detection of inflowing gas towards three star-forming galaxies at z   2}",
      journal = {\mnras},
         year = 2023,
        month = aug,
       volume = {523},
       number = {4},
        pages = {5624-5634},
          doi = {10.1093/mnras/stad1615},
archivePrefix = {arXiv},
       eprint = {2305.16394},
 primaryClass = {astro-ph.GA},
       adsurl = {https://ui.adsabs.harvard.edu/abs/2023MNRAS.523.5624W}
}

@ARTICLE{Chen_overdensity,
       author = {{Chen}, Zuyi and {Stark}, Daniel P. and {Mason}, Charlotte A. and {Tang}, Mengtao and {Whitler}, Lily and {Lu}, Ting-Yi and {Topping}, Michael W.},
        title = "{The Impact of Galaxy Overdensities and Ionized Bubbles on Ly{\ensuremath{\alpha}} Emission at z {\ensuremath{\sim}} 7.0─8.5}",
      journal = {\apj},
         year = 2026,
        month = apr,
       volume = {1001},
       number = {2},
          eid = {236},
        pages = {236},
          doi = {10.3847/1538-4357/ae4220},
archivePrefix = {arXiv},
       eprint = {2505.24080},
 primaryClass = {astro-ph.GA},
       adsurl = {https://ui.adsabs.harvard.edu/abs/2026ApJ..1001..236C}
}

@ARTICLE{jones_dustinthewind,
       author = {{Jones}, Tucker and {Stark}, Daniel P. and {Ellis}, Richard S.},
        title = "{Dust in the Wind: Composition and Kinematics of Galaxy Outflows at the Peak Epoch of Star Formation}",
      journal = {\apj},
         year = 2018,
        month = aug,
       volume = {863},
       number = {2},
          eid = {191},
        pages = {191},
          doi = {10.3847/1538-4357/aad37f},
archivePrefix = {arXiv},
       eprint = {1805.01484},
 primaryClass = {astro-ph.GA},
       adsurl = {https://ui.adsabs.harvard.edu/abs/2018ApJ...863..191J}
}

@ARTICLE{Vasan2025,
       author = {Keerthi {Vasan G. C.} and {Jones}, Tucker and {Shajib}, Anowar J. and {Rhoades}, Sunny and {Chen}, Yuguang and {Sanders}, Ryan L. and {Stark}, Daniel P. and {Ellis}, Richard S. and {Leethochawalit}, Nicha and {Kacprzak}, Glenn G. and {Barone}, Tania M. and {Glazebrook}, Karl and {Tran}, Kim-Vy H. and {Skobe}, Hannah and {Mortensen}, Kris and {Barisic}, Ivana},
        title = "{Spatially Resolved Galactic Winds at Cosmic Noon: Outflow Kinematics and Mass Loading in a Lensed Star-forming Galaxy at z = 1.87}",
      journal = {\apj},
         year = 2025,
        month = mar,
       volume = {981},
       number = {2},
          eid = {105},
        pages = {105},
          doi = {10.3847/1538-4357/ada95b},
archivePrefix = {arXiv},
       eprint = {2402.00942},
 primaryClass = {astro-ph.GA},
       adsurl = {https://ui.adsabs.harvard.edu/abs/2025ApJ...981..105V}
}

@ARTICLE{du2018,
       author = {{Du}, Xinnan and {Shapley}, Alice E. and {Reddy}, Naveen A. and {Jones}, Tucker and {Stark}, Daniel P. and {Steidel}, Charles C. and {Strom}, Allison L. and {Rudie}, Gwen C. and {Erb}, Dawn K. and {Ellis}, Richard S. and {Pettini}, Max},
        title = "{The Redshift Evolution of Rest-UV Spectroscopic Properties in Lyman-break Galaxies at z {\ensuremath{\sim}} 2-4}",
      journal = {\apj},
         year = 2018,
        month = jun,
       volume = {860},
       number = {1},
          eid = {75},
        pages = {75},
          doi = {10.3847/1538-4357/aabfcf},
archivePrefix = {arXiv},
       eprint = {1803.05912},
 primaryClass = {astro-ph.GA},
       adsurl = {https://ui.adsabs.harvard.edu/abs/2018ApJ...860...75D}
}

@ARTICLE{heckman1990,
       author = {{Heckman}, Timothy M. and {Armus}, Lee and {Miley}, George K.},
        title = "{On the Nature and Implications of Starburst-driven Galactic Superwinds}",
      journal = {\apjs},
         year = 1990,
        month = dec,
       volume = {74},
        pages = {833},
          doi = {10.1086/191522},
       adsurl = {https://ui.adsabs.harvard.edu/abs/1990ApJS...74..833H}
}

@ARTICLE{peroux2020,
       author = {{P{\'e}roux}, C{\'e}line and {Howk}, J. Christopher},
        title = "{The Cosmic Baryon and Metal Cycles}",
      journal = {\araa},
         year = 2020,
        month = aug,
       volume = {58},
        pages = {363-406},
          doi = {10.1146/annurev-astro-021820-120014},
archivePrefix = {arXiv},
       eprint = {2011.01935},
 primaryClass = {astro-ph.GA},
       adsurl = {https://ui.adsabs.harvard.edu/abs/2020ARA&A..58..363P}
}

@ARTICLE{shapley2003,
       author = {{Shapley}, Alice E. and {Steidel}, Charles C. and {Pettini}, Max and {Adelberger}, Kurt L.},
        title = "{Rest-Frame Ultraviolet Spectra of z\raisebox{-0.5ex}\textasciitilde3 Lyman Break Galaxies}",
      journal = {\apj},
         year = 2003,
        month = may,
       volume = {588},
       number = {1},
        pages = {65-89},
          doi = {10.1086/373922},
archivePrefix = {arXiv},
       eprint = {astro-ph/0301230},
 primaryClass = {astro-ph},
       adsurl = {https://ui.adsabs.harvard.edu/abs/2003ApJ...588...65S}
}

@ARTICLE{jones2012,
       author = {{Jones}, Tucker and {Stark}, Daniel P. and {Ellis}, Richard S.},
        title = "{Keck Spectroscopy of Faint 3 < z < 7 Lyman Break Galaxies. III. The Mean Ultraviolet Spectrum at z \raisebox{-0.5ex}\textasciitilde= 4}",
      journal = {\apj},
         year = 2012,
        month = may,
       volume = {751},
       number = {1},
          eid = {51},
        pages = {51},
          doi = {10.1088/0004-637X/751/1/51},
archivePrefix = {arXiv},
       eprint = {1111.5102},
 primaryClass = {astro-ph.CO},
       adsurl = {https://ui.adsabs.harvard.edu/abs/2012ApJ...751...51J}
}

@ARTICLE{tumlinson_CGM_review,
       author = {{Tumlinson}, Jason and {Peeples}, Molly S. and {Werk}, Jessica K.},
        title = "{The Circumgalactic Medium}",
      journal = {\araa},
         year = 2017,
        month = aug,
       volume = {55},
       number = {1},
        pages = {389-432},
          doi = {10.1146/annurev-astro-091916-055240},
archivePrefix = {arXiv},
       eprint = {1709.09180},
 primaryClass = {astro-ph.GA},
       adsurl = {https://ui.adsabs.harvard.edu/abs/2017ARA&A..55..389T}
}

@ARTICLE{nicha2016-escape-fraction,
       author = {{Leethochawalit}, Nicha and {Jones}, Tucker A. and {Ellis}, Richard S. and {Stark}, Daniel P. and {Zitrin}, Adi},
        title = "{Absorption-line Spectroscopy of Gravitationally Lensed Galaxies: Further Constraints on the Escape Fraction of Ionizing Photons at High Redshift}",
      journal = {\apj},
         year = 2016,
        month = nov,
       volume = {831},
       number = {2},
          eid = {152},
        pages = {152},
          doi = {10.3847/0004-637X/831/2/152},
archivePrefix = {arXiv},
       eprint = {1606.05309},
 primaryClass = {astro-ph.GA},
       adsurl = {https://ui.adsabs.harvard.edu/abs/2016ApJ...831..152L}
}

@ARTICLE{tng50,
       author = {{Nelson}, Dylan and {Pillepich}, Annalisa and {Springel}, Volker and {Pakmor}, R{\"u}diger and {Weinberger}, Rainer and {Genel}, Shy and {Torrey}, Paul and {Vogelsberger}, Mark and {Marinacci}, Federico and {Hernquist}, Lars},
        title = "{First results from the TNG50 simulation: galactic outflows driven by supernovae and black hole feedback}",
      journal = {\mnras},
         year = 2019,
        month = dec,
       volume = {490},
       number = {3},
        pages = {3234-3261},
          doi = {10.1093/mnras/stz2306},
archivePrefix = {arXiv},
       eprint = {1902.05554},
 primaryClass = {astro-ph.GA},
       adsurl = {https://ui.adsabs.harvard.edu/abs/2019MNRAS.490.3234N}
}

@ARTICLE{pandya2021,
       author = {{Pandya}, Viraj and {Fielding}, Drummond and {Angl{\'e}s-Alc{\'a}zar}, Daniel and {Somerville}, Rachel S. and {Bryan}, Greg L. and {Hayward}, Christopher C. and {Stern}, Jonathan and {Kim}, Chang-Goo and {Quataert}, Eliot and {Forbes}, John C. and {Faucher-Gigu{\`e}re}, Claude-Andr{\'e} and {Feldmann}, Robert and {Hafen}, Zachary and {Hopkins}, Philip F. and {Kere{\v{s}}}, Du{\v{s}}an and {Murray}, Norman and {Wetzel}, Andrew},
        title = "{Characterizing mass, momentum, energy and metal outflow rates of multi-phase galactic winds in the FIRE-2 cosmological simulations}",
      journal = {arXiv e-prints},
         year = 2021,
        month = mar,
          eid = {arXiv:2103.06891},
        pages = {arXiv:2103.06891},
archivePrefix = {arXiv},
       eprint = {2103.06891},
 primaryClass = {astro-ph.GA},
       adsurl = {https://ui.adsabs.harvard.edu/abs/2021arXiv210306891P}
}

@ARTICLE{steidel2010,
       author = {{Steidel}, Charles C. and {Erb}, Dawn K. and {Shapley}, Alice E. and {Pettini}, Max and {Reddy}, Naveen and {Bogosavljevi{\'c}}, Milan and {Rudie}, Gwen C. and {Rakic}, Olivera},
        title = "{The Structure and Kinematics of the Circumgalactic Medium from Far-ultraviolet Spectra of z \raisebox{-0.5ex}\textasciitilde= 2-3 Galaxies}",
      journal = {\apj},
         year = 2010,
        month = jul,
       volume = {717},
       number = {1},
        pages = {289-322},
          doi = {10.1088/0004-637X/717/1/289},
archivePrefix = {arXiv},
       eprint = {1003.0679},
 primaryClass = {astro-ph.CO},
       adsurl = {https://ui.adsabs.harvard.edu/abs/2010ApJ...717..289S}
}

@ARTICLE{muratov2015,
       author = {{Muratov}, Alexander L. and {Kere{\v{s}}}, Du{\v{s}}an and {Faucher-Gigu{\`e}re}, Claude-Andr{\'e} and {Hopkins}, Philip F. and {Quataert}, Eliot and {Murray}, Norman},
        title = "{Gusty, gaseous flows of FIRE: galactic winds in cosmological simulations with explicit stellar feedback}",
      journal = {\mnras},
         year = 2015,
        month = dec,
       volume = {454},
       number = {3},
        pages = {2691-2713},
          doi = {10.1093/mnras/stv2126},
archivePrefix = {arXiv},
       eprint = {1501.03155},
 primaryClass = {astro-ph.GA},
       adsurl = {https://ui.adsabs.harvard.edu/abs/2015MNRAS.454.2691M}
}

@ARTICLE{pettini-cb58-2002,
       author = {{Pettini}, Max and {Rix}, Samantha A. and {Steidel}, Charles C. and {Adelberger}, Kurt L. and {Hunt}, Matthew P. and {Shapley}, Alice E.},
        title = "{New Observations of the Interstellar Medium in the Lyman Break Galaxy MS 1512-cB58}",
      journal = {\apj},
         year = 2002,
        month = apr,
       volume = {569},
       number = {2},
        pages = {742-757},
          doi = {10.1086/339355},
archivePrefix = {arXiv},
       eprint = {astro-ph/0110637},
 primaryClass = {astro-ph},
       adsurl = {https://ui.adsabs.harvard.edu/abs/2002ApJ...569..742P}
}

@ARTICLE{jones2013,
       author = {{Jones}, Tucker A. and {Ellis}, Richard S. and {Schenker}, Matthew A. and {Stark}, Daniel P.},
        title = "{Keck Spectroscopy of Gravitationally Lensed z \raisebox{-0.5ex}\textasciitilde= 4 Galaxies: Improved Constraints on the Escape Fraction of Ionizing Photons}",
      journal = {\apj},
         year = 2013,
        month = dec,
       volume = {779},
       number = {1},
          eid = {52},
        pages = {52},
          doi = {10.1088/0004-637X/779/1/52},
archivePrefix = {arXiv},
       eprint = {1304.7015},
 primaryClass = {astro-ph.CO},
       adsurl = {https://ui.adsabs.harvard.edu/abs/2013ApJ...779...52J}
}

@ARTICLE{Paardekooper2015,
       author = {{Paardekooper}, Jan-Pieter and {Khochfar}, Sadegh and {Dalla Vecchia}, Claudio},
        title = "{The First Billion Years project: the escape fraction of ionizing photons in the epoch of reionization}",
      journal = {\mnras},
         year = 2015,
        month = aug,
       volume = {451},
       number = {3},
        pages = {2544-2563},
          doi = {10.1093/mnras/stv1114},
archivePrefix = {arXiv},
       eprint = {1501.01967},
 primaryClass = {astro-ph.CO},
       adsurl = {https://ui.adsabs.harvard.edu/abs/2015MNRAS.451.2544P}
}

@ARTICLE{Kimm2014,
       author = {{Kimm}, Taysun and {Cen}, Renyue},
        title = "{Escape Fraction of Ionizing Photons during Reionization: Effects due to Supernova Feedback and Runaway OB Stars}",
      journal = {\apj},
         year = 2014,
        month = jun,
       volume = {788},
       number = {2},
          eid = {121},
        pages = {121},
          doi = {10.1088/0004-637X/788/2/121},
archivePrefix = {arXiv},
       eprint = {1405.0552},
 primaryClass = {astro-ph.GA},
       adsurl = {https://ui.adsabs.harvard.edu/abs/2014ApJ...788..121K}
}

@ARTICLE{FRESCO_slitless_SFR,
       author = {{Matharu}, Jasleen and {Nelson}, Erica J. and {Brammer}, Gabriel and {Oesch}, Pascal A. and {Allen}, Natalie and {Shivaei}, Irene and {Naidu}, Rohan P. and {Chisholm}, John and {Covelo-Paz}, Alba and {Fudamoto}, Yoshinobu and {Giovinazzo}, Emma and {Herard-Demanche}, Thomas and {Kerutt}, Josephine and {Kramarenko}, Ivan and {Marchesini}, Danilo and {Meyer}, Romain A. and {Prieto-Lyon}, Gonzalo and {Reddy}, Naveen and {Shuntov}, Marko and {Weibel}, Andrea and {Wuyts}, Stijn and {Xiao}, Mengyuan},
        title = "{A first look at spatially resolved star formation at 4.8 < z < 6.5 with JWST FRESCO NIRCam slitless spectroscopy}",
      journal = {\aap},
         year = 2024,
        month = oct,
       volume = {690},
          eid = {A64},
        pages = {A64},
          doi = {10.1051/0004-6361/202450522},
archivePrefix = {arXiv},
       eprint = {2404.17629},
 primaryClass = {astro-ph.GA},
       adsurl = {https://ui.adsabs.harvard.edu/abs/2024A&A...690A..64M}
}

@ARTICLE{Yoshiaki2024_sizemorphology,
       author = {{Ono}, Yoshiaki and {Harikane}, Yuichi and {Ouchi}, Masami and {Nakajima}, Kimihiko and {Isobe}, Yuki and {Shibuya}, Takatoshi and {Nakane}, Minami and {Umeda}, Hiroya and {Xu}, Yi and {Zhang}, Yechi},
        title = "{Census for the rest-frame optical and UV morphologies of galaxies at z = 4-10: First phase of inside-out galaxy formation}",
      journal = {\pasj},
         year = 2024,
        month = apr,
       volume = {76},
       number = {2},
        pages = {219-250},
          doi = {10.1093/pasj/psae004},
archivePrefix = {arXiv},
       eprint = {2309.02790},
 primaryClass = {astro-ph.GA},
       adsurl = {https://ui.adsabs.harvard.edu/abs/2024PASJ...76..219O}
}

@ARTICLE{Yung2020,
       author = {{Yung}, L.~Y. Aaron and {Somerville}, Rachel S. and {Finkelstein}, Steven L. and {Popping}, Gerg{\"o} and {Dav{\'e}}, Romeel and {Venkatesan}, Aparna and {Behroozi}, Peter and {Ferguson}, Harry C.},
        title = "{Semi-analytic forecasts for JWST - IV. Implications for cosmic reionization and LyC escape fraction}",
      journal = {\mnras},
         year = 2020,
        month = aug,
       volume = {496},
       number = {4},
        pages = {4574-4592},
          doi = {10.1093/mnras/staa1800},
archivePrefix = {arXiv},
       eprint = {2001.08751},
 primaryClass = {astro-ph.GA},
       adsurl = {https://ui.adsabs.harvard.edu/abs/2020MNRAS.496.4574Y}
}

@ARTICLE{Mason2015_UVluminosity,
       author = {{Mason}, Charlotte A. and {Trenti}, Michele and {Treu}, Tommaso},
        title = "{The Galaxy UV Luminosity Function before the Epoch of Reionization}",
      journal = {\apj},
         year = 2015,
        month = nov,
       volume = {813},
       number = {1},
          eid = {21},
        pages = {21},
          doi = {10.1088/0004-637X/813/1/21},
archivePrefix = {arXiv},
       eprint = {1508.01204},
 primaryClass = {astro-ph.GA},
       adsurl = {https://ui.adsabs.harvard.edu/abs/2015ApJ...813...21M}
}

@ARTICLE{Kimm2019,
       author = {{Kimm}, Taysun and {Blaizot}, J{\'e}r{\'e}my and {Garel}, Thibault and {Michel-Dansac}, L{\'e}o and {Katz}, Harley and {Rosdahl}, Joakim and {Verhamme}, Anne and {Haehnelt}, Martin},
        title = "{Understanding the escape of LyC and Ly{\ensuremath{\alpha}} photons from turbulent clouds}",
      journal = {\mnras},
         year = 2019,
        month = jun,
       volume = {486},
       number = {2},
        pages = {2215-2237},
          doi = {10.1093/mnras/stz989},
archivePrefix = {arXiv},
       eprint = {1901.05990},
 primaryClass = {astro-ph.GA},
       adsurl = {https://ui.adsabs.harvard.edu/abs/2019MNRAS.486.2215K}
}

@ARTICLE{Ma2020_FIRE2,
       author = {{Ma}, Xiangcheng and {Quataert}, Eliot and {Wetzel}, Andrew and {Hopkins}, Philip F. and {Faucher-Gigu{\`e}re}, Claude-Andr{\'e} and {Kere{\v{s}}}, Du{\v{s}}an},
        title = "{No missing photons for reionization: moderate ionizing photon escape fractions from the FIRE-2 simulations}",
      journal = {\mnras},
         year = 2020,
        month = oct,
       volume = {498},
       number = {2},
        pages = {2001-2017},
          doi = {10.1093/mnras/staa2404},
archivePrefix = {arXiv},
       eprint = {2003.05945},
 primaryClass = {astro-ph.GA},
       adsurl = {https://ui.adsabs.harvard.edu/abs/2020MNRAS.498.2001M}
}

@ARTICLE{ClaudeAndre2023,
       author = {{Faucher-Gigu{\`e}re}, Claude-Andr{\'e} and {Oh}, S. Peng},
        title = "{Key Physical Processes in the Circumgalactic Medium}",
      journal = {\araa},
         year = 2023,
        month = aug,
       volume = {61},
        pages = {131-195},
          doi = {10.1146/annurev-astro-052920-125203},
archivePrefix = {arXiv},
       eprint = {2301.10253},
 primaryClass = {astro-ph.GA},
       adsurl = {https://ui.adsabs.harvard.edu/abs/2023ARA&A..61..131F}
}

@article{Endsley2024,
    author = {Endsley, Ryan and Stark, Daniel P and Whitler, Lily and Topping, Michael W and Johnson, Benjamin D and Robertson, Brant and Tacchella, Sandro and Alberts, Stacey and Baker, William M and Bhatawdekar, Rachana and Boyett, Kristan and Bunker, Andrew J and Cameron, Alex J and Carniani, Stefano and Charlot, Stephane and Chen, Zuyi and Chevallard, Jacopo and Curtis-Lake, Emma and Danhaive, A Lola and Egami, Eiichi and Eisenstein, Daniel J and Hainline, Kevin and Helton, Jakob M and Ji, Zhiyuan and Looser, Tobias J and Maiolino, Roberto and Nelson, Erica and Puskás, Dávid and Rieke, George and Rieke, Marcia and Rix, Hans-Walter and Sandles, Lester and Saxena, Aayush and Simmonds, Charlotte and Smit, Renske and Sun, Fengwu and Williams, Christina C and Willmer, Christopher N A and Willott, Chris and Witstok, Joris},
    title = {The star-forming and ionizing properties of dwarf z ~ 6–9 galaxies in JADES: insights on bursty star formation and ionized bubble growth},
    journal = {Monthly Notices of the Royal Astronomical Society},
    volume = {533},
    number = {1},
    pages = {1111-1142},
    year = {2024},
    month = {09},
    issn = {0035-8711},
    doi = {10.1093/mnras/stae1857},
    url = {https://doi.org/10.1093/mnras/stae1857},
    eprint = {https://academic.oup.com/mnras/article-pdf/533/1/1111/58840165/stae1857.pdf},
}

@ARTICLE{Carniani2024,
       author = {{Carniani}, Stefano and {Hainline}, Kevin and {D'Eugenio}, Francesco and {Eisenstein}, Daniel J. and {Jakobsen}, Peter and {Witstok}, Joris and {Johnson}, Benjamin D. and {Chevallard}, Jacopo and {Maiolino}, Roberto and {Helton}, Jakob M. and {Willott}, Chris and {Robertson}, Brant and {Alberts}, Stacey and {Arribas}, Santiago and {Baker}, William M. and {Bhatawdekar}, Rachana and {Boyett}, Kristan and {Bunker}, Andrew J. and {Cameron}, Alex J. and {Cargile}, Phillip A. and {Charlot}, St{\'e}phane and {Curti}, Mirko and {Curtis-Lake}, Emma and {Egami}, Eiichi and {Giardino}, Giovanna and {Isaak}, Kate and {Ji}, Zhiyuan and {Jones}, Gareth C. and {Kumari}, Nimisha and {Maseda}, Michael V. and {Parlanti}, Eleonora and {P{\'e}rez-Gonz{\'a}lez}, Pablo G. and {Rawle}, Tim and {Rieke}, George and {Rieke}, Marcia and {Del Pino}, Bruno Rodr{\'\i}guez and {Saxena}, Aayush and {Scholtz}, Jan and {Smit}, Renske and {Sun}, Fengwu and {Tacchella}, Sandro and {{\"U}bler}, Hannah and {Venturi}, Giacomo and {Williams}, Christina C. and {Willmer}, Christopher N.~A.},
        title = "{Spectroscopic confirmation of two luminous galaxies at a redshift of 14}",
      journal = {\nat},
         year = 2024,
        month = sep,
       volume = {633},
       number = {8029},
        pages = {318-322},
          doi = {10.1038/s41586-024-07860-9},
archivePrefix = {arXiv},
       eprint = {2405.18485},
 primaryClass = {astro-ph.GA},
       adsurl = {https://ui.adsabs.harvard.edu/abs/2024Natur.633..318C}
}

@ARTICLE{Witstok2024,
       author = {{Witstok}, Joris and {Smit}, Renske and {Saxena}, Aayush and {Jones}, Gareth C. and {Helton}, Jakob M. and {Sun}, Fengwu and {Maiolino}, Roberto and {Kumari}, Nimisha and {Stark}, Daniel P. and {Bunker}, Andrew J. and {Arribas}, Santiago and {Baker}, William M. and {Bhatawdekar}, Rachana and {Boyett}, Kristan and {Cameron}, Alex J. and {Carniani}, Stefano and {Charlot}, Stephane and {Chevallard}, Jacopo and {Curti}, Mirko and {Curtis-Lake}, Emma and {Eisenstein}, Daniel J. and {Endsley}, Ryan and {Hainline}, Kevin and {Ji}, Zhiyuan and {Johnson}, Benjamin D. and {Looser}, Tobias J. and {Nelson}, Erica and {Perna}, Michele and {Rix}, Hans-Walter and {Robertson}, Brant E. and {Sandles}, Lester and {Scholtz}, Jan and {Simmonds}, Charlotte and {Tacchella}, Sandro and {{\"U}bler}, Hannah and {Williams}, Christina C. and {Willmer}, Christopher N.~A. and {Willott}, Chris},
        title = "{Inside the bubble: exploring the environments of reionisation-era Lyman-{\ensuremath{\alpha}} emitting galaxies with JADES and FRESCO}",
      journal = {\aap},
         year = 2024,
        month = feb,
       volume = {682},
          eid = {A40},
        pages = {A40},
          doi = {10.1051/0004-6361/202347176},
archivePrefix = {arXiv},
       eprint = {2306.04627},
 primaryClass = {astro-ph.GA},
       adsurl = {https://ui.adsabs.harvard.edu/abs/2024A&A...682A..40W}
}

@ARTICLE{johan_richard_lensmodel_abel2744,
       author = {{Richard}, Johan and {Jauzac}, Mathilde and {Limousin}, Marceau and {Jullo}, Eric and {Cl{\'e}ment}, Benjamin and {Ebeling}, Harald and {Kneib}, Jean-Paul and {Atek}, Hakim and {Natarajan}, Priya and {Egami}, Eiichi and {Livermore}, Rachael and {Bower}, Richard},
        title = "{Mass and magnification maps for the Hubble Space Telescope Frontier Fields clusters: implications for high-redshift studies}",
      journal = {\mnras},
         year = 2014,
        month = oct,
       volume = {444},
       number = {1},
        pages = {268-289},
          doi = {10.1093/mnras/stu1395},
archivePrefix = {arXiv},
       eprint = {1405.3303},
 primaryClass = {astro-ph.CO},
       adsurl = {https://ui.adsabs.harvard.edu/abs/2014MNRAS.444..268R}
}

@ARTICLE{Kawamata_lensmodel_abell2744,
       author = {{Kawamata}, Ryota and {Oguri}, Masamune and {Ishigaki}, Masafumi and {Shimasaku}, Kazuhiro and {Ouchi}, Masami},
        title = "{Precise Strong Lensing Mass Modeling of Four Hubble Frontier Field Clusters and a Sample of Magnified High-redshift Galaxies}",
      journal = {\apj},
         year = 2016,
        month = mar,
       volume = {819},
       number = {2},
          eid = {114},
        pages = {114},
          doi = {10.3847/0004-637X/819/2/114},
archivePrefix = {arXiv},
       eprint = {1510.06400},
 primaryClass = {astro-ph.GA},
       adsurl = {https://ui.adsabs.harvard.edu/abs/2016ApJ...819..114K}
}

@ARTICLE{Jones2023_GLASS,
       author = {{Jones}, Tucker and {Sanders}, Ryan and {Chen}, Yuguang and {Wang}, Xin and {Morishita}, Takahiro and {Roberts-Borsani}, Guido and {Treu}, Tommaso and {Dressler}, Alan and {Merlin}, Emiliano and {Paris}, Diego and {Santini}, Paola and {Bergamini}, Pietro and {Henry}, A. and {Huntzinger}, Erin and {Nanayakkara}, Themiya and {Boyett}, Kristan and {Bradac}, Marusa and {Brammer}, Gabriel and {Calabr{\'o}}, Antonello and {Glazebrook}, Karl and {Grasha}, Kathryn and {Mascia}, Sara and {Pentericci}, Laura and {Trenti}, Michele and {Vulcani}, Benedetta},
        title = "{Early Results from GLASS-JWST. XXI. Rapid Asembly of a Galaxy at z = 6.23 Revealed by Its C/O Abundance}",
      journal = {\apjl},
         year = 2023,
        month = jul,
       volume = {951},
       number = {1},
          eid = {L17},
        pages = {L17},
          doi = {10.3847/2041-8213/acd938},
archivePrefix = {arXiv},
       eprint = {2301.07126},
 primaryClass = {astro-ph.GA},
       adsurl = {https://ui.adsabs.harvard.edu/abs/2023ApJ...951L..17J}
}

@ARTICLE{Ishigaki_2016_ID17,
       author = {{Ishigaki}, Masafumi and {Ouchi}, Masami and {Harikane}, Yuichi},
        title = "{A Very Compact Dense Galaxy Overdensity with {\ensuremath{\delta}} ≃ 130 Identified at z {\ensuremath{\sim}} 8: Implications for Early Protocluster and Cluster Core Formation}",
      journal = {\apj},
         year = 2016,
        month = may,
       volume = {822},
       number = {1},
          eid = {5},
        pages = {5},
          doi = {10.3847/0004-637X/822/1/5},
archivePrefix = {arXiv},
       eprint = {1509.01751},
 primaryClass = {astro-ph.GA},
       adsurl = {https://ui.adsabs.harvard.edu/abs/2016ApJ...822....5I}
}

@ARTICLE{Fujimoto2024_UNCOVER,
       author = {{Fujimoto}, Seiji and {Wang}, Bingjie and {Weaver}, John R. and {Kokorev}, Vasily and {Atek}, Hakim and {Bezanson}, Rachel and {Labbe}, Ivo and {Brammer}, Gabriel and {Greene}, Jenny E. and {Chemerynska}, Iryna and {Dayal}, Pratika and {de Graaff}, Anna and {Furtak}, Lukas J. and {Oesch}, Pascal A. and {Setton}, David J. and {Price}, Sedona H. and {Miller}, Tim B. and {Williams}, Christina C. and {Whitaker}, Katherine E. and {Zitrin}, Adi and {Cutler}, Sam E. and {Leja}, Joel and {Pan}, Richard and {Coe}, Dan and {van Dokkum}, Pieter and {Feldmann}, Robert and {Fudamoto}, Yoshinobu and {Goulding}, Andy D. and {Khullar}, Gourav and {Marchesini}, Danilo and {Maseda}, Michael and {Nanayakkara}, Themiya and {Nelson}, Erica J. and {Smit}, Renske and {Stefanon}, Mauro and {Weibel}, Andrea},
        title = "{UNCOVER: A NIRSpec Census of Lensed Galaxies at z = 8.50─13.08 Probing a High-AGN Fraction and Ionized Bubbles in the Shadow}",
      journal = {\apj},
         year = 2024,
        month = dec,
       volume = {977},
       number = {2},
          eid = {250},
        pages = {250},
          doi = {10.3847/1538-4357/ad9027},
archivePrefix = {arXiv},
       eprint = {2308.11609},
 primaryClass = {astro-ph.GA},
       adsurl = {https://ui.adsabs.harvard.edu/abs/2024ApJ...977..250F}
}

@ARTICLE{Charlot_Bruzual_2003,
       author = {{Bruzual}, G. and {Charlot}, S.},
        title = "{Stellar population synthesis at the resolution of 2003}",
      journal = {\mnras},
         year = 2003,
        month = oct,
       volume = {344},
       number = {4},
        pages = {1000-1028},
          doi = {10.1046/j.1365-8711.2003.06897.x},
archivePrefix = {arXiv},
       eprint = {astro-ph/0309134},
 primaryClass = {astro-ph},
       adsurl = {https://ui.adsabs.harvard.edu/abs/2003MNRAS.344.1000B}
}

@ARTICLE{Castellano2023_glass,
       author = {{Castellano}, Marco and {Fontana}, Adriano and {Treu}, Tommaso and {Merlin}, Emiliano and {Santini}, Paola and {Bergamini}, Pietro and {Grillo}, Claudio and {Rosati}, Piero and {Acebron}, Ana and {Leethochawalit}, Nicha and {Paris}, Diego and {Bonchi}, Andrea and {Belfiori}, Davide and {Calabr{\`o}}, Antonello and {Correnti}, Matteo and {Nonino}, Mario and {Polenta}, Gianluca and {Trenti}, Michele and {Boyett}, Kristan and {Brammer}, G. and {Broadhurst}, Tom and {Caminha}, Gabriel B. and {Chen}, Wenlei and {Filippenko}, Alexei V. and {Fortuni}, Flaminia and {Glazebrook}, Karl and {Mascia}, Sara and {Mason}, Charlotte A. and {Menci}, Nicola and {Meneghetti}, Massimo and {Mercurio}, Amata and {Metha}, Benjamin and {Morishita}, Takahiro and {Nanayakkara}, Themiya and {Pentericci}, Laura and {Roberts-Borsani}, Guido and {Roy}, Namrata and {Vanzella}, Eros and {Vulcani}, Benedetta and {Yang}, Lilan and {Wang}, Xin},
        title = "{Early Results from GLASS-JWST. XIX. A High Density of Bright Galaxies at z {\ensuremath{\approx}} 10 in the A2744 Region}",
      journal = {\apjl},
         year = 2023,
        month = may,
       volume = {948},
       number = {2},
          eid = {L14},
        pages = {L14},
          doi = {10.3847/2041-8213/accea5},
archivePrefix = {arXiv},
       eprint = {2212.06666},
 primaryClass = {astro-ph.GA},
       adsurl = {https://ui.adsabs.harvard.edu/abs/2023ApJ...948L..14C}
}

@ARTICLE{Castellano2016_astrodeep,
       author = {{Castellano}, M. and {Amor{\'\i}n}, R. and {Merlin}, E. and {Fontana}, A. and {McLure}, R.~J. and {M{\'a}rmol-Queralt{\'o}}, E. and {Mortlock}, A. and {Parsa}, S. and {Dunlop}, J.~S. and {Elbaz}, D. and {Balestra}, I. and {Boucaud}, A. and {Bourne}, N. and {Boutsia}, K. and {Brammer}, G. and {Bruce}, V.~A. and {Buitrago}, F. and {Capak}, P. and {Cappelluti}, N. and {Ciesla}, L. and {Comastri}, A. and {Cullen}, F. and {Derriere}, S. and {Faber}, S.~M. and {Giallongo}, E. and {Grazian}, A. and {Grillo}, C. and {Mercurio}, A. and {Micha{\l}owski}, M.~J. and {Nonino}, M. and {Paris}, D. and {Pentericci}, L. and {Pilo}, S. and {Rosati}, P. and {Santini}, P. and {Schreiber}, C. and {Shu}, X. and {Wang}, T.},
        title = "{The ASTRODEEP Frontier Fields catalogues. II. Photometric redshifts and rest frame properties in Abell-2744 and MACS-J0416}",
      journal = {\aap},
         year = 2016,
        month = may,
       volume = {590},
          eid = {A31},
        pages = {A31},
          doi = {10.1051/0004-6361/201527514},
archivePrefix = {arXiv},
       eprint = {1603.02461},
 primaryClass = {astro-ph.GA},
       adsurl = {https://ui.adsabs.harvard.edu/abs/2016A&A...590A..31C}
}

@ARTICLE{Valentino2025,
       author = {{Valentino}, F. and {Heintz}, K.~E. and {Brammer}, G. and {Ito}, K. and {Kokorev}, V. and {Whitaker}, K.~E. and {Gallazzi}, A. and {de Graaff}, A. and {Weibel}, A. and {Frye}, B.~L. and {Kamieneski}, P.~S. and {Jin}, S. and {Ceverino}, D. and {Faisst}, A. and {Farcy}, M. and {Fujimoto}, S. and {Gillman}, S. and {Gottumukkala}, R. and {Hamadouche}, M. and {Harrington}, K.~C. and {Hirschmann}, M. and {Jespersen}, C.~K. and {Kakimoto}, T. and {Kubo}, M. and {Lagos}, C. d. P. and {Lee}, M. and {Magdis}, G.~E. and {Man}, A.~W.~S. and {Onodera}, M. and {Rizzo}, F. and {Shimakawa}, R. and {Setton}, D.~J. and {Tanaka}, M. and {Toft}, S. and {Wu}, P.-F. and {Zhu}, P.},
        title = "{Gas outflows in two recently quenched galaxies at z = 4 and 7}",
      journal = {\aap},
         year = 2025,
        month = jul,
       volume = {699},
          eid = {A358},
        pages = {A358},
          doi = {10.1051/0004-6361/202553908},
archivePrefix = {arXiv},
       eprint = {2503.01990},
 primaryClass = {astro-ph.GA},
       adsurl = {https://ui.adsabs.harvard.edu/abs/2025A&A...699A.358V}
}

@ARTICLE{JWST_primal_Survey,
       author = {{Heintz}, K.~E. and {Brammer}, G.~B. and {Watson}, D. and {Oesch}, P.~A. and {Keating}, L.~C. and {Hayes}, M.~J. and {Abdurro'uf} and {Arellano-C{\'o}rdova}, K.~Z. and {Carnall}, A.~C. and {Christiansen}, C.~R. and {Cullen}, F. and {Dav{\'e}}, R. and {Dayal}, P. and {Ferrara}, A. and {Finlator}, K. and {Fynbo}, J.~P.~U. and {Flury}, S.~R. and {Gelli}, V. and {Gillman}, S. and {Gottumukkala}, R. and {Gould}, K. and {Greve}, T.~R. and {Hardin}, S.~E. and {Hsiao}, T.~Y.-Y. and {Hutter}, A. and {Jakobsson}, P. and {Killi}, M. and {Khosravaninezhad}, N. and {Laursen}, P. and {Lee}, M.~M. and {Magdis}, G.~E. and {Matthee}, J. and {Naidu}, R.~P. and {Narayanan}, D. and {Pollock}, C. and {Prescott}, M.~K.~M. and {Rusakov}, V. and {Shuntov}, M. and {Sneppen}, A. and {Smit}, R. and {Tanvir}, N.~R. and {Terp}, C. and {Toft}, S. and {Valentino}, F. and {Vijayan}, A.~P. and {Weaver}, J.~R. and {Wise}, J.~H. and {Witstok}, J.},
        title = "{The JWST-PRIMAL archival survey: A JWST/NIRSpec reference sample for the physical properties and Lyman-{\ensuremath{\alpha}} absorption and emission of {\ensuremath{\sim}}600 galaxies at z = 5.0 {\ensuremath{-}} 13.4}",
      journal = {\aap},
         year = 2025,
        month = jan,
       volume = {693},
          eid = {A60},
        pages = {A60},
          doi = {10.1051/0004-6361/202450243},
archivePrefix = {arXiv},
       eprint = {2404.02211},
 primaryClass = {astro-ph.GA},
       adsurl = {https://ui.adsabs.harvard.edu/abs/2025A&A...693A..60H}
}

@ARTICLE{Inoue2014,
       author = {{Inoue}, Akio K. and {Shimizu}, Ikkoh and {Iwata}, Ikuru and {Tanaka}, Masayuki},
        title = "{An updated analytic model for attenuation by the intergalactic medium}",
      journal = {\mnras},
         year = 2014,
        month = aug,
       volume = {442},
       number = {2},
        pages = {1805-1820},
          doi = {10.1093/mnras/stu936},
archivePrefix = {arXiv},
       eprint = {1402.0677},
 primaryClass = {astro-ph.CO},
       adsurl = {https://ui.adsabs.harvard.edu/abs/2014MNRAS.442.1805I}
}

@ARTICLE{Snapp-Kolas_dwarfs_z2,
       author = {{Snapp-Kolas}, Christopher and {Siana}, Brian and {Gburek}, Timothy and {Alavi}, Anahita and {Emami}, Najmeh and {Richard}, Johan and {Stark}, Daniel P.},
        title = "{The rest-UV spectral properties of dwarf galaxies at z \raisebox{-0.5ex}\textasciitilde 2}",
      journal = {\mnras},
         year = 2025,
        month = may,
       volume = {539},
       number = {1},
        pages = {34-44},
          doi = {10.1093/mnras/staf451},
archivePrefix = {arXiv},
       eprint = {2401.05498},
 primaryClass = {astro-ph.GA},
       adsurl = {https://ui.adsabs.harvard.edu/abs/2025MNRAS.539...34S}
}

@ARTICLE{Kakiichi_lyC_escape,
       author = {{Kakiichi}, Koki and {Gronke}, Max},
        title = "{Radiation Hydrodynamics of Turbulent H II Regions in Molecular Clouds: A Physical Origin of LyC Leakage and the Associated Ly{\ensuremath{\alpha}} Spectra}",
      journal = {\apj},
         year = 2021,
        month = feb,
       volume = {908},
       number = {1},
          eid = {30},
        pages = {30},
          doi = {10.3847/1538-4357/abc2d9},
archivePrefix = {arXiv},
       eprint = {1905.02480},
 primaryClass = {astro-ph.GA},
       adsurl = {https://ui.adsabs.harvard.edu/abs/2021ApJ...908...30K}
}

@ARTICLE{pysersic_imad_pasha,
       author = {{Pasha}, Imad and {Miller}, Tim B.},
        title = "{pysersic: A Python package for determining galaxy structural properties via Bayesian inference, accelerated with jax}",
      journal = {The Journal of Open Source Software},
         year = 2023,
        month = sep,
       volume = {8},
       number = {89},
          eid = {5703},
        pages = {5703},
          doi = {10.21105/joss.05703},
archivePrefix = {arXiv},
       eprint = {2306.05454},
 primaryClass = {astro-ph.GA},
       adsurl = {https://ui.adsabs.harvard.edu/abs/2023JOSS....8.5703P}
}

@ARTICLE{Bron_duvet_outflows,
       author = {{Reichardt Chu}, Bronwyn and {Fisher}, Deanne B. and {Chisholm}, John and {Berg}, Danielle and {Bolatto}, Alberto and {Cameron}, Alex J. and {Fielding}, Drummond B. and {Herrera-Camus}, Rodrigo and {Kacprzak}, Glenn G. and {Li}, Miao and {McLeod}, Anna F. and {McPherson}, Daniel K. and {Nielsen}, Nikole M. and {Rickards Vaught}, Ryan J. and {Ridolfo}, Sophia G. and {Sandstrom}, Karin},
        title = "{DUVET: sub-kiloparsec resolved star formation driven outflows in a sample of local starbursting disc galaxies}",
      journal = {\mnras},
         year = 2025,
        month = jan,
       volume = {536},
       number = {2},
        pages = {1799-1821},
          doi = {10.1093/mnras/stae2705},
archivePrefix = {arXiv},
       eprint = {2402.17830},
 primaryClass = {astro-ph.GA},
       adsurl = {https://ui.adsabs.harvard.edu/abs/2025MNRAS.536.1799R}
}

@ARTICLE{momz14_rohan,
       author = {{Naidu}, Rohan P. and {Oesch}, Pascal A. and {Brammer}, Gabriel and {Weibel}, Andrea and {Li}, Yijia and {Matthee}, Jorryt and {Chisholm}, John and {Pollock}, Clara L. and {Heintz}, Kasper E. and {Johnson}, Benjamin D. and {Shen}, Xuejian and {Hviding}, Raphael E. and {Leja}, Joel and {Tacchella}, Sandro and {Ganguly}, Arpita and {Witten}, Callum and {Atek}, Hakim and {Belli}, Sirio and {Bose}, Sownak and {Bouwens}, Rychard and {Dayal}, Pratika and {Decarli}, Roberto and {de Graaff}, Anna and {Fudamoto}, Yoshinobu and {Giovinazzo}, Emma and {Greene}, Jenny E. and {Illingworth}, Garth and {Inoue}, Akio K. and {Kane}, Sarah G. and {Labbe}, Ivo and {Leonova}, Ecaterina and {Marques-Chaves}, Rui and {Meyer}, Romain A. and {Nelson}, Erica J. and {Roberts-Borsani}, Guido and {Schaerer}, Daniel and {Simcoe}, Robert A. and {Stefanon}, Mauro and {Sugahara}, Yuma and {Toft}, Sune and {van der Wel}, Arjen and {van Dokkum}, Pieter and {Walter}, Fabian and {Watson}, Darach and {Weaver}, John R. and {Whitaker}, Katherine E.},
        title = "{A Cosmic Miracle: A Remarkably Luminous Galaxy at $z_{\rm{spec}}=14.44$ Confirmed with JWST}",
      journal = {arXiv e-prints},
         year = 2025,
        month = may,
          eid = {arXiv:2505.11263},
        pages = {arXiv:2505.11263},
          doi = {10.48550/arXiv.2505.11263},
archivePrefix = {arXiv},
       eprint = {2505.11263},
 primaryClass = {astro-ph.GA},
       adsurl = {https://ui.adsabs.harvard.edu/abs/2025arXiv250511263N}
}

@ARTICLE{bunker_gnz11,
       author = {{Bunker}, Andrew J. and {Saxena}, Aayush and {Cameron}, Alex J. and {Willott}, Chris J. and {Curtis-Lake}, Emma and {Jakobsen}, Peter and {Carniani}, Stefano and {Smit}, Renske and {Maiolino}, Roberto and {Witstok}, Joris and {Curti}, Mirko and {D'Eugenio}, Francesco and {Jones}, Gareth C. and {Ferruit}, Pierre and {Arribas}, Santiago and {Charlot}, Stephane and {Chevallard}, Jacopo and {Giardino}, Giovanna and {de Graaff}, Anna and {Looser}, Tobias J. and {L{\"u}tzgendorf}, Nora and {Maseda}, Michael V. and {Rawle}, Tim and {Rix}, Hans-Walter and {Del Pino}, Bruno Rodr{\'\i}guez and {Alberts}, Stacey and {Egami}, Eiichi and {Eisenstein}, Daniel J. and {Endsley}, Ryan and {Hainline}, Kevin and {Hausen}, Ryan and {Johnson}, Benjamin D. and {Rieke}, George and {Rieke}, Marcia and {Robertson}, Brant E. and {Shivaei}, Irene and {Stark}, Daniel P. and {Sun}, Fengwu and {Tacchella}, Sandro and {Tang}, Mengtao and {Williams}, Christina C. and {Willmer}, Christopher N.~A. and {Baker}, William M. and {Baum}, Stefi and {Bhatawdekar}, Rachana and {Bowler}, Rebecca and {Boyett}, Kristan and {Chen}, Zuyi and {Circosta}, Chiara and {Helton}, Jakob M. and {Ji}, Zhiyuan and {Kumari}, Nimisha and {Lyu}, Jianwei and {Nelson}, Erica and {Parlanti}, Eleonora and {Perna}, Michele and {Sandles}, Lester and {Scholtz}, Jan and {Suess}, Katherine A. and {Topping}, Michael W. and {{\"U}bler}, Hannah and {Wallace}, Imaan E.~B. and {Whitler}, Lily},
        title = "{JADES NIRSpec Spectroscopy of GN-z11: Lyman-{\ensuremath{\alpha}} emission and possible enhanced nitrogen abundance in a z = 10.60 luminous galaxy}",
      journal = {\aap},
         year = 2023,
        month = sep,
       volume = {677},
          eid = {A88},
        pages = {A88},
          doi = {10.1051/0004-6361/202346159},
archivePrefix = {arXiv},
       eprint = {2302.07256},
 primaryClass = {astro-ph.GA},
       adsurl = {https://ui.adsabs.harvard.edu/abs/2023A&A...677A..88B}
}

@ARTICLE{Morishita2024_sizeanalysis,
       author = {{Morishita}, Takahiro and {Stiavelli}, Massimo and {Chary}, Ranga-Ram and {Trenti}, Michele and {Bergamini}, Pietro and {Chiaberge}, Marco and {Leethochawalit}, Nicha and {Roberts-Borsani}, Guido and {Shen}, Xuejian and {Treu}, Tommaso},
        title = "{Enhanced Subkiloparsec-scale Star Formation: Results from a JWST Size Analysis of 341 Galaxies at 5 < z < 14}",
      journal = {\apj},
         year = 2024,
        month = mar,
       volume = {963},
       number = {1},
          eid = {9},
        pages = {9},
          doi = {10.3847/1538-4357/ad1404},
archivePrefix = {arXiv},
       eprint = {2308.05018},
 primaryClass = {astro-ph.GA},
       adsurl = {https://ui.adsabs.harvard.edu/abs/2024ApJ...963....9M}
}

@ARTICLE{Witten_protocluster_GLASS,
       author = {{Witten}, Callum and {Oesch}, Pascal A. and {McClymont}, William and {Meyer}, Romain A. and {Fudamoto}, Yoshinobu and {Sijacki}, Debora and {Laporte}, Nicolas and {Bennett}, Jake S. and {Simmonds}, Charlotte and {Giovinazzo}, Emma and {Danhaive}, A. Lola and {Ciesla}, Laure and {Carvajal-Bohorquez}, Cristian and {Trebitsch}, Maxime},
        title = "{Before its time: a remarkably evolved protocluster core at z=7.88}",
      journal = {arXiv e-prints},
         year = 2025,
        month = jul,
          eid = {arXiv:2507.06284},
        pages = {arXiv:2507.06284},
          doi = {10.48550/arXiv.2507.06284},
archivePrefix = {arXiv},
       eprint = {2507.06284},
 primaryClass = {astro-ph.GA},
       adsurl = {https://ui.adsabs.harvard.edu/abs/2025arXiv250706284W}
}

@ARTICLE{oke_gunn_ab_magnitude,
       author = {{Oke}, J.~B. and {Gunn}, J.~E.},
        title = "{Secondary standard stars for absolute spectrophotometry.}",
      journal = {\apj},
         year = 1983,
        month = mar,
       volume = {266},
        pages = {713-717},
          doi = {10.1086/160817},
       adsurl = {https://ui.adsabs.harvard.edu/abs/1983ApJ...266..713O}
}

@ARTICLE{fire2_stochastic_SFR,
       author = {{Sun}, Guochao and {Faucher-Gigu{\`e}re}, Claude-Andr{\'e} and {Hayward}, Christopher C. and {Shen}, Xuejian and {Wetzel}, Andrew and {Cochrane}, Rachel K.},
        title = "{Bursty Star Formation Naturally Explains the Abundance of Bright Galaxies at Cosmic Dawn}",
      journal = {\apjl},
         year = 2023,
        month = oct,
       volume = {955},
       number = {2},
          eid = {L35},
        pages = {L35},
          doi = {10.3847/2041-8213/acf85a},
archivePrefix = {arXiv},
       eprint = {2307.15305},
 primaryClass = {astro-ph.GA},
       adsurl = {https://ui.adsabs.harvard.edu/abs/2023ApJ...955L..35S}
}

@ARTICLE{beagle_2016,
       author = {{Chevallard}, Jacopo and {Charlot}, St{\'e}phane},
        title = "{Modelling and interpreting spectral energy distributions of galaxies with BEAGLE}",
      journal = {\mnras},
         year = 2016,
        month = oct,
       volume = {462},
       number = {2},
        pages = {1415-1443},
          doi = {10.1093/mnras/stw1756},
archivePrefix = {arXiv},
       eprint = {1603.03037},
 primaryClass = {astro-ph.GA},
       adsurl = {https://ui.adsabs.harvard.edu/abs/2016MNRAS.462.1415C}
}

@ARTICLE{Endsley_2025_diversity_sfh,
       author = {{Endsley}, Ryan and {Chisholm}, John and {Stark}, Daniel P. and {Topping}, Michael W. and {Whitler}, Lily},
        title = "{The Burstiness of Star Formation at z {\ensuremath{\sim}} 6: A Huge Diversity in the Recent Star Formation Histories of Very UV-faint Galaxies}",
      journal = {\apj},
         year = 2025,
        month = jul,
       volume = {987},
       number = {2},
          eid = {189},
        pages = {189},
          doi = {10.3847/1538-4357/addc74},
archivePrefix = {arXiv},
       eprint = {2410.01905},
 primaryClass = {astro-ph.GA},
       adsurl = {https://ui.adsabs.harvard.edu/abs/2025ApJ...987..189E}
}

@ARTICLE{topping_aurora_electrondensity,
       author = {{Topping}, Michael W. and {Sanders}, Ryan L. and {Shapley}, Alice E. and {Pahl}, Anthony J. and {Reddy}, Naveen A. and {Stark}, Daniel P. and {Berg}, Danielle A. and {Clarke}, Leonardo and {Cullen}, Fergus and {Dunlop}, James S. and {Ellis}, Richard S. and {Schreiber}, N.~M. F{\"o}rster and {Illingworth}, Garth D. and {Jones}, Tucker and {Narayanan}, Desika and {Pettini}, Max and {Schaerer}, Daniel},
        title = "{The AURORA survey: the evolution of multiphase electron densities at high redshift}",
      journal = {\mnras},
         year = 2025,
        month = aug,
       volume = {541},
       number = {2},
        pages = {1707-1721},
          doi = {10.1093/mnras/staf903},
archivePrefix = {arXiv},
       eprint = {2502.08712},
 primaryClass = {astro-ph.GA},
       adsurl = {https://ui.adsabs.harvard.edu/abs/2025MNRAS.541.1707T}
}

@ARTICLE{uncover_survey_first_release,
       author = {{Price}, Sedona H. and {Bezanson}, Rachel and {Labbe}, Ivo and {Furtak}, Lukas J. and {de Graaff}, Anna and {Greene}, Jenny E. and {Kokorev}, Vasily and {Setton}, David J. and {Suess}, Katherine A. and {Brammer}, Gabriel and {Cutler}, Sam E. and {Leja}, Joel and {Pan}, Richard and {Wang}, Bingjie and {Weaver}, John R. and {Whitaker}, Katherine E. and {Atek}, Hakim and {Burgasser}, Adam J. and {Chemerynska}, Iryna and {Dayal}, Pratika and {Feldmann}, Robert and {F{\"o}rster Schreiber}, Natascha M. and {Fudamoto}, Yoshinobu and {Fujimoto}, Seiji and {Glazebrook}, Karl and {Goulding}, Andy D. and {Khullar}, Gourav and {Kriek}, Mariska and {Marchesini}, Danilo and {Maseda}, Michael V. and {Miller}, Tim B. and {Muzzin}, Adam and {Nanayakkara}, Themiya and {Nelson}, Erica and {Oesch}, Pascal A. and {Shipley}, Heath and {Smit}, Renske and {Taylor}, Edward N. and {Dokkum}, Pieter van and {Williams}, Christina C. and {Zitrin}, Adi},
        title = "{The UNCOVER Survey: First Release of Ultradeep JWST/NIRSpec PRISM Spectra for {\ensuremath{\sim}}700 Galaxies from z {\ensuremath{\sim}} 0.3{\textendash}13 in A2744}",
      journal = {\apj},
         year = 2025,
        month = mar,
       volume = {982},
       number = {1},
          eid = {51},
        pages = {51},
          doi = {10.3847/1538-4357/adaec1},
archivePrefix = {arXiv},
       eprint = {2408.03920},
 primaryClass = {astro-ph.GA},
       adsurl = {https://ui.adsabs.harvard.edu/abs/2025ApJ...982...51P}
}

@ARTICLE{furtak_lensmodel,
       author = {{Furtak}, Lukas J. and {Zitrin}, Adi and {Weaver}, John R. and {Atek}, Hakim and {Bezanson}, Rachel and {Labb{\'e}}, Ivo and {Whitaker}, Katherine E. and {Leja}, Joel and {Price}, Sedona H. and {Brammer}, Gabriel B. and {Wang}, Bingjie and {Marchesini}, Danilo and {Pan}, Richard and {Dayal}, Pratika and {van Dokkum}, Pieter and {Feldmann}, Robert and {Fujimoto}, Seiji and {Franx}, Marijn and {Khullar}, Gourav and {Nelson}, Erica J. and {Mowla}, Lamiya A.},
        title = "{UNCOVERing the extended strong lensing structures of Abell 2744 with the deepest JWST imaging}",
      journal = {\mnras},
         year = 2023,
        month = aug,
       volume = {523},
       number = {3},
        pages = {4568-4582},
          doi = {10.1093/mnras/stad1627},
archivePrefix = {arXiv},
       eprint = {2212.04381},
 primaryClass = {astro-ph.GA},
       adsurl = {https://ui.adsabs.harvard.edu/abs/2023MNRAS.523.4568F}
}

@ARTICLE{density_modeulated_SFE_rachel_somerville,
       author = {{Somerville}, Rachel S. and {Yung}, L.~Y. Aaron and {Lancaster}, Lachlan and {Menon}, Shyam and {Sommovigo}, Laura and {Finkelstein}, Steven L.},
        title = "{Density-modulated star formation efficiency: implications for the observed abundance of ultraviolet luminous galaxies at z > 10}",
      journal = {\mnras},
         year = 2025,
        month = dec,
       volume = {544},
       number = {4},
        pages = {3774-3798},
          doi = {10.1093/mnras/staf1824},
archivePrefix = {arXiv},
       eprint = {2505.05442},
 primaryClass = {astro-ph.GA},
       adsurl = {https://ui.adsabs.harvard.edu/abs/2025MNRAS.544.3774S}
}

@ARTICLE{Pahl_2020,
       author = {{Pahl}, Anthony J. and {Shapley}, Alice and {Faisst}, Andreas L. and {Capak}, Peter L. and {Du}, Xinnan and {Reddy}, Naveen A. and {Laursen}, Peter and {Topping}, Michael W.},
        title = "{The redshift evolution of rest-UV spectroscopic properties to z {\ensuremath{\sim}} 5}",
      journal = {\mnras},
         year = 2020,
        month = apr,
       volume = {493},
       number = {3},
        pages = {3194-3211},
          doi = {10.1093/mnras/staa355},
archivePrefix = {arXiv},
       eprint = {1910.04179},
 primaryClass = {astro-ph.GA},
       adsurl = {https://ui.adsabs.harvard.edu/abs/2020MNRAS.493.3194P}
}

@ARTICLE{Mengtao_z_9_14,
       author = {{Tang}, Mengtao and {Stark}, Daniel P. and {Mason}, Charlotte A. and {Gelli}, Viola and {Chen}, Zuyi and {Topping}, Michael W.},
        title = "{The JWST Spectroscopic Properties of Galaxies at $z=9-14$}",
      journal = {arXiv e-prints},
         year = 2025,
        month = jul,
          eid = {arXiv:2507.08245},
        pages = {arXiv:2507.08245},
          doi = {10.48550/arXiv.2507.08245},
archivePrefix = {arXiv},
       eprint = {2507.08245},
 primaryClass = {astro-ph.GA},
       adsurl = {https://ui.adsabs.harvard.edu/abs/2025arXiv250708245T}
}

@ARTICLE{sunny_stellarkin,
       author = {{Rhoades}, Sunny and {Jones}, Tucker and {Vasan G.~C.}, Keerthi and {Chen}, Yuguang and {Leethochawalit}, Nicha and {Ellis}, Richard and {Shajib}, Anowar J. and {Glazebrook}, Karl and {Mortensen}, Kris and {Sanders}, Ryan L.},
        title = "{Resolved Stellar and Nebular Kinematics of a Star-forming Galaxy at z {\ensuremath{\sim}} 2}",
      journal = {\apj},
         year = 2025,
        month = sep,
       volume = {991},
       number = {1},
          eid = {86},
        pages = {86},
          doi = {10.3847/1538-4357/adfa22},
archivePrefix = {arXiv},
       eprint = {2503.22039},
 primaryClass = {astro-ph.GA},
       adsurl = {https://ui.adsabs.harvard.edu/abs/2025ApJ...991...86R}
}

@ARTICLE{pancakez_kelsey,
       author = {{Glazer}, Kelsey S. and {Jones}, Tucker and {Chen}, Yuguang and {Sanders}, Ryan L. and {Brada{\v{c}}}, Maru{\v{s}}a and {Pahl}, Anthony J. and {Shapley}, Alice E. and {Ellis}, Richard S. and {Topping}, Michael W. and {Reddy}, Naveen A.},
        title = "{Stacking PANCAKEZ: Spectroscopic Analysis with NIRSpec Stacks in the Epoch of Reionization. Weak Interstellar Medium Absorption and Implications for Ionizing Photon Escape at z {\ensuremath{\sim}} 7}",
      journal = {\apj},
         year = 2025,
        month = oct,
       volume = {992},
       number = {2},
          eid = {191},
        pages = {191},
          doi = {10.3847/1538-4357/ae0194},
archivePrefix = {arXiv},
       eprint = {2504.21080},
 primaryClass = {astro-ph.GA},
       adsurl = {https://ui.adsabs.harvard.edu/abs/2025ApJ...992..191G}
}

@article{vidal-garciaModellingUltravioletlineDiagnostics2017,
  title = {Modelling ultraviolet-line diagnostics of stars, the ionized and the neutral interstellar medium in star-forming galaxies},
  author = {Vidal-García, A. and Charlot, S. and Bruzual, G. and Hubeny, I.},
  year = 2017,
  month = sep,
  journal = {\mnras},
  volume = {470},
  pages = {3532-3556},
  doi = {10.1093/mnras/stx1324}
}

@article{platConstraintsProductionEscape2019,
  title = {Constraints on the Production and Escape of Ionizing Radiation from the Emission-Line Spectra of Metal-Poor Star-Forming Galaxies},
  author = {Plat, A. and Charlot, S. and Bruzual, G. and Feltre, A. and {Vidal-Garc{\'i}a}, A. and Morisset, C. and Chevallard, J. and Todt, H.},
  year = 2019,
  month = nov,
  journal = {Monthly Notices of the Royal Astronomical Society},
  volume = {490},
  pages = {978--1009},
  issn = {0035-8711},
  doi = {10.1093/mnras/stz2616},
  urldate = {2020-03-21}
}

@article{senchynaDirectConstraintsExtremely2022,
  title = {Direct {{Constraints}} on the {{Extremely Metal-poor Massive Stars Underlying Nebular C IV Emission}} from {{Ultra-deep HST}}/{{COS Ultraviolet Spectroscopy}}},
  author = {Senchyna, Peter and Stark, Daniel P. and Charlot, St{\'e}phane and Plat, Adele and Chevallard, Jacopo and Chen, Zuyi and Jones, Tucker and Sanders, Ryan L. and Rudie, Gwen C. and Cooper, Thomas J. and Bruzual, Gustavo},
  year = 2022,
  month = may,
  journal = {The Astrophysical Journal},
  volume = {930},
  pages = {105},
  issn = {0004-637X},
  doi = {10.3847/1538-4357/ac5d38},
  urldate = {2022-10-05}
}

@article{gordonQuantitativeComparisonSmall2003,
  title = {A Quantitative Comparison of the Small Magellanic Cloud, Large Magellanic Cloud, and Milky Way Ultraviolet to Near-Infrared Extinction Curves},
  author = {Gordon, K. D. and Clayton, G. C. and Misselt, K. A. and Landolt, A. U. and Wolff, M. J.},
  year = 2003,
  month = sep,
  journal = {\apj},
  volume = {594},
  pages = {279-293},
  doi = {10.1086/376774}
}

@article{miralda-escudeReionizationIntergalacticMedium1998,
  title = {Reionization of the {{Intergalactic Medium}} and the {{Damping Wing}} of the {{Gunn-Peterson Trough}}},
  author = {{Miralda-Escud{\'e}}, Jordi},
  year = 1998,
  month = jul,
  journal = {The Astrophysical Journal},
  volume = {501},
  number = {1},
  pages = {15--22},
  issn = {0004-637X},
  doi = {10.1086/305799},
  urldate = {2026-02-25},
  langid = {english}
}

@article{barkanaDidUniverseReionize2002,
  title = {Did the Universe Reionize at Redshift Six?},
  author = {Barkana, Rennan},
  year = 2002,
  month = mar,
  journal = {New Astronomy, Volume 7, Issue 2, p. 85-100.},
  volume = {7},
  number = {2},
  pages = {85},
  issn = {1384-1076},
  doi = {10.1016/S1384-1076(01)00091-4},
  urldate = {2026-02-25},
  langid = {english}
}

@article{pahlUncontaminatedMeasurementEscaping2021,
  title = {An Uncontaminated Measurement of the Escaping {{Lyman}} Continuum at z   3},
  author = {Pahl, Anthony J. and Shapley, Alice and Steidel, Charles C. and Chen, Yuguang and Reddy, Naveen A.},
  year = 2021,
  month = aug,
  journal = {Monthly Notices of the Royal Astronomical Society},
  volume = {505},
  pages = {2447--2467},
  issn = {0035-8711},
  doi = {10.1093/mnras/stab1374},
  urldate = {2022-09-06}
}

@article{jungConstraintsLymanContinuum2024,
  title = {Constraints on the {{Lyman Continuum Escape}} from {{Low-mass Lensed Galaxies}} at 1.3 {$\leq$} z {$\leq$} 3.0},
  author = {Jung, Intae and Ferguson, Henry C. and Hayes, Matthew J. and Henry, Alaina and Jaskot, Anne E. and Schaerer, Daniel and Sharon, Keren and Amor{\'i}n, Ricardo O. and Atek, Hakim and Bayliss, Matthew B. and Dahle, H{\aa}kon and Finkelstein, Steven L. and Grazian, Andrea and Guaita, Lucia and {\"O}stlin, G{\"o}ran and Pentericci, Laura and Ravindranath, Swara and Scarlata, Claudia and Teplitz, Harry I. and Verhamme, Anne},
  year = 2024,
  month = aug,
  journal = {The Astrophysical Journal},
  volume = {971},
  number = {2},
  pages = {175},
  issn = {0004-637X},
  doi = {10.3847/1538-4357/ad554d},
  urldate = {2026-05-12},
  langid = {english}
}

@article{steidelKeckLymanContinuum2018,
  title = {The {{Keck Lyman Continuum Spectroscopic Survey}} ({{KLCS}}): {{The Emergent Ionizing Spectrum}} of {{Galaxies}} at z\~  3},
  shorttitle = {The {{Keck Lyman Continuum Spectroscopic Survey}} ({{KLCS}})},
  author = {Steidel, Charles C. and Bogosavljevi{\'c}, Milan and Shapley, Alice E. and Reddy, Naveen A. and Rudie, Gwen C. and Pettini, Max and Trainor, Ryan F. and Strom, Allison L.},
  year = 2018,
  month = dec,
  journal = {The Astrophysical Journal},
  volume = {869},
  number = {2},
  pages = {123},
  doi = {10.3847/1538-4357/aaed28},
  urldate = {2019-08-28},
  langid = {english}
}

@article{chisholmConstrainingMetallicitiesAges2019,
  title = {Constraining the {{Metallicities}}, {{Ages}}, {{Star Formation Histories}}, and {{Ionizing Continua}} of {{Extragalactic Massive Star Populations}}},
  author = {Chisholm, J. and Rigby, J. R. and Bayliss, M. and Berg, D. A. and Dahle, H. and Gladders, M. and Sharon, K.},
  year = 2019,
  month = sep,
  journal = {The Astrophysical Journal},
  volume = {882},
  number = {2},
  pages = {182},
  doi = {10.3847/1538-4357/ab3104},
  urldate = {2019-10-03},
  langid = {english}
}

@article{finkelsteinConditionsReionizingUniverse2019,
  title = {Conditions for {{Reionizing}} the {{Universe}} with a {{Low Galaxy Ionizing Photon Escape Fraction}}},
  author = {Finkelstein, Steven L. and D'Aloisio, Anson and Paardekooper, Jan-Pieter and {Russell Ryan Jr.} and Behroozi, Peter and Finlator, Kristian and Livermore, Rachael and Sanderbeck, Phoebe R. Upton and Vecchia, Claudio Dalla and Khochfar, Sadegh},
  year = 2019,
  month = jul,
  journal = {The Astrophysical Journal},
  volume = {879},
  number = {1},
  pages = {36},
  issn = {1538-4357},
  doi = {10.3847/1538-4357/ab1ea8},
  urldate = {2019-09-11}
}

@article{atekMostPhotonsThat2024,
  title = {Most of the Photons That Reionized the {{Universe}} Came from Dwarf Galaxies},
  author = {Atek, Hakim and Labb{\'e}, Ivo and Furtak, Lukas J. and Chemerynska, Iryna and Fujimoto, Seiji and Setton, David J. and Miller, Tim B. and Oesch, Pascal and Bezanson, Rachel and Price, Sedona H. and Dayal, Pratika and Zitrin, Adi and Kokorev, Vasily and Weaver, John R. and Brammer, Gabriel and van Dokkum, Pieter and Williams, Christina C. and Cutler, Sam E. and Feldmann, Robert and Fudamoto, Yoshinobu and Greene, Jenny E. and Leja, Joel and Maseda, Michael V. and Muzzin, Adam and Pan, Richard and Papovich, Casey and Nelson, Erica J. and Nanayakkara, Themiya and Stark, Daniel P. and Stefanon, Mauro and Suess, Katherine A. and Wang, Bingjie and Whitaker, Katherine E.},
  year = 2024,
  month = feb,
  journal = {Nature},
  volume = {626},
  number = {8001},
  pages = {975--978},
  publisher = {Nature Publishing Group},
  issn = {1476-4687},
  doi = {10.1038/s41586-024-07043-6},
  urldate = {2024-03-01},
  copyright = {2024 The Author(s), under exclusive licence to Springer Nature Limited},
  langid = {english}
}

@misc{starkObservationsFirstGalaxies2025,
  title = {Observations of the {{First Galaxies}} in the {{Era}} of {{JWST}}},
  author = {Stark, Daniel P. and Topping, Michael W. and Endsley, Ryan and Tang, Mengtao},
  year = 2025,
  month = jan,
  number = {arXiv:2501.17078},
  eprint = {2501.17078},
  primaryclass = {astro-ph},
  publisher = {arXiv},
  doi = {10.48550/arXiv.2501.17078},
  urldate = {2025-01-30},
  archiveprefix = {arXiv}
}

@article{gelliImpactMassdependentStochasticity2024,
  title = {The {{Impact}} of {{Mass-dependent Stochasticity}} at {{Cosmic Dawn}}},
  author = {Gelli, Viola and Mason, Charlotte and Hayward, Christopher C.},
  year = 2024,
  month = nov,
  journal = {The Astrophysical Journal, Volume 975, Issue 2, id.192, 11 pp.},
  volume = {975},
  number = {2},
  pages = {192},
  issn = {0004-637X},
  doi = {10.3847/1538-4357/ad7b36},
  urldate = {2026-05-20},
  langid = {english}
}

@article{erbFeedbackLowmassGalaxies2015,
  title = {Feedback in low-mass galaxies in the early Universe},
  author = {Erb, D. K.},
  year = 2015,
  month = jul,
  journal = {\nat},
  volume = {523},
  pages = {169-176},
  doi = {10.1038/nature14454}
}

@article{begleyVANDELSSurveyMeasurement2022,
  title = {The {{VANDELS}} Survey: A Measurement of the Average {{Lyman-continuum}} Escape Fraction of Star-Forming Galaxies at z = 3.5},
  shorttitle = {The {{VANDELS}} Survey},
  author = {Begley, R. and Cullen, F. and McLure, R. J. and Dunlop, J. S. and Hall, A. and Carnall, A. C. and Hamadouche, M. L. and McLeod, D. J. and Amor{\'i}n, R. and Calabr{\`o}, A. and Fontana, A. and Fynbo, J. P. U. and Guaita, L. and Hathi, N. P. and Hibon, P. and Ji, Z. and Llerena, M. and Pentericci, L. and {Saldana-Lopez}, A. and Schaerer, D. and Talia, M. and Vanzella, E. and Zamorani, G.},
  year = 2022,
  month = jul,
  journal = {Monthly Notices of the Royal Astronomical Society, Volume 513, Issue 3, pp.3510-3525},
  volume = {513},
  number = {3},
  pages = {3510},
  issn = {0035-8711},
  doi = {10.1093/mnras/stac1067},
  urldate = {2026-05-21},
  langid = {english}
}

@article{tacchellaRedshiftindependentEfficiencyModel2018,
  title = {A {{Redshift-independent Efficiency Model}}: {{Star Formation}} and {{Stellar Masses}} in {{Dark Matter Halos}} at z {$\greaterequivlnt$} 4},
  shorttitle = {A {{Redshift-independent Efficiency Model}}},
  author = {Tacchella, Sandro and Bose, Sownak and Conroy, Charlie and Eisenstein, Daniel J. and Johnson, Benjamin D.},
  year = 2018,
  month = dec,
  journal = {The Astrophysical Journal, Volume 868, Issue 2, article id. 92, {$<$}NUMPAGES{$>$}28{$<$}/NUMPAGES{$>$} pp. (2018).},
  volume = {868},
  number = {2},
  pages = {92},
  issn = {0004-637X},
  doi = {10.3847/1538-4357/aae8e0},
  urldate = {2026-05-21},
  langid = {english}
}

@article{naiduRapidReionizationOligarchs2020,
  title = {Rapid {{Reionization}} by the {{Oligarchs}}: {{The Case}} for {{Massive}}, {{UV-bright}}, {{Star-forming Galaxies}} with {{High Escape Fractions}}},
  shorttitle = {Rapid {{Reionization}} by the {{Oligarchs}}},
  author = {Naidu, Rohan P. and Tacchella, Sandro and Mason, Charlotte A. and Bose, Sownak and Oesch, Pascal A. and Conroy, Charlie},
  year = 2020,
  month = apr,
  journal = {The Astrophysical Journal},
  volume = {892},
  pages = {109},
  issn = {0004-637X},
  doi = {10.3847/1538-4357/ab7cc9},
  urldate = {2024-01-23}
}

@article{jaskotIonizingRadiationEscape2025,
  title = {Ionizing {{Radiation Escape}} from {{Low-Redshift Galaxies}} and {{Its Connection}} to {{Cosmic Reionization}}},
  author = {Jaskot, Anne E.},
  year = 2025,
  month = aug,
  journal = {Annual Review of Astronomy and Astrophysics, Volume 63, Issue 1, pp. 45-82, 37 pp.},
  volume = {63},
  number = {1},
  pages = {45},
  issn = {0066-4146},
  doi = {10.1146/annurev-astro-111324-074935},
  urldate = {2026-01-07},
  langid = {english}
}

@article{robertsonGalaxyFormationReionization2022,
  title = {Galaxy {{Formation}} and {{Reionization}}: {{Key Unknowns}} and {{Expected Breakthroughs}} by the {{James Webb Space Telescope}}},
  shorttitle = {Galaxy {{Formation}} and {{Reionization}}},
  author = {Robertson, Brant E.},
  year = 2022,
  month = aug,
  journal = {Annual Review of Astronomy and Astrophysics},
  volume = {60},
  pages = {121--158},
  issn = {0066-4146},
  doi = {10.1146/annurev-astro-120221-044656},
  urldate = {2023-03-04},
  langid = {english}
}

@article{morishitaDiverseOxygenAbundance2024,
  title = {Diverse {{Oxygen Abundance}} in {{Early Galaxies Unveiled}} by {{Auroral Line Analysis}} with {{JWST}}},
  author = {Morishita, Takahiro and Stiavelli, Massimo and Grillo, Claudio and Rosati, Piero and Schuldt, Stefan and Trenti, Michele and Bergamini, Pietro and Boyett, Kit and Chary, Ranga-Ram and Leethochawalit, Nicha and {Roberts-Borsani}, Guido and Treu, Tommaso and Vanzella, Eros},
  year = 2024,
  month = aug,
  journal = {The Astrophysical Journal, Volume 971, Issue 1, id.43, 14 pp.},
  volume = {971},
  number = {1},
  pages = {43},
  issn = {0004-637X},
  doi = {10.3847/1538-4357/ad5290},
  urldate = {2026-05-21},
  langid = {english}
}

@article{liInsightsMetalEnrichment2025,
  title = {Insights on Metal Enrichment and Environmental Effects at z {$\approx$} 5-7 with {{JWST ASPIRE}}/{{EIGER}} and the Chemical Evolution Model},
  author = {Li, Zihao and Kakiichi, Koki and Christensen, Lise and Cai, Zheng and Dekel, Avishai and Fan, Xiaohui and Farina, Emanuele Paolo and Jun, Hyunsung D. and Li, Zhaozhou and Li, Mingyu and Pudoka, Maria and Sun, Fengwu and Trebitsch, Maxime and Walter, Fabian and Wang, Feige and Yang, Jinyi and Zhang, Huanian and Zou, Siwei},
  year = 2025,
  month = nov,
  journal = {Astronomy \&amp; Astrophysics, Volume 703, id.A106, 21 pp.},
  volume = {703},
  pages = {A106},
  issn = {0004-6361},
  doi = {10.1051/0004-6361/202555372},
  urldate = {2026-05-21},
  langid = {english}
}

@misc{dekelEfficientFormationMassive2023,
  title = {Efficient {{Formation}} of {{Massive Galaxies}} at {{Cosmic Dawn}} by {{Feedback-Free Starbursts}}},
  author = {Dekel, Avishai and Sarkar, Kartick S. and Birnboim, Yuval and Mandelker, Nir and Li, Zhaozhou},
  year = 2023,
  month = mar,
  journal = {arXiv e-prints},
  doi = {10.48550/arXiv.2303.04827},
  urldate = {2023-03-20}
}

@article{boylan-kolchinAcceleratedDarkMatter2025,
  title = {Accelerated by Dark Matter: A High-Redshift Pathway to Efficient Galaxy-Scale Star Formation},
  shorttitle = {Accelerated by Dark Matter},
  author = {{Boylan-Kolchin}, Michael},
  year = 2025,
  month = apr,
  journal = {Monthly Notices of the Royal Astronomical Society},
  volume = {538},
  number = {4},
  pages = {3210--3218},
  issn = {0035-8711},
  doi = {10.1093/mnras/staf471},
  urldate = {2026-05-28},
  langid = {english}
}
\bibliographystyle{aasjournal}

\appendix

\section{Systemic redshift and ISM absorption line profiles} \label{app:spectra}
\begin{figure*}
    \centering
    {\Large \textbf{SPURS-A2744-7, $z_{\rm spec}=9.311$}}\\
    \vspace{0.2in}
    {\large{Nebular emission lines}}\\

    \includegraphics[width=2.8in]{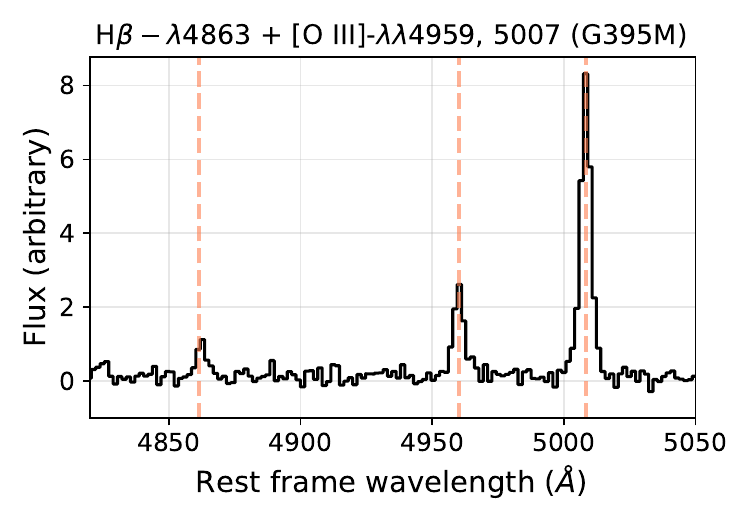} 
    \includegraphics[width=2.1in]{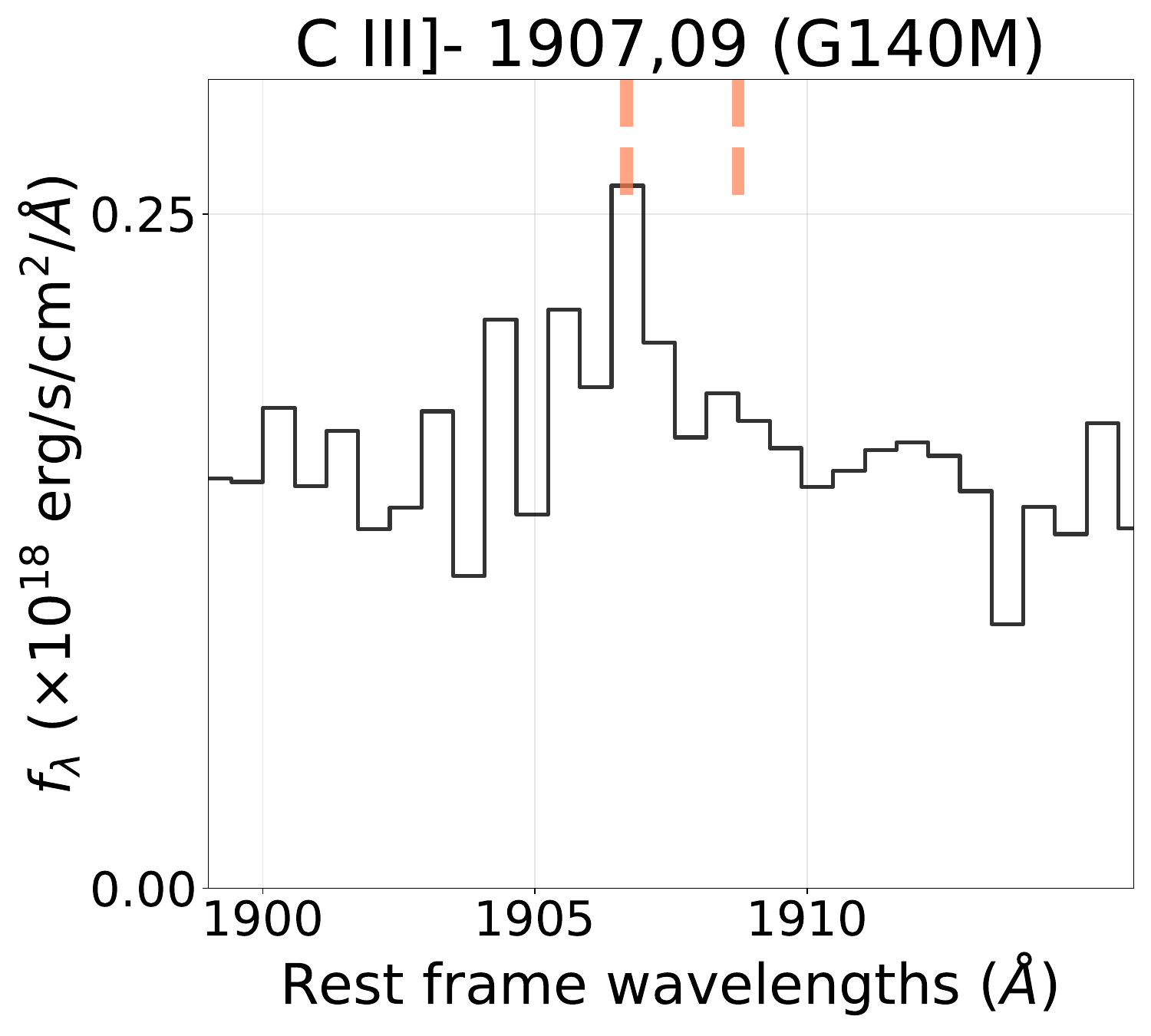}\\
    
    \vspace{0.2in}

    \begin{minipage}[t]{0.48\textwidth}
        \centering
        \large{Low-ionization absorption lines}\\

        \includegraphics[width=0.49\linewidth]{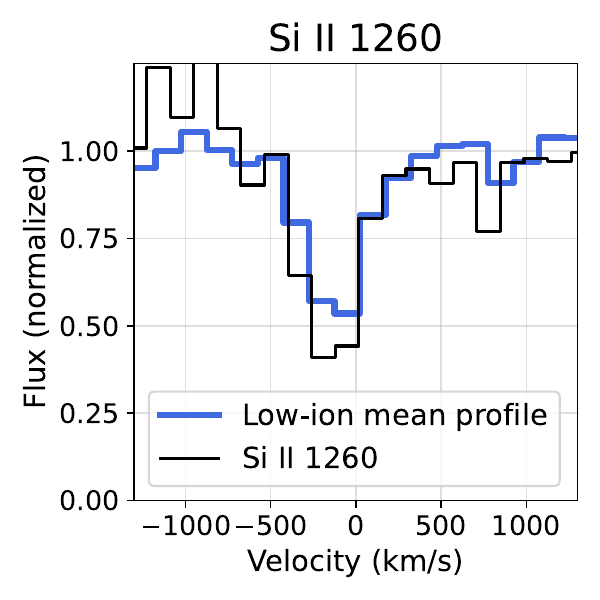}
        \includegraphics[width=0.49\linewidth]{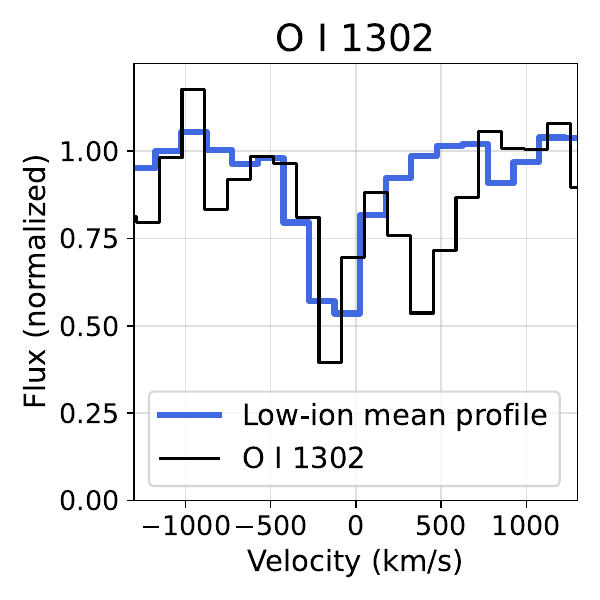}\\

        \includegraphics[width=0.49\linewidth]{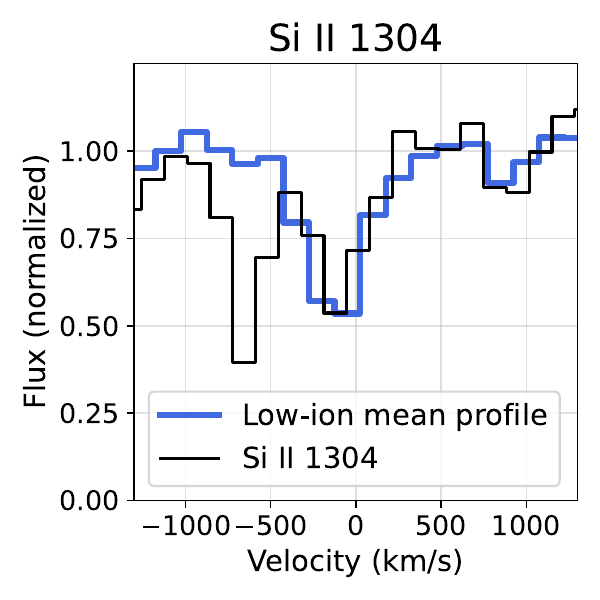}
        \includegraphics[width=0.49\linewidth]{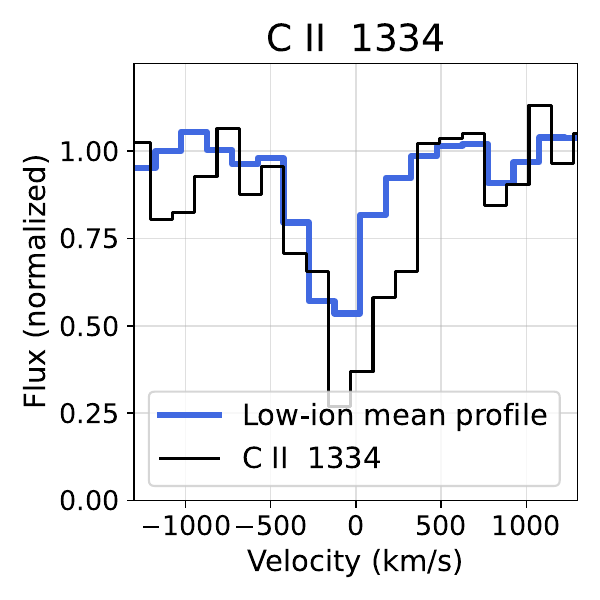}\\

        \includegraphics[width=0.49\linewidth]{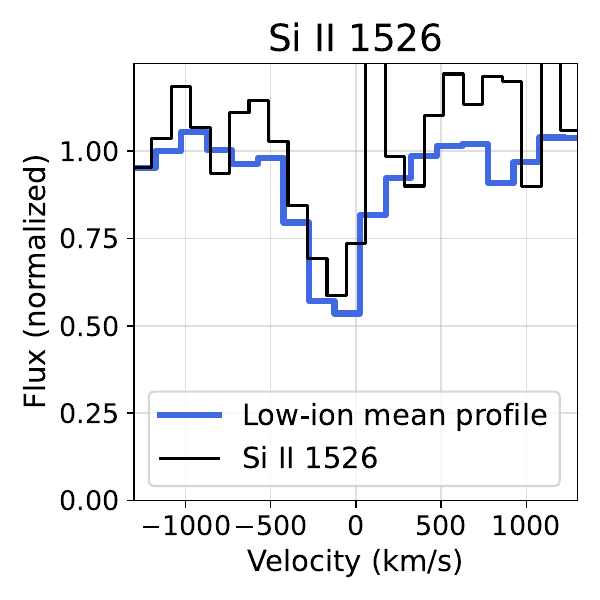}
        \includegraphics[width=0.49\linewidth]{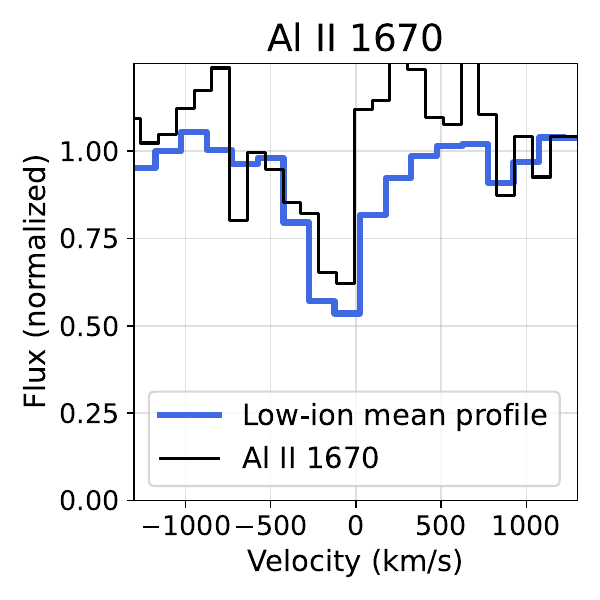}
    \end{minipage}
    \hfill
    \begin{minipage}[t]{0.48\textwidth}
        \centering
        \large{High-ionization absorption lines}\\

        \includegraphics[width=0.49\linewidth]{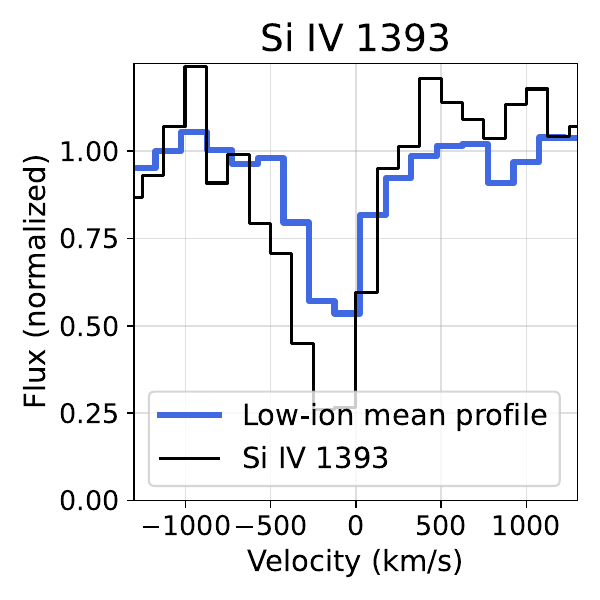}
        \includegraphics[width=0.49\linewidth]{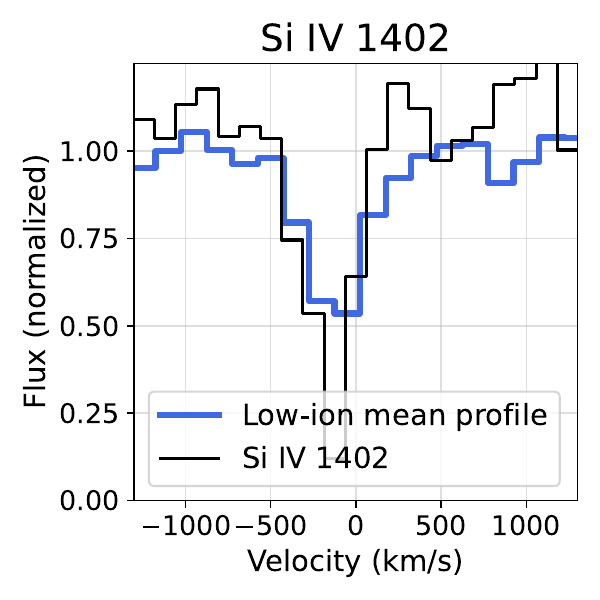}\\

        \includegraphics[width=0.49\linewidth]{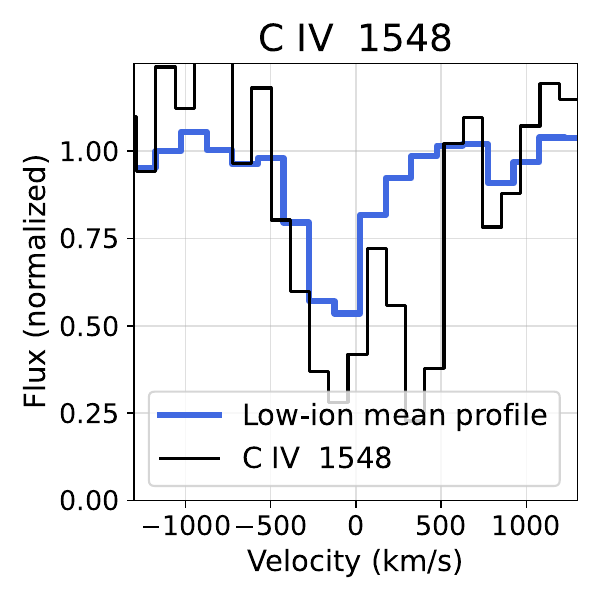}
        \includegraphics[width=0.49\linewidth]{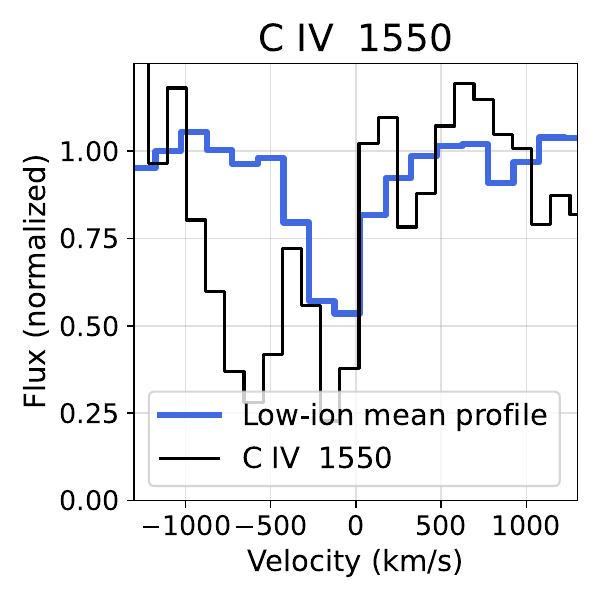}\\  

        \includegraphics[width=0.49\linewidth]{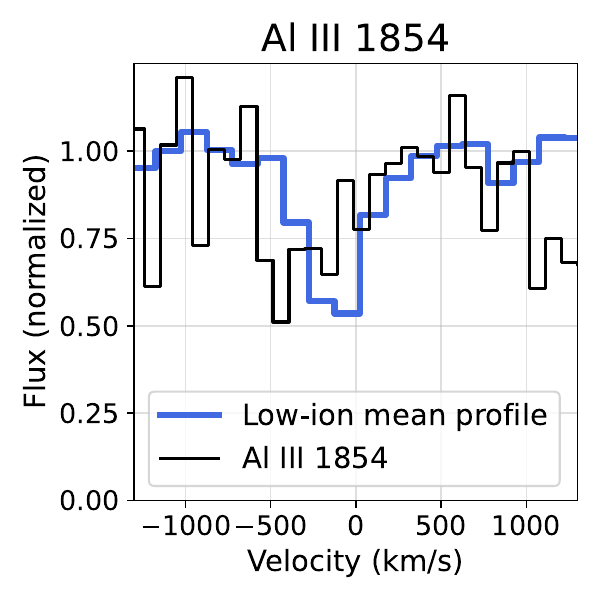}
        \includegraphics[width=0.49\linewidth]{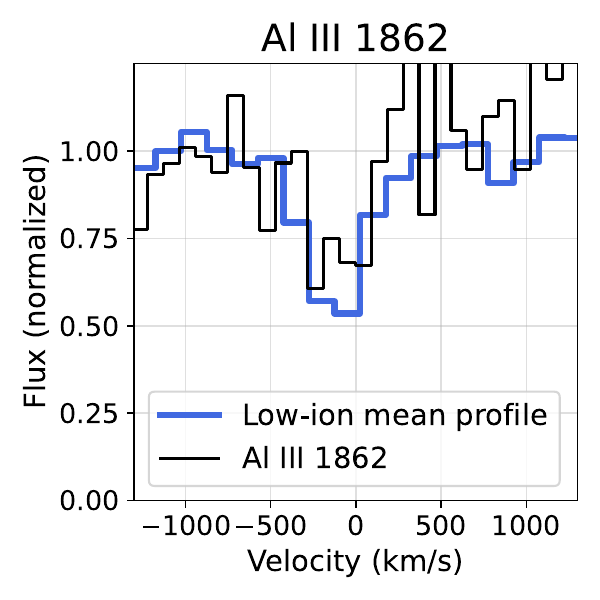}
    \end{minipage}

    \caption{\emph{Top left panel}: G395M spectra of [\ion{O}{3}]-$\lambda\lambda$4959,5007 emission line doublet used to determine the systemic redshift for the galaxies in this work. The dashed orange lines denote the rest frame wavelengths. 
    \emph{Top right panel}: Close up of the  \ion{C}{3}]-$\lambda\lambda1907,09$ doublet covered in the G140M spectra.
    We find that the redshifts from \ion{C}{3}] doublet agrees well with those obtained from optical nebular emission lines.
    \emph{Bottom panels}: Individual low- and high-ionization ISM absorption lines are shown in black in each panel. The mean low-ion ISM absorption profile is shown in blue in all the panels.}
    \label{fig:ism-profiles_7}
\end{figure*}

\begin{figure*}
    \centering
    {\Large \textbf{SPURS-A2744-17, $z_{\rm spec}=7.878$}}\\
    \vspace{0.2in}
    \large{Nebular emission lines}\\

    \includegraphics[width=2.8in]{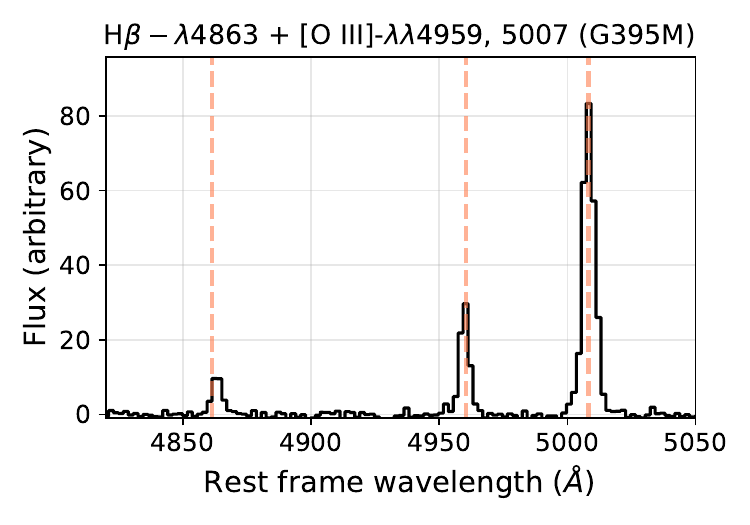} 
    \includegraphics[width=2.1in]{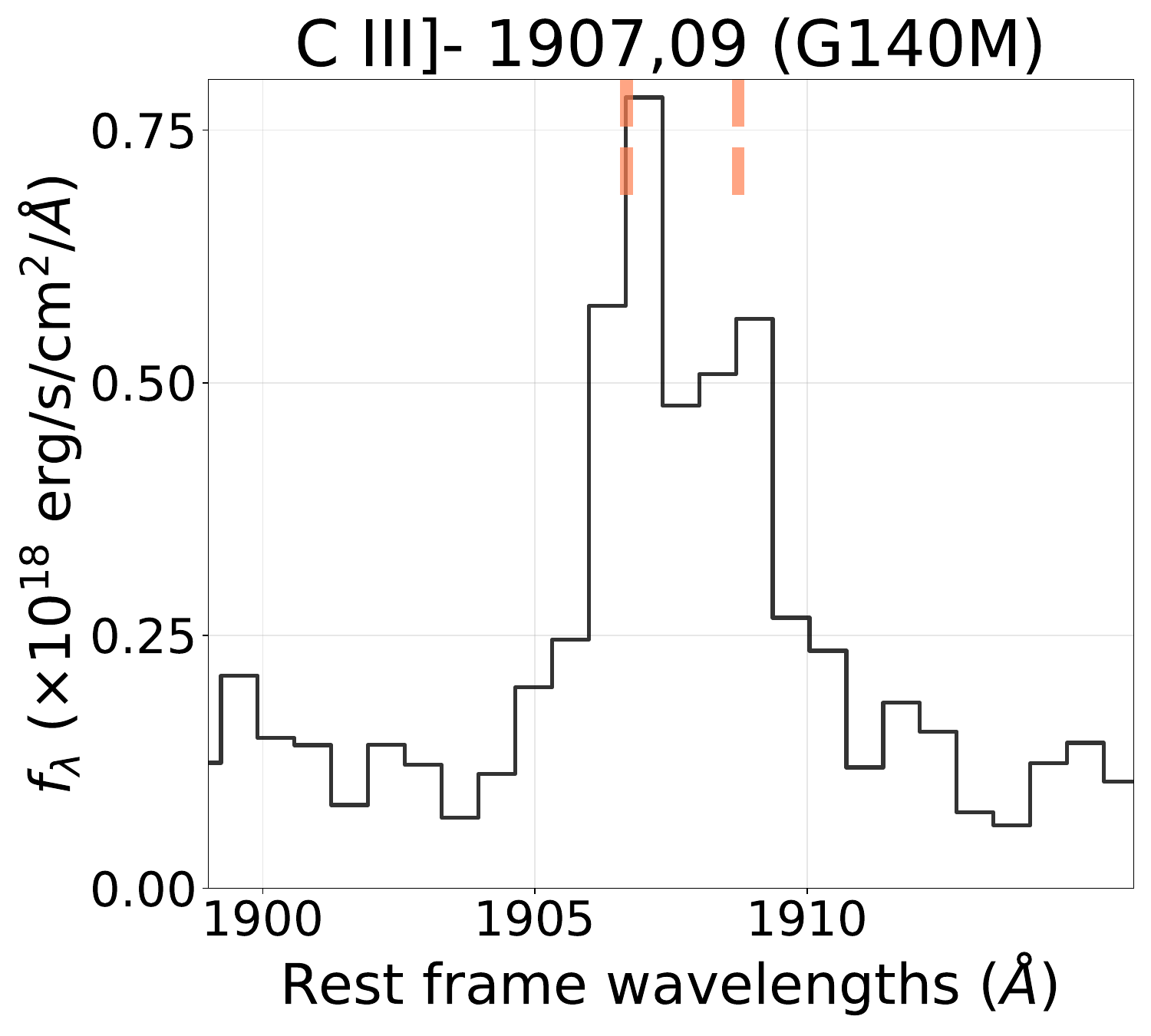}\\
    
    \vspace{0.2in}

    \begin{minipage}[t]{0.48\textwidth}
        \centering
        \large{Low-ionization absorption lines}\\

        \includegraphics[width=0.49\linewidth]{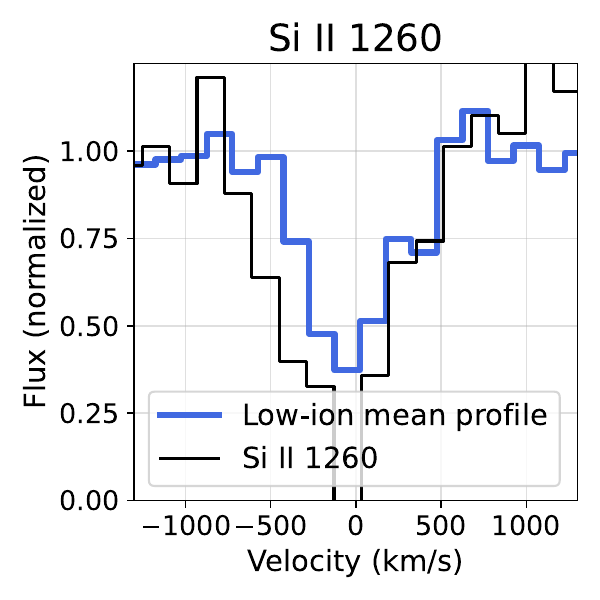}
        \includegraphics[width=0.49\linewidth]{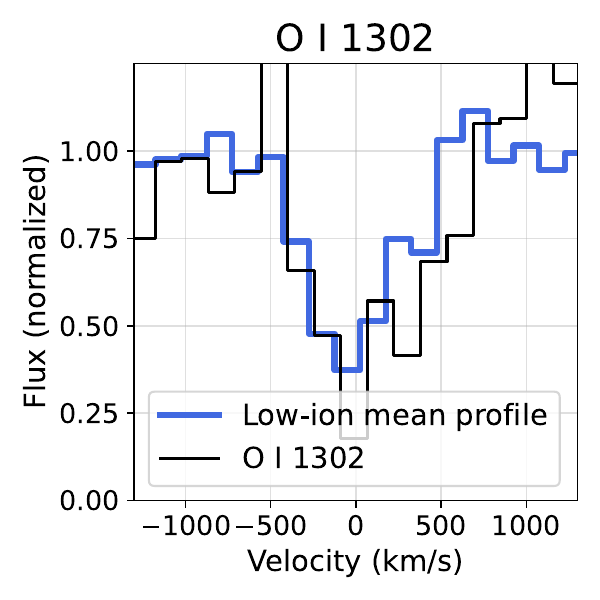}\\

        \includegraphics[width=0.49\linewidth]{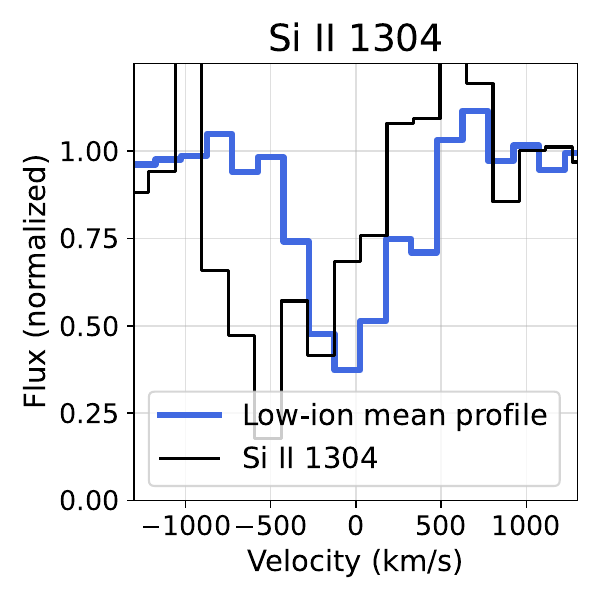}
        \includegraphics[width=0.49\linewidth]{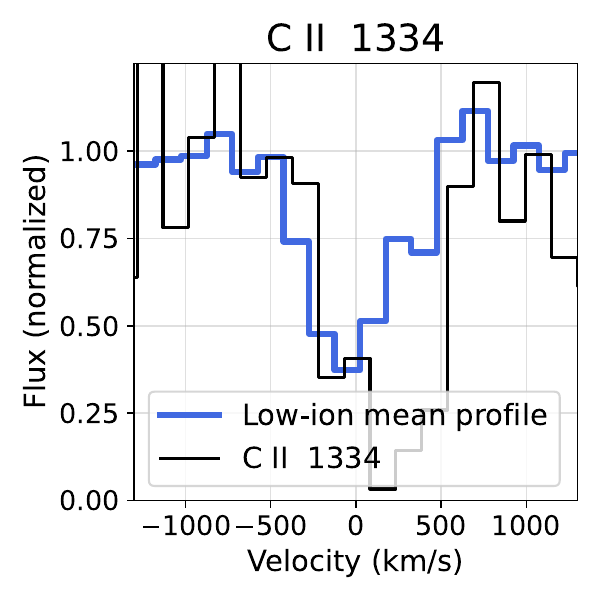}\\

        \includegraphics[width=0.49\linewidth]{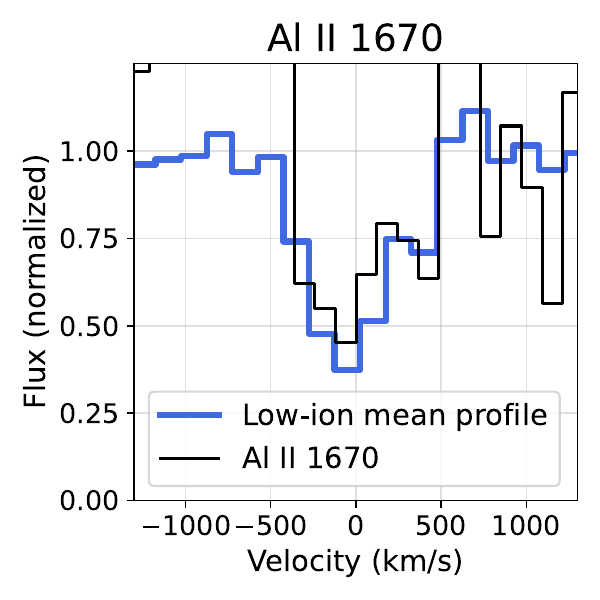}
    \end{minipage}
    \hfill
    \begin{minipage}[t]{0.48\textwidth}
        \centering
        \large{High-ionization absorption lines}\\

        \includegraphics[width=0.49\linewidth]{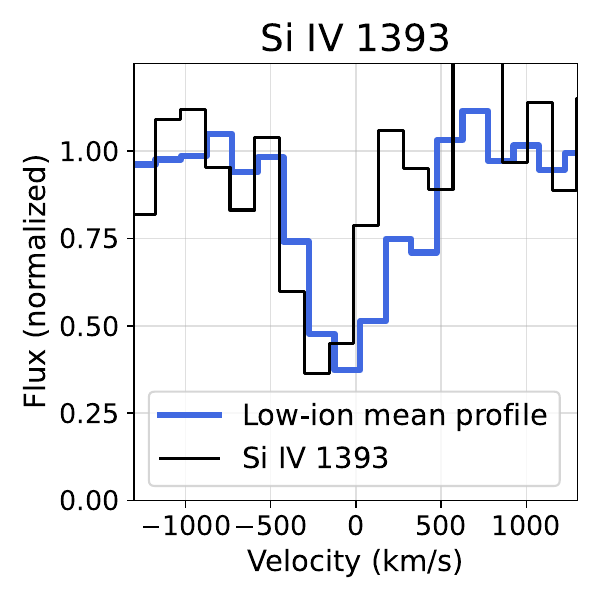}
        \includegraphics[width=0.49\linewidth]{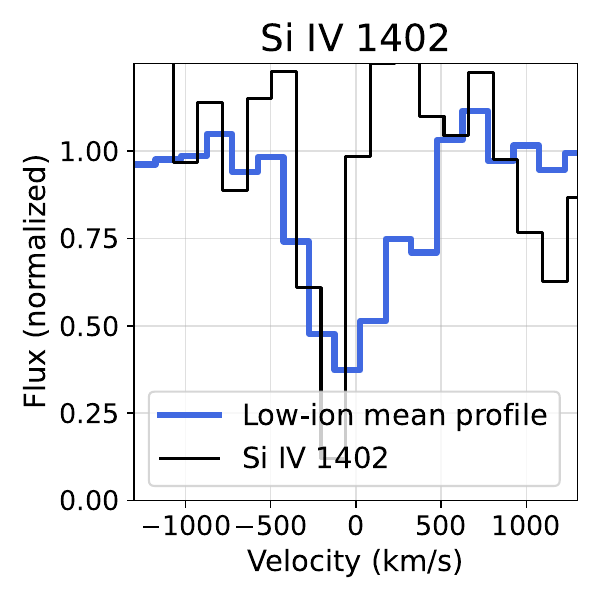}
    \end{minipage}

    \caption{Same as Figure~\ref{fig:ism-profiles_7} for SPURS-A2744-17}
    \label{fig:ism-profiles_17}
\end{figure*}

\begin{figure*}
    \centering
    {\Large \textbf{SPURS-A2744-24, $z_{\rm spec}=7.287$}}\\
    \vspace{0.2in}
    \large{Nebular emission lines}\\

    \includegraphics[width=2.8in]{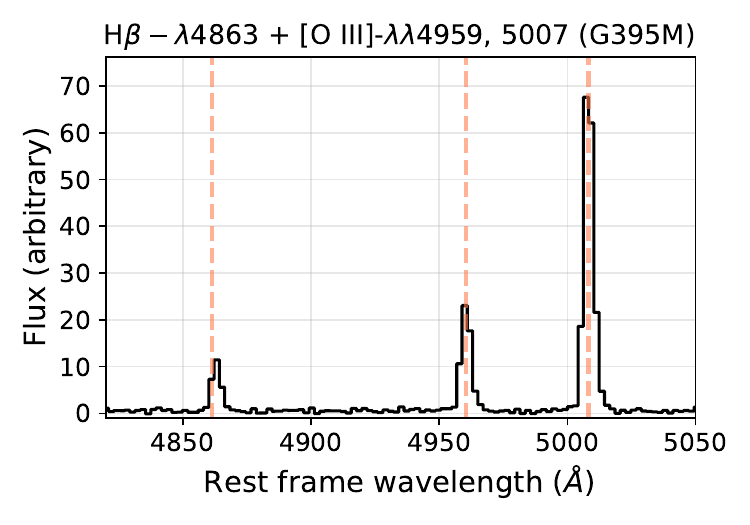}
    \includegraphics[width=2.1in]{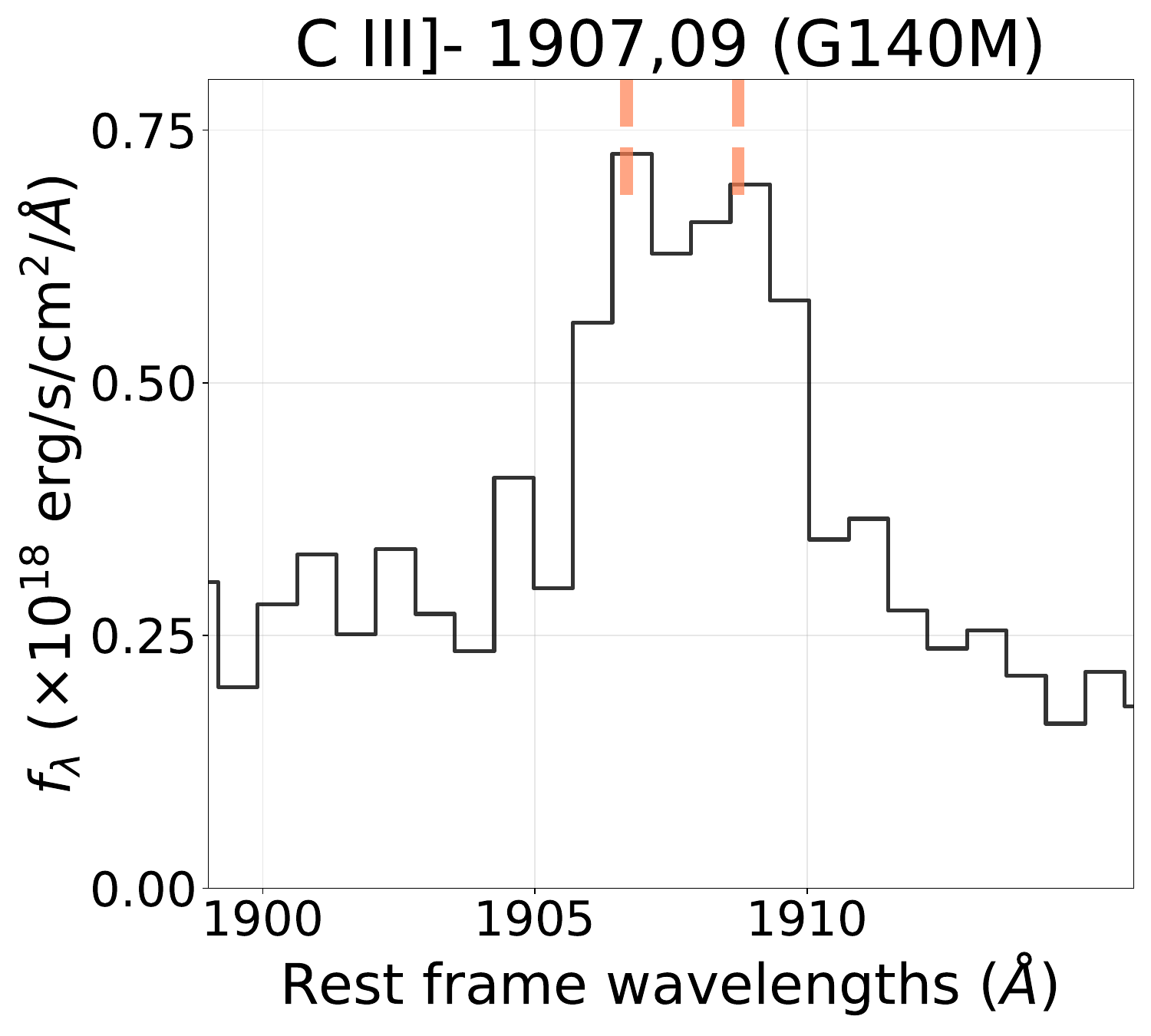}\\
    \vspace{0.2in}

    \begin{minipage}[t]{0.48\textwidth}
        \centering
        \large{Low-ionization absorption lines}\\

        \includegraphics[width=0.49\linewidth]{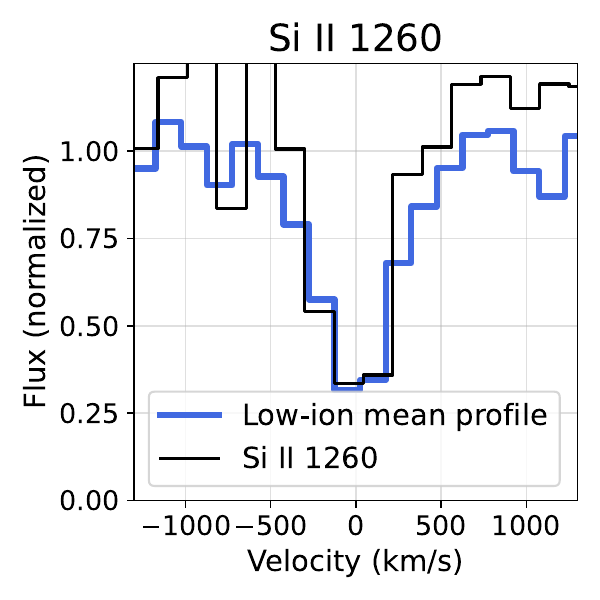}
        \includegraphics[width=0.49\linewidth]{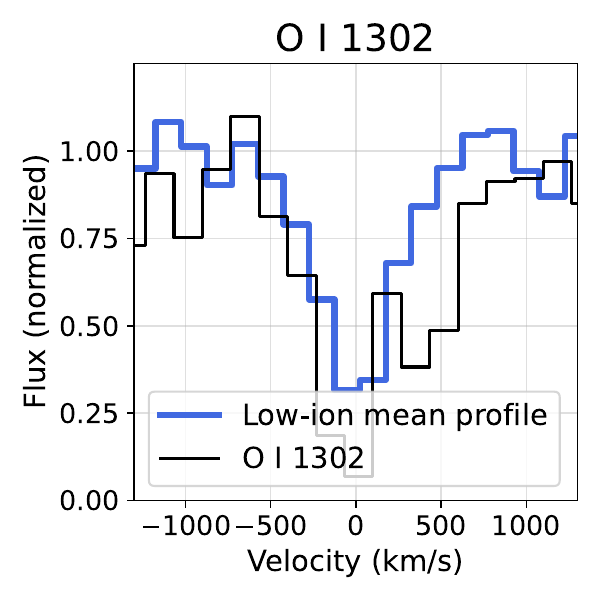}\\

        \includegraphics[width=0.49\linewidth]{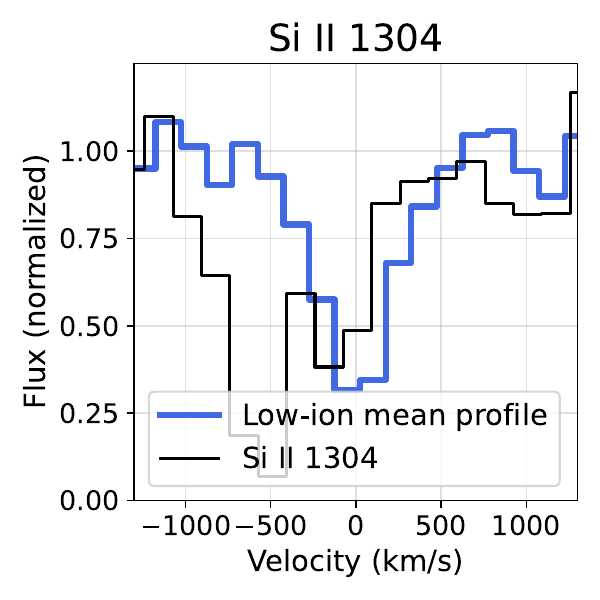}
        \includegraphics[width=0.49\linewidth]{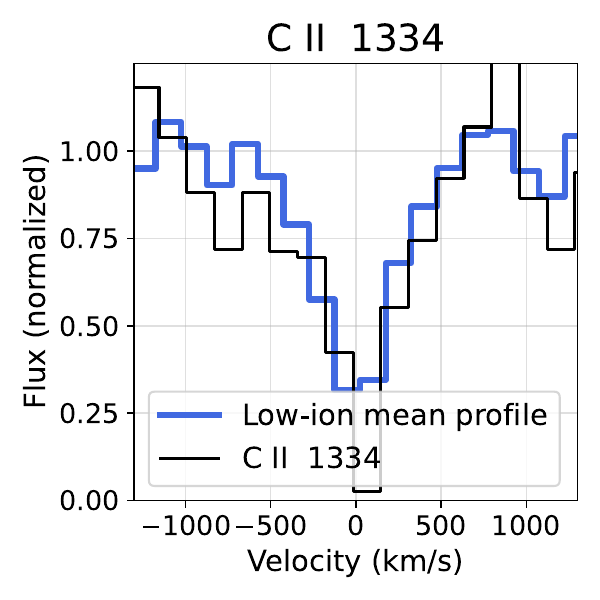}\\

        \includegraphics[width=0.49\linewidth]{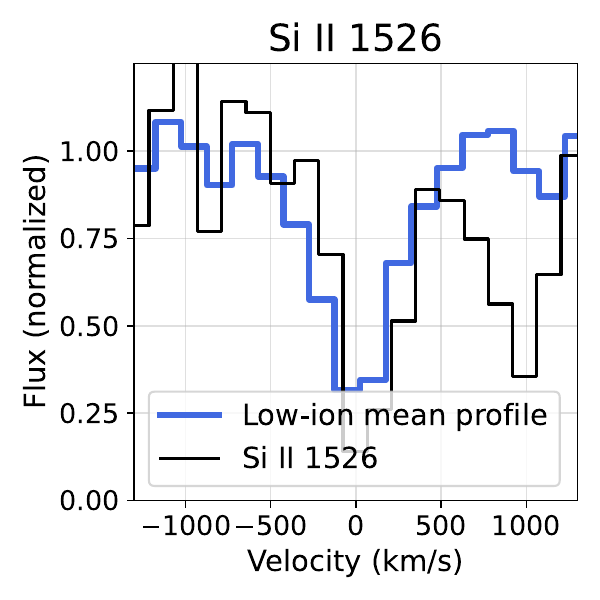}
        \includegraphics[width=0.49\linewidth]{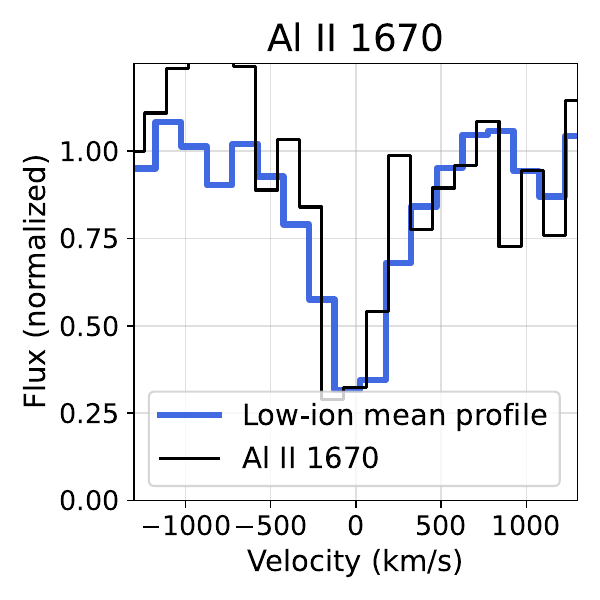}
    \end{minipage}
    \hfill
    \begin{minipage}[t]{0.48\textwidth}
        \centering
        \large{High-ionization absorption lines}\\

        \includegraphics[width=0.49\linewidth]{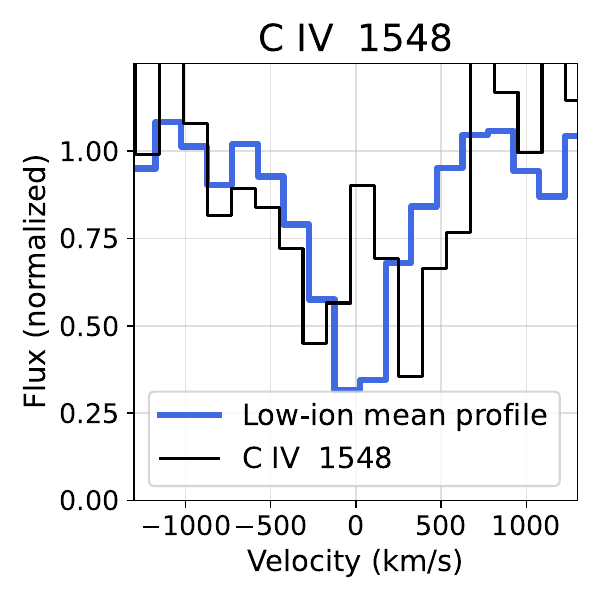}
        \includegraphics[width=0.49\linewidth]{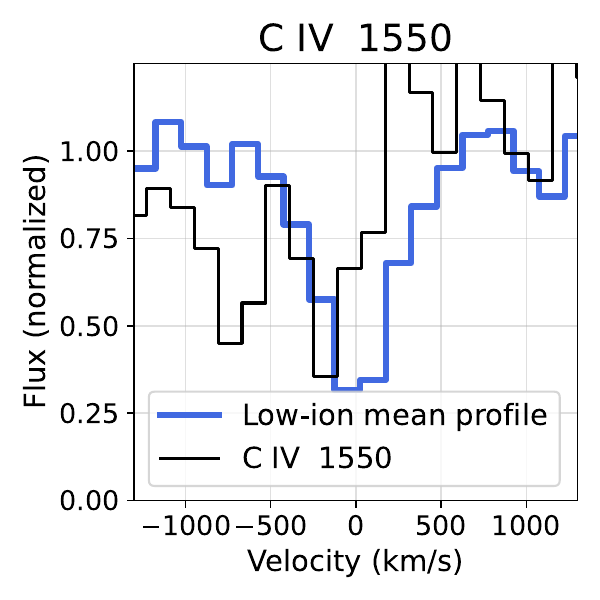}
    \end{minipage}

    \caption{Same as Figure~\ref{fig:ism-profiles_7} for SPURS-A2744-24}
    \label{fig:ism-profiles_24}
\end{figure*}

\begin{figure*}
    \centering
    {\Large \textbf{SPURS-A2744-384, $z_{\rm spec}=6.134$}}\\
    \vspace{0.2in}
    \large{Nebular emission lines}\\

    \includegraphics[width=2.8in]{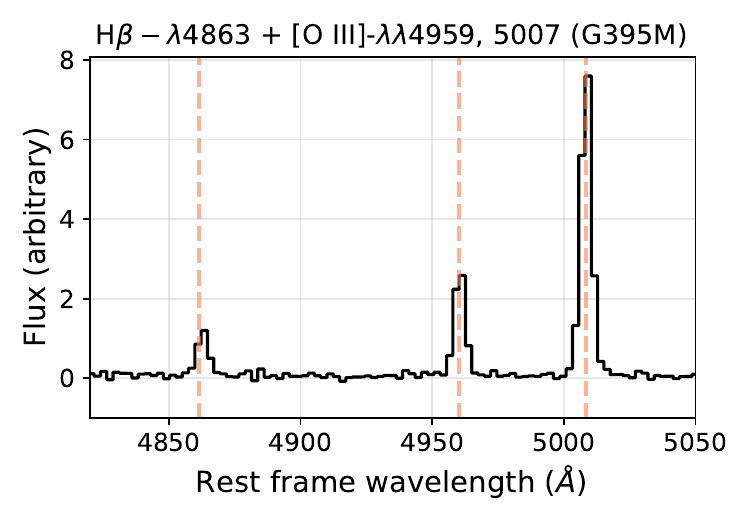}
    \includegraphics[width=2.1in]{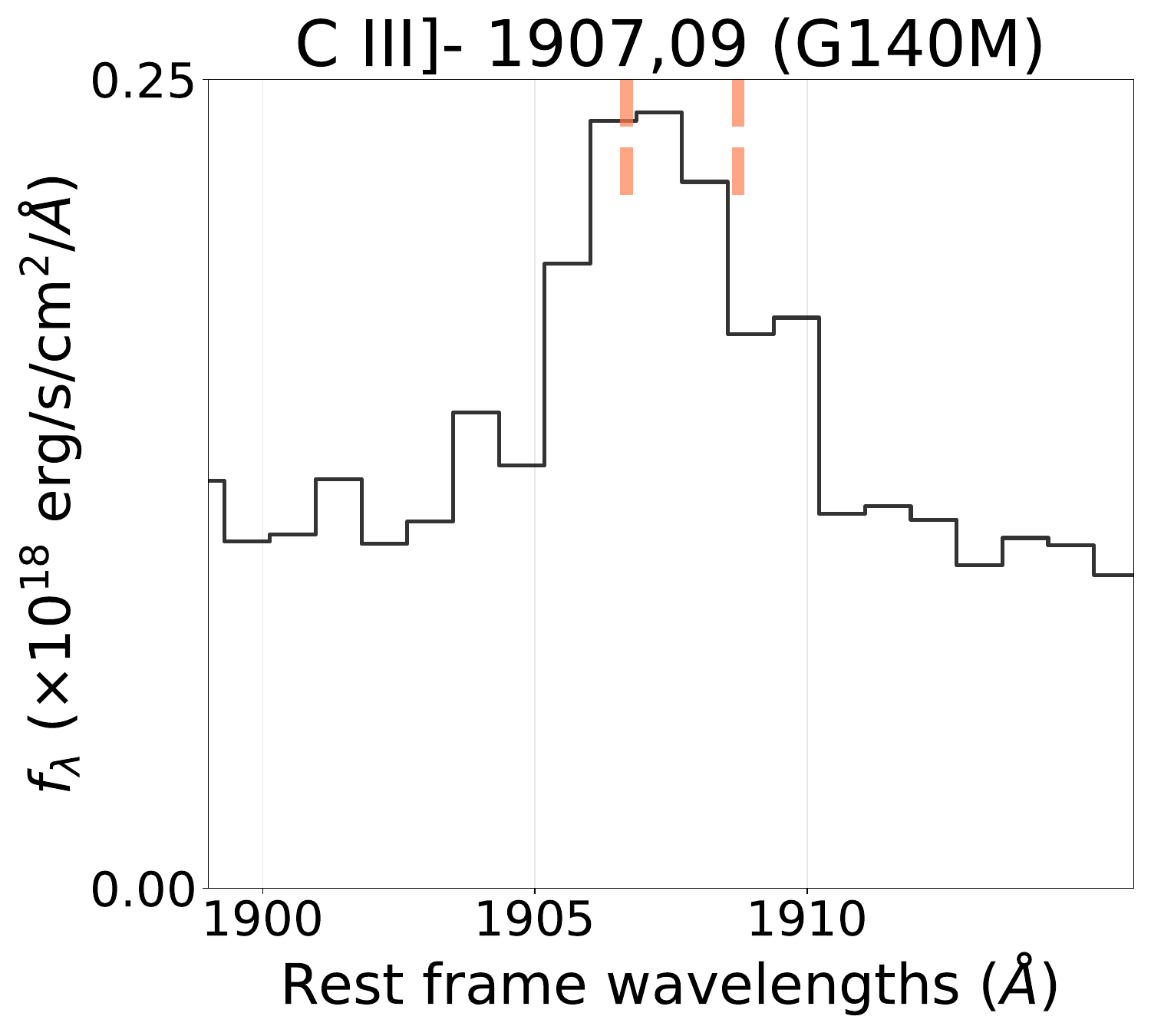}\\
    \vspace{0.2in}

    \begin{minipage}[t]{0.48\textwidth}
        \centering
        \large{Low-ionization absorption lines}\\

        \includegraphics[width=0.49\linewidth]{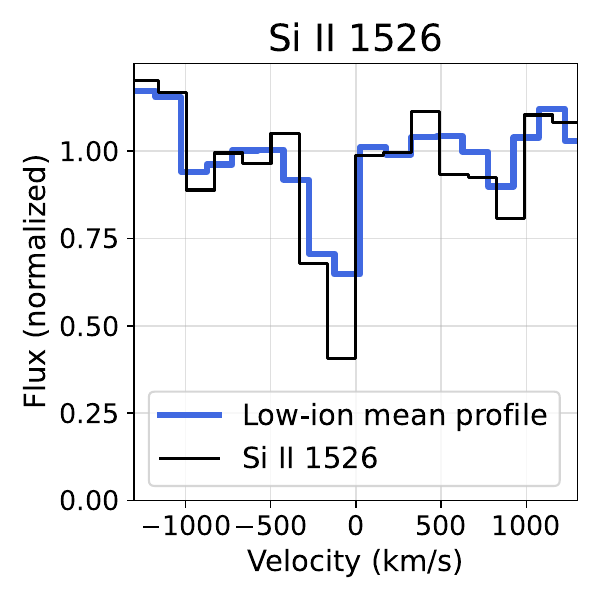}
        \includegraphics[width=0.49\linewidth]{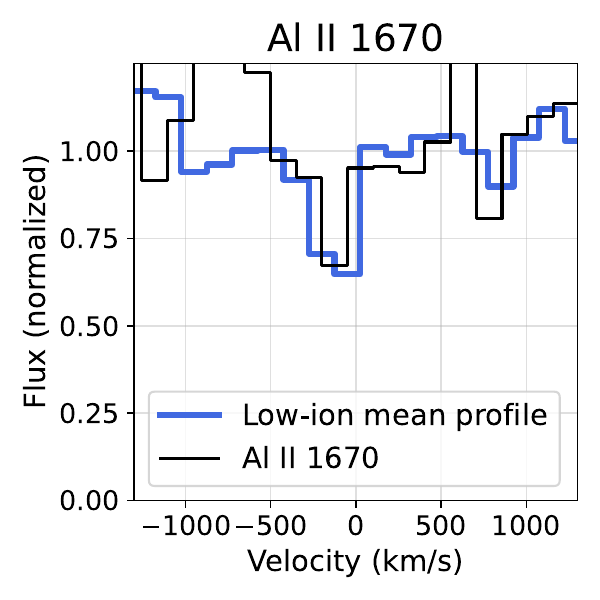}\\

        \includegraphics[width=0.49\linewidth]{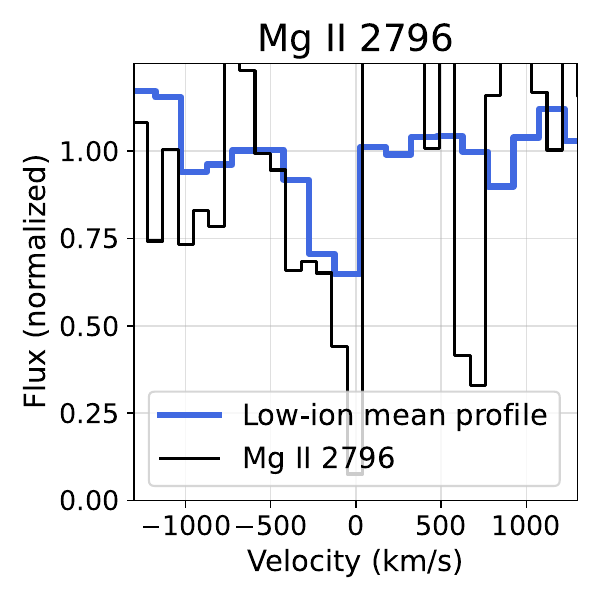}
        \includegraphics[width=0.49\linewidth]{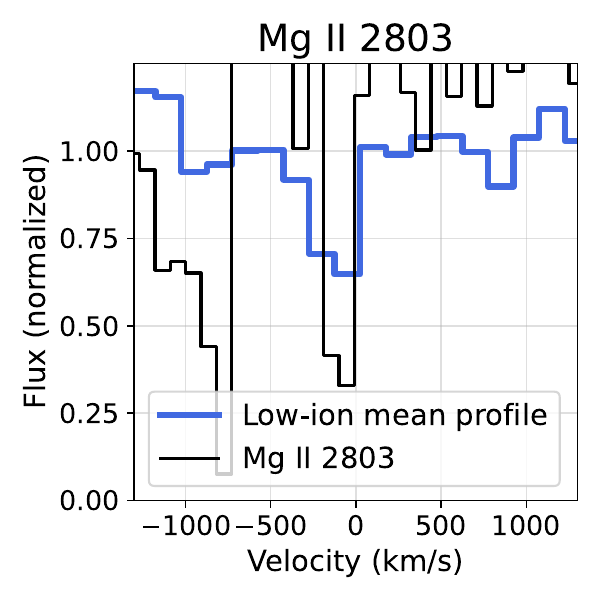}
    \end{minipage}
    \hfill
    \begin{minipage}[t]{0.48\textwidth}
        \centering
        \large{High-ionization absorption lines}\\

        \includegraphics[width=0.49\linewidth]{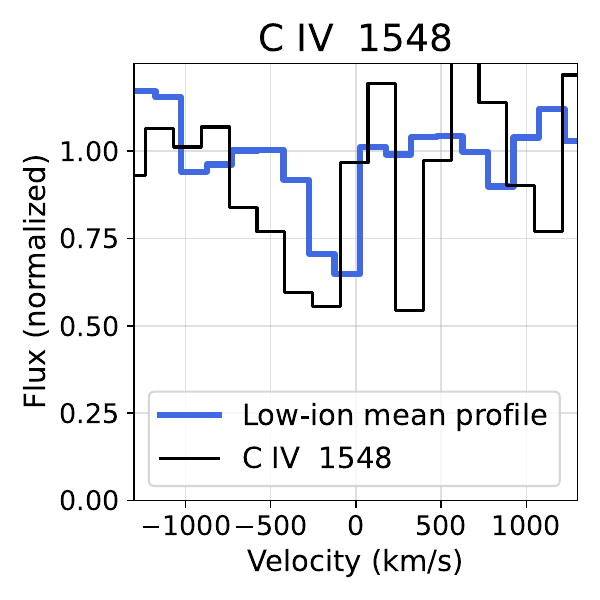}
        \includegraphics[width=0.49\linewidth]{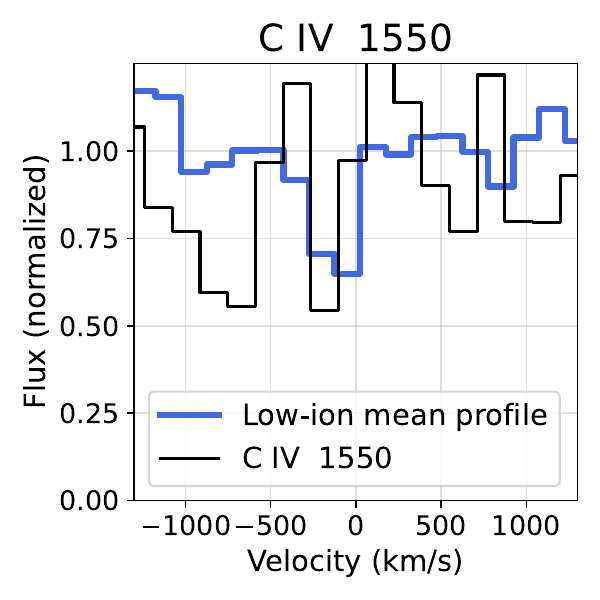}
    \end{minipage}

    \caption{Same as Figure~\ref{fig:ism-profiles_7} for SPURS-A2744-384}
    \label{fig:ism-profiles_384}
\end{figure*}

\begin{figure*}
    \centering
    {\Large \textbf{SPURS-A2744-439, $z_{\rm spec}=5.773$}}\\
    \vspace{0.2in}
    \large{Nebular emission lines}\\

    \includegraphics[width=2.8in]{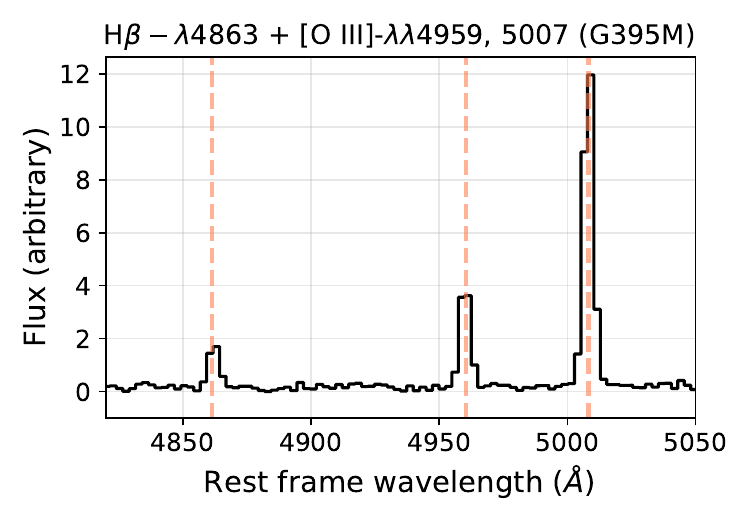}
    \includegraphics[width=2.1in]{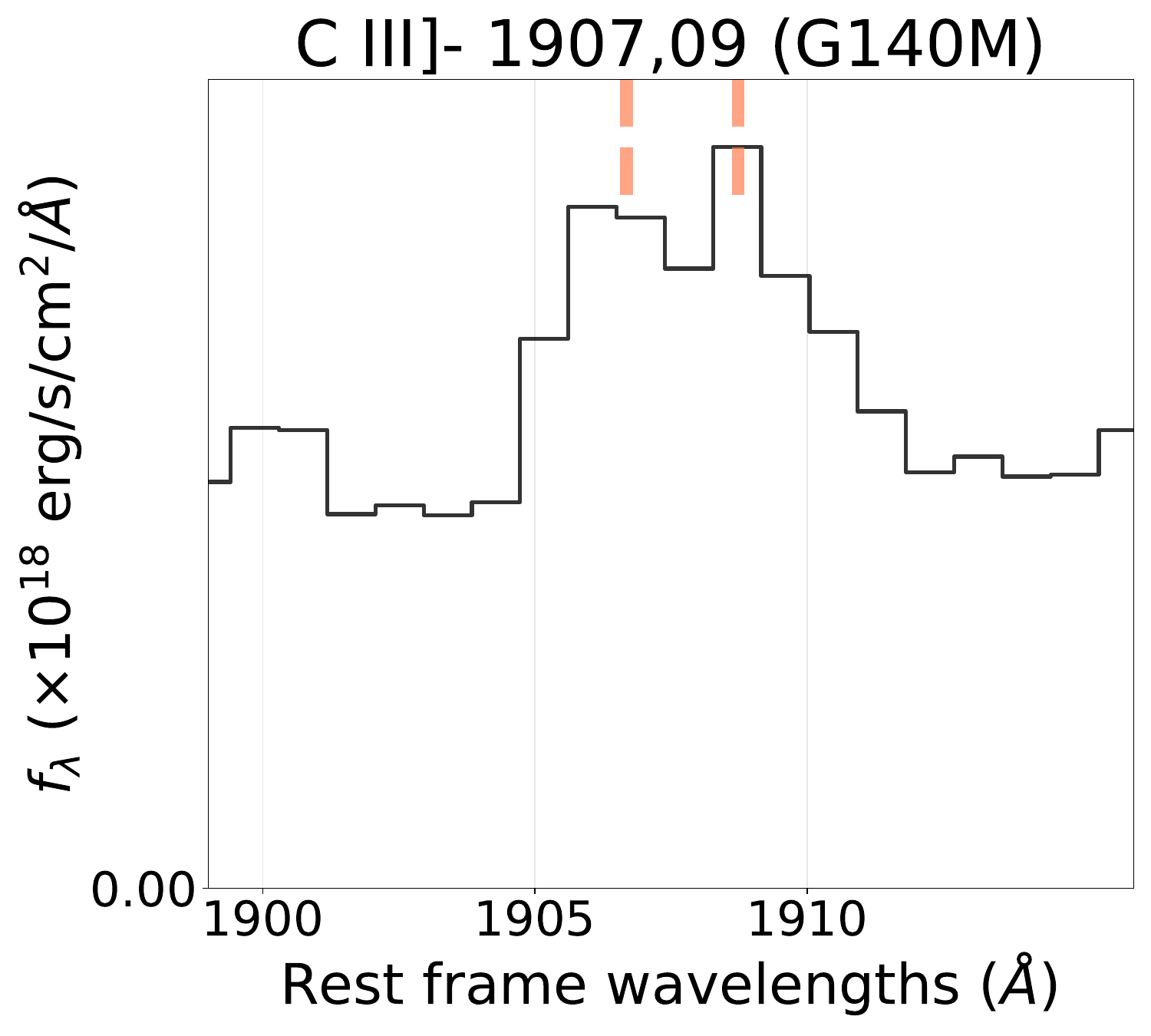}\\
    \vspace{0.2in}

    \begin{minipage}[t]{0.48\textwidth}
        \centering
        \large{Low-ionization absorption lines}\\

        \includegraphics[width=0.49\linewidth]{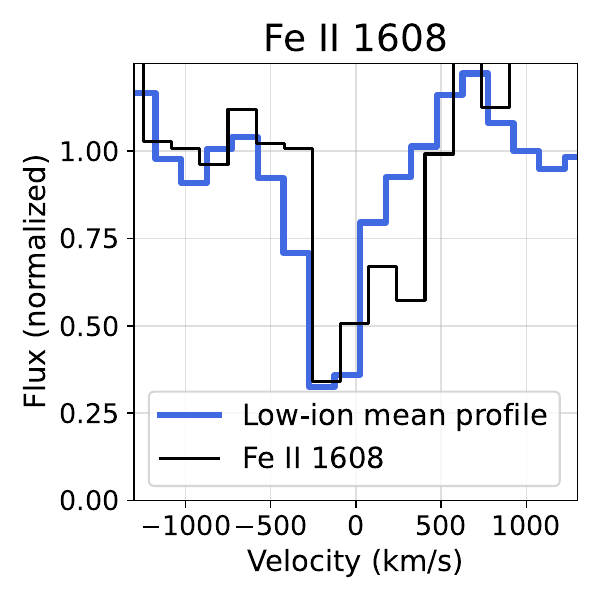}
        \includegraphics[width=0.49\linewidth]{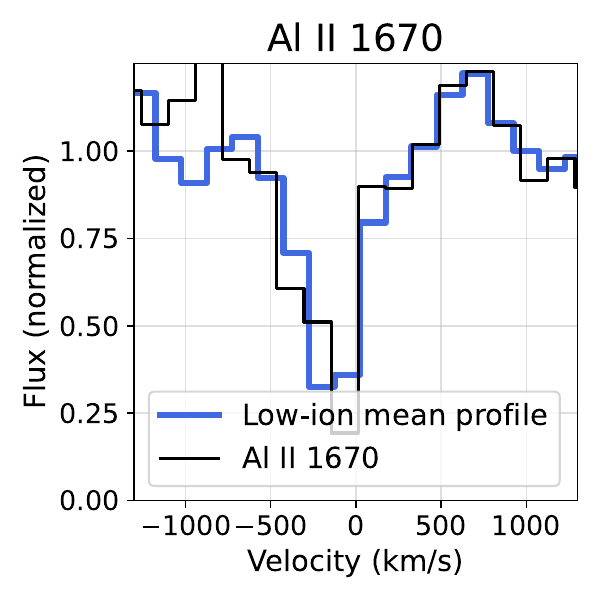}\\

        \includegraphics[width=0.49\linewidth]{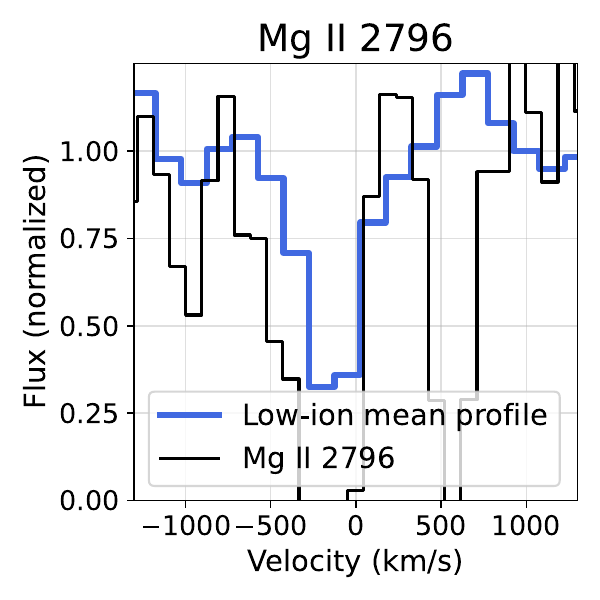}
        \includegraphics[width=0.49\linewidth]{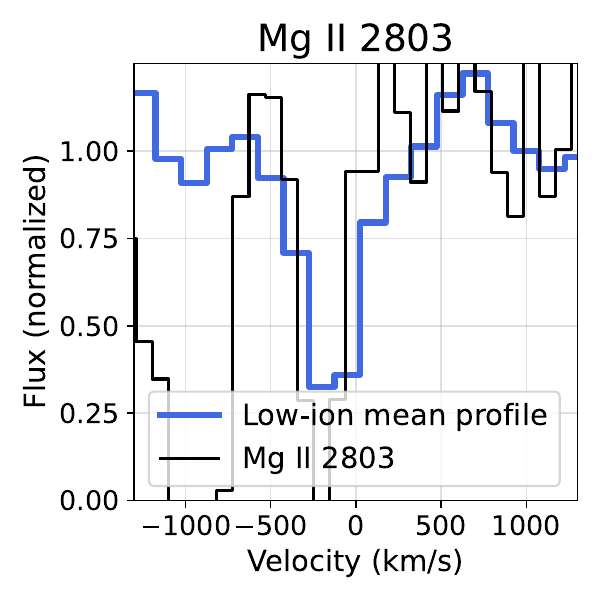}
    \end{minipage}
    \hfill
    \begin{minipage}[t]{0.48\textwidth}
        \centering
        \large{High-ionization absorption lines}\\

        \includegraphics[width=0.49\linewidth]{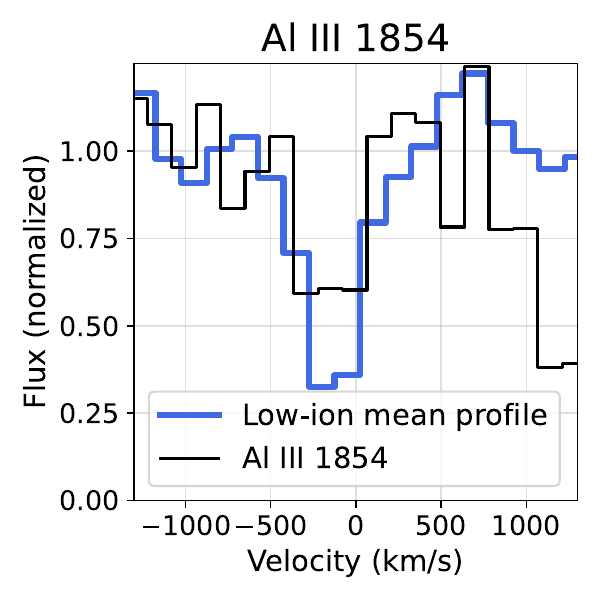}
        \includegraphics[width=0.49\linewidth]{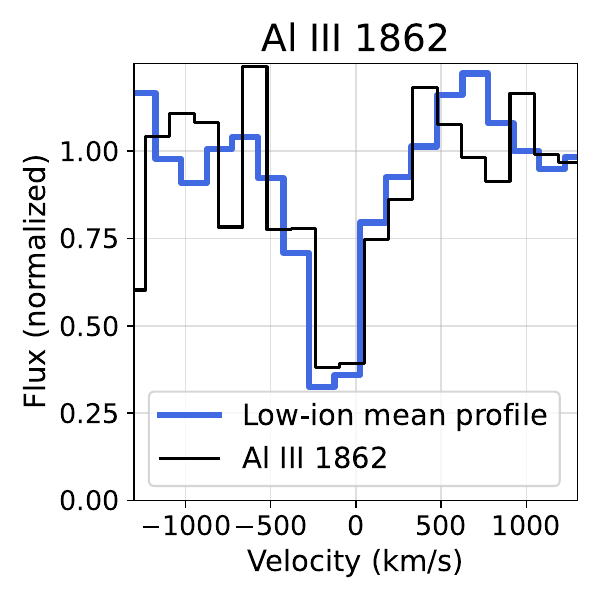}
    \end{minipage}

    \caption{Same as Figure~\ref{fig:ism-profiles_7} for SPURS-A2744-439}
    \label{fig:ism-profiles_439}
\end{figure*}

\begin{figure*}
    \centering
    {\Large \textbf{SPURS-A2744-1069, $z_{\rm spec}=5.145$}}\\
    \vspace{0.2in}
    \large{Nebular emission lines}\\

    \includegraphics[width=2.8in]{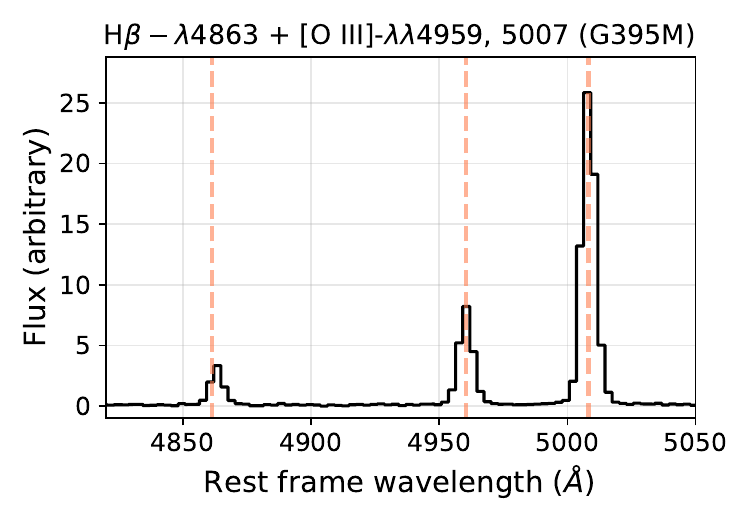}
    \includegraphics[width=2.1in]{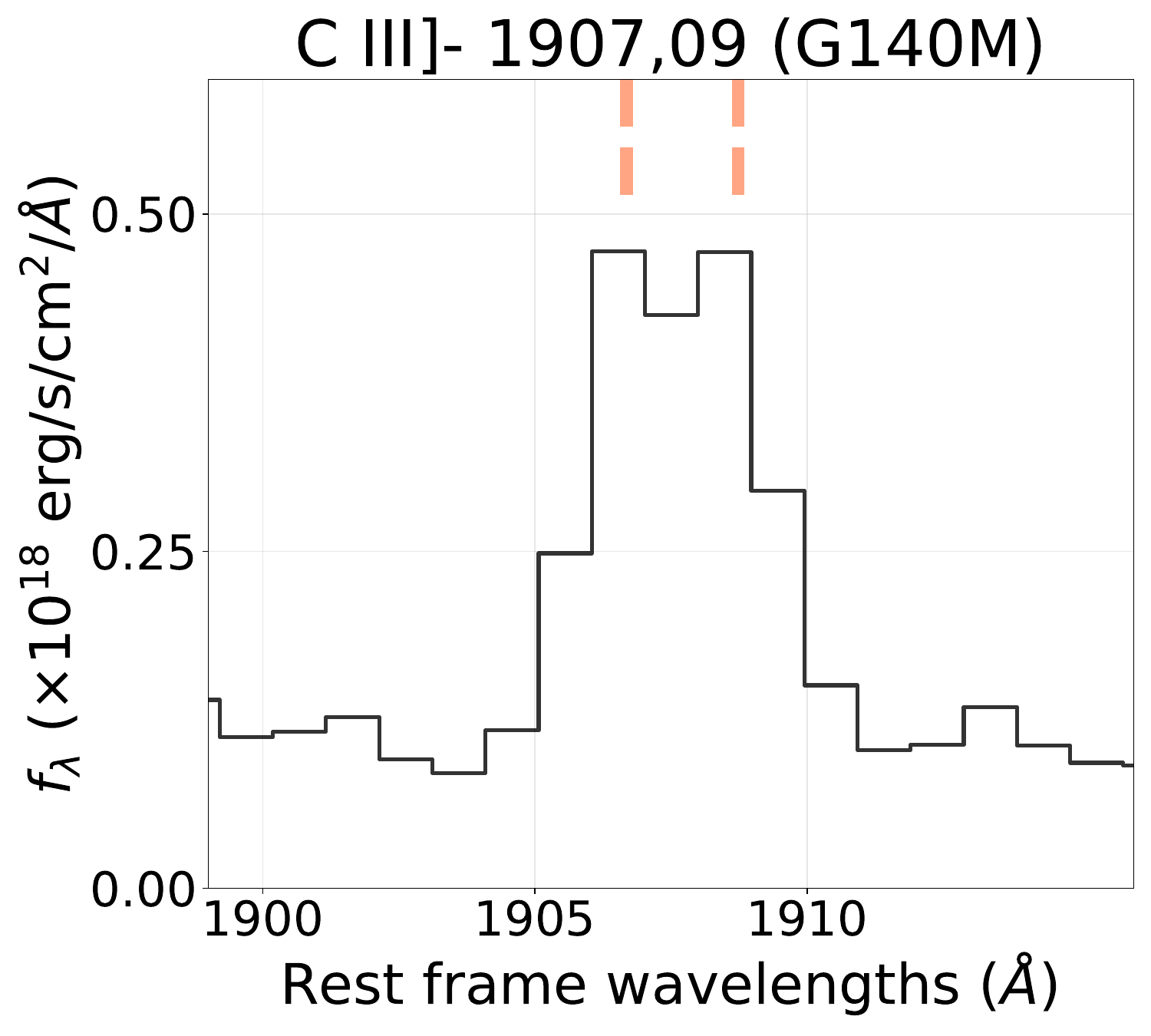}\\
    \vspace{0.2in}

    \begin{minipage}[t]{0.48\textwidth}
        \centering
        \large{Low-ionization absorption lines}\\

        \includegraphics[width=0.49\linewidth]{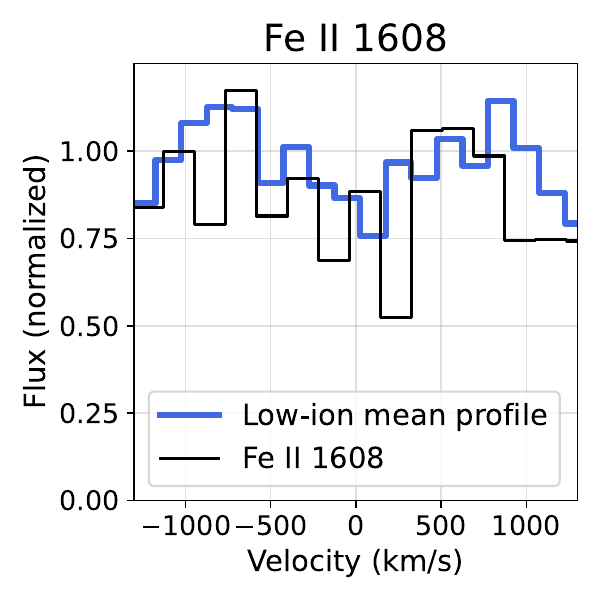}
        \includegraphics[width=0.49\linewidth]{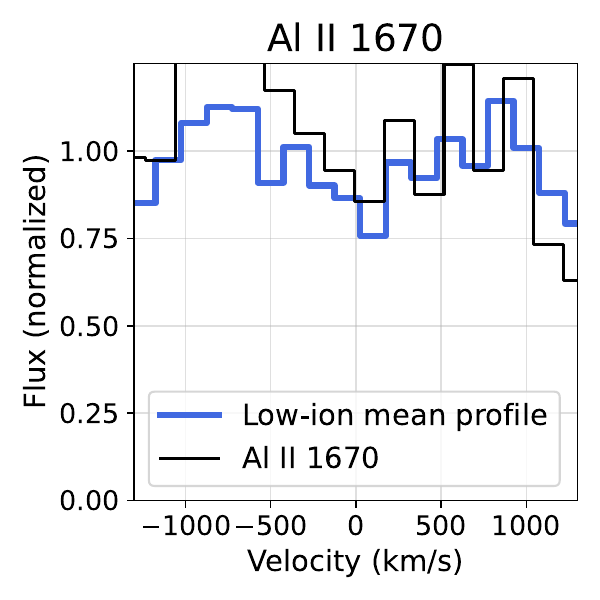}\\

        \includegraphics[width=0.49\linewidth]{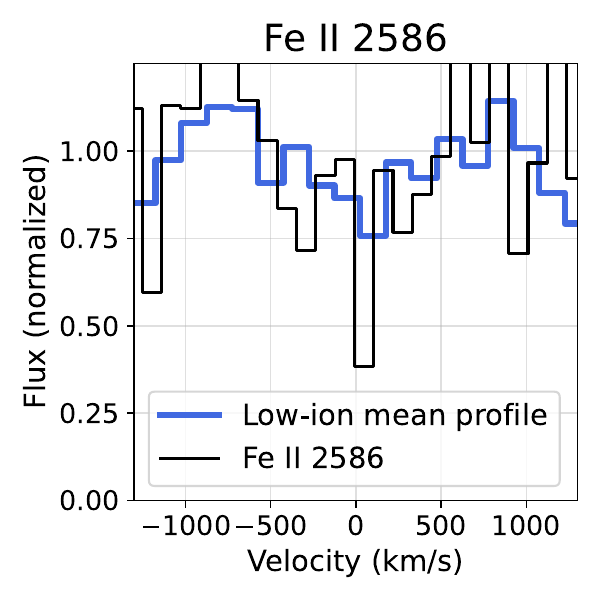}
        \includegraphics[width=0.49\linewidth]{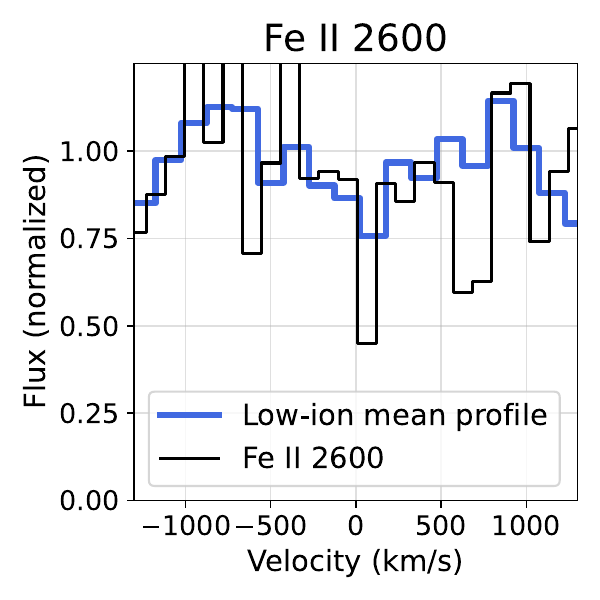}
    \end{minipage}

    \caption{Same as Figure~\ref{fig:ism-profiles_7} for SPURS-A2744-1069}
    \label{fig:ism-profiles_1069}
\end{figure*}

In this appendix, we provide additional information regarding the construction of the mean low- and high-ionization ISM absorption line profiles presented in Figures~\ref{fig:ism-profiles}, \ref{fig:low-ion-high-ion-comparison}.
Table~\ref{tab:lines-used} lists the low- and high-ionization absorption lines used to produce the mean absorption profile.
Figures~\ref{fig:ism-profiles_7} -- ~\ref{fig:ism-profiles_1069} plot the individual low- and high-ionization absorption profiles that were used to construct the respective mean absorption profiles. 
The upper panel of each Figure shows the G395M spectra of  [\ion{O}{3}]-$\lambda\lambda$4959,5007 emission line doublet, which is used to calculate each galaxy's systemic redshift (Section~\ref{subsec:redshift-determination}).

\begin{deluxetable*}{cccc} \label{tab:lines-used}
\tablecaption{Table of low and high-ionization absorption lines used to construct our mean ISM absorption line}
\tablehead{\colhead{objid} & \colhead{Low-ionization lines used} & \colhead{High-ionization lines used} & \colhead{Reference}}
\startdata
7 & \ion{Si}{2}-$\lambda$1260, \ion{O}{1}-$\lambda$1302, \ion{Si}{2}-$\lambda$1304, & \ion{Si}{4}-$\lambda\lambda$1393,1402, \ion{C}{4}-$\lambda\lambda$1548,50,  &  Figure~\ref{fig:ism-profiles_7}\\
  & \ion{C}{2}-$\lambda$1334, \ion{Si}{2}-$\lambda$1526, \ion{Al}{2}-$\lambda$1670  & \ion{Al}{3}-$\lambda\lambda$1854,62 & \\ 
17 & \ion{Si}{2}-$\lambda$1260, \ion{O}{1}-$\lambda$1302, \ion{Si}{2}-$\lambda$1304,  & \ion{Si}{4}-$\lambda\lambda$1393,1402 & Figure~\ref{fig:ism-profiles_17}\\
  &  \ion{C}{2}-$\lambda$1334, \ion{Al}{2}-$\lambda$1670 & \\ 
24 & \ion{Si}{2}-$\lambda$1260, \ion{O}{1}-$\lambda$1302, \ion{Si}{2}-$\lambda$1304,  & \ion{C}{4}-$\lambda\lambda$1548,50  & Figure~\ref{fig:ism-profiles_24}\\
  & \ion{C}{2}-$\lambda$1334, \ion{Si}{2}-$\lambda$1526, \ion{Al}{2}-$\lambda$1670  &  & \\ 
384 & \ion{Si}{2}-$\lambda$1526, \ion{Al}{2}-$\lambda$1670, \ion{Mg}{2}-$\lambda\lambda$2796,2803 & \ion{C}{4}-$\lambda\lambda$1548,50 &Figure~\ref{fig:ism-profiles_384}\\
439 & \ion{Fe}{2}-$\lambda$1608, \ion{Al}{2}-$\lambda$1670, \ion{Mg}{2}-$\lambda\lambda$2796,2803 & \ion{Al}{3}-$\lambda\lambda$1854,62 &Figure~\ref{fig:ism-profiles_439}\\
1069 & \ion{Al}{2}-$\lambda$1670, \ion{Fe}{2}-$\lambda$1608, \ion{Fe}{2}-2586, \ion{Fe}{2}-2600 & - & Figure~\ref{fig:ism-profiles_1069}\\
\enddata
\end{deluxetable*}

\end{document}